\documentclass[12pt]{article}
\usepackage{latexsym,amsmath,amssymb,epsfig,graphicx}
\newcommand{\be}{\begin{equation}}
\newcommand{\ee}{\end{equation}}
\newcommand{\bea}{\begin{eqnarray}}
\newcommand{\eea}{\end{eqnarray}}
\newcommand{\kma}{\; ,}
\newcommand{\pkt}{\; .}
\newcommand{\diag}{{\rm diag}}
\newcommand{\calm}{{\cal M}}
\newcommand{\calq}{{\cal Q}} 

\newcommand{\call}{{\cal L}}
\newcommand{\calf}{{\cal F}}

\newcommand{\cali}{{\cal I}}
\newcommand{\calj}{{\cal J}}

\newcommand{\bfk}{{\bf k}}
\newcommand{\bfq}{{\bf q}}
\newcommand{\bfx}{{\bf x}}

\newcommand{\tr}{{\rm Tr}}
\newcommand{\eqn}[1]{(\ref{#1})}

\begin{document}
\begin{titlepage}
\begin{flushright}
hep-th/yymmnn \\
June 2008
\end{flushright}
\vspace{8mm}
\begin{center}
{\Large \bf
One-loop corrections to the string tension of the vortex
in the Abelian Higgs model.}
\\\vspace{8mm}
{\large  J\"urgen Baacke\footnote{e-mail:~
juergen.baacke@tu-dortmund.de}} \\
{  Fachbereich Physik, Technische Universit\"at Dortmund \\
D - 44221 Dortmund, Germany
}\\
\vspace{4mm}
and\\
\vspace{4mm}
{\large  Nina Kevlishvili\footnote{e-mail:~
nkevli@fe.infn.it}} \\
{  Dipartimento di Fisica, Universita degli studi di Ferrara
\\ I-44100 Ferrara,
Italy\\
INFN, Sezione di Ferrara, I-44100 Ferrara, Italy \\
Andronikashvili Institute of Physics, GAS, 0177 Tbilisi, Georgia
}\\
\vspace{5mm}
\bf{Abstract}
\end{center}

We present an exact numerical computation of the one-loop
correction of the string tension for the Nielsen-Olesen vortex
in the Abelian Higgs model. The computations proceeds via
the computation of the Euclidean Green's function for the gauge, Higgs
and Faddeev-Popov fields using mode functions, and taking the 
appropriate trace.
Renormalization is an essential part of this computation, it is done by
removing leading order contributions from the numerical
results so as to make these finite, and to add the divergent
parts back, after suitable regularization and renormalization.
We encounter and solve some problems which are specific to
gauge theories and topological solutions. The corrections to the
energy are found to be sizeable, but still smaller than the classical
energy as long as $g^2$ is smaller than unity.

\end{titlepage}


\section{Introduction}
\label{intro}
Extended classical solutions that can be interpreted as strings
exist in various realistic and semi-realistic models of particle
physics \cite{Achucarro:1999it,Schaposnik:2006xt}. Their possible r\^ole
in cosmology has been evocated long ago
\cite{Kibble:1976sj}, see \cite{Vilenkin:1984ib,Hindmarsh:1994re}
for reviews.
In confrontation with the recent wealth of cosmological
observations they present a  very interesting
and  active, but still controversial field of research, see e.g. 
\cite{Vilenkin:2005jg,Sakellariadou:2006qs}.

In the present investigation we consider the string that 
is made up by the vortex solution of the Abelian Higgs model 
in $3+1$ dimensions, well-known from
superconductivity \cite{Abrikosov:1957}, and commonly denoted 
in particle physics as the
Nielsen-Olesen vortex \cite{Nielsen:1973cs}. The $1+1$ dimensional
version of this solution represents an instanton solution,
that has been widely considered in the context of baryon number violation.

There have been various investigations of the one-loop
correction to the string tension. The fermionic corrections to the
energy of the Nielsen-Olesen
vortex have been computed exactly in Ref. \cite{Bordag:2003at}. 
Fermionic corrections to strings in $2+1$ and $3+1$ dimensions have also 
been considered by various other authors 
\cite{Groves:1999ks,Bordag:2002sa,Vassilevich:2003xk,
Graham:2004jb}. 
Such calculations
may be important in the context of the instability
of the electroweak string \cite{Goodband:1995nr}

Here we consider the energy corrections 
arising from the gauge-Higgs and Faddeev-Popov sector. There are
some previous investigations, using heat kernel techniques
\cite{AlonsoIzquierdo:2007ds,Izquierdo:2005af}. Here we attempt an
exact computation, using techniques
 that have been developed previously in Refs. \cite{Baacke:1989sb,
Baacke:1991nh,Baacke:1991sa}
and applied in various semiclassical computations, of one-loop
energy corrections and one-loop prefactors to transition rates.
The computations have three essential ingredients: the use of mode
functions (``Jost functions'', see e.g., \cite{deAlfaro:1965}) 
in order to compute exact results,
the use of perturbative subtractions, so as to make these results finite,
and the computation of the subtracted parts using a covariant 
regularization and renormalization scheme. Similar approaches have been
used recently by other authors, see e.g. 
\cite{Bordag:1996fv,Graham:2002xq,Graham:2002fi,Graham:2002fw}.
In a gauge theory as considered here
one finds some complications due to the fact that there are cancellations of
divergences between graphs of a different number of vertices. This will
discussed in detail in the context of renormalization. 

Whether the energy corrections are small or big depends, 
in the present case, of the
gauge coupling. The classical string tension 
is proportional to $v^2=m_W^2/g^2$, where $g$ is
the gauge coupling, while the corrections are proportional
to $m_W^2$ multiplied by a function of
$\xi=m_H/m_W$. So the corrections are necessarily small 
relative to the classical
string tension if $g$ is sufficiently small. If ever the corrections are big
then this signals the breakdown of the semiclassical method. 
In intermediate situations one may have recourse to  a Hartree type
approximation by including the back-reaction of the quantum fluctuations 
to the classical solution. The methods we use here 
are suitable for such investigations
\cite{Baacke:2004xk,Baacke:2006kv}, 
or even for self-consistent calculations without classical
solution \cite{Baacke:1998nm}

The text is organized as follows:
In section \ref{basics} we present the model, the classical vortex solution
and the classical string tension.
In section \ref{flucsandtension} we relate
in general the 
fluctuation operator to the one-loop correction to the string tension,
and we explicitly derive the fluctuation operator.
Its partial wave reduction is presented in section \ref{partialwaves}.
This is the basis for the actual numerical computation of the 
one-loop string tension, which is described in section \ref{effectivetension}.
Renormalization is discussed in some detail in section \ref{Renorm}.
In section \ref{numerics} we give some details of the numerical
implementation and present the results. We conclude with a summary in
section \ref{summary}. Some technical details are discussed in
Appendices A-G.


\section{Basic relations}
\setcounter{equation}{0}
\par
\label{basics}
The Abelian Higgs model in (3+1) dimensions is
defined by the Lagrange density 
\begin{equation}
  {\cal L}=-\frac{1}{4}F_{\mu\nu}F^{\mu\nu}
+\frac{1}{2}(D_\mu\phi)^*D^\mu\phi-\frac{\lambda}
{4}\left(|\phi|^2-v^2\right)^2\; .  
\end{equation}
Here $\phi$ is a complex scalar field and
\bea
F_{\mu\nu}&=&\partial_\mu A_\nu-\partial_\nu A_\mu\kma \\
D_\mu&=&\partial_\mu-igA_\mu 
\pkt\eea 
The particle spectrum consists of Higgs bosons of mass
$m_H^2=2\lambda v^2$ and vector bosons of mass $m_W^2=g^2v^2$.
The model allows for vortex type solutions, representing
strings with a magnetic flux, the Nielsen-Olesen vortices \cite{
Abrikosov:1957,Nielsen:1973cs,deVega:1976mi,
Schaposnik:1978wq}.
The cylindrically symmetric
ansatz for this solution is given by 
\footnote{ We use Euclidean notation for the transverse components, so
$A^\perp_1\equiv A^1=-A_1$ etc. } 
\begin{eqnarray}
A_i^{\rm cl,\perp}(x,y,z)&=&\frac{\epsilon_{ij} x^\perp_j}
{gr^2}A(r) \;\;\;i=1,2 \kma\\
\phi^{cl}(x,y,z)&=&vf(r)e^{i\varphi(x)} \; \kma
\end{eqnarray}
where $r=\sqrt{x^2+y^2}$ and $\varphi$ is the polar angle. 
Furthermore $A_3^{\rm cl}=A_0^{\rm cl}=0$. 
In order to have a purely real Higgs field one performs a
gauge transformation
\bea \label{gaugetr}
\phi &\to& e^{-i\varphi}\phi \kma\\
 A^\perp_i&\to& A^\perp_i-\nabla^\perp_i\varphi/g 
\end{eqnarray}
to obtain the instanton fields in the singular gauge
\begin{eqnarray}
A^{{\rm cl}\perp}_i (x,y,z)&=&\frac{\varepsilon_{ij}x^\perp_j} 
{gr^2}\left[A(r)+1\right] \;\;\; i=1,2 \\
\phi^{cl}(x,y,z)&=&vf(r) \; .
\end{eqnarray}
With this ansatz the energy per unit length, or string
tension $\sigma$ takes the form
\begin{eqnarray}
\sigma_{cl}&=&\pi v^2
 \int^{\infty}_{0}\!\!\! dr \left\{\frac{1}{rm_W^2}\left[
\frac{dA(r)}{dr}\right]^2\!\!+r\left[\frac{df(r)}{dr}\right]^2\!\!
+\frac{f^2(r)}{r}\left[A(r)+1\right]^2\!\! \right. \nonumber \\ 
&+&\left. \frac{rm^2_H}{4}\left[
f^2(r)-1\right]^2\!\right\}\pkt
\end{eqnarray}
The magnetic flux is given by
\begin{eqnarray}
\Phi_M&=&\int d^2x \,B_3=-\int dx~dy F_{12} 
\pkt \end{eqnarray}
Explicitly we find
\be
\Phi_M=\int d^2 x (\nabla^\perp_1 A^{\rm cl \perp}_2-\nabla^\perp_2 
A^{\rm cl \perp}_1)=
\int d\phi r dr \left[\frac{-1}{gr}A'(r)\right]=\frac{2\pi}{g}\left[
A(0)-A(\infty)\right]
\pkt 
\ee
For the case $m_H=m_W$ an exact solution to the
variational equation is known \cite{deVega:1976mi},  for which
the classical string tension takes the value $\sigma_{cl}=\pi v^2$.
We here will consider the general case $m_W\neq m_H$, for which
the classical equations of motion
\begin{eqnarray}
\left\{\frac{\partial^2}{\partial r^2}+
\frac{1}{r}\frac{\partial}{\partial r}-\frac{\left[A(r)+1\right]^2}{r^2}-
\frac{m^2_H}{2}\left[f^2(r)-1\right]
\right\}f(r)&=&0 \kma \\
\left\{\frac{\partial^2}
{\partial r^2}-\frac{1}{r}
\frac{\partial}{\partial r}-m^2_W f^2(r)\right\}
\left[A(r)+1\right]&=&0
\end{eqnarray}
have to be solved numerically.
 
Imposing the boundary conditions on the profile functions
\begin{equation}  \label{rb}
\begin{array}{rcccccr}
A(r)&\stackrel{\scriptscriptstyle{r\to 0}}
{\longrightarrow}& const\cdot r^2                       
&,&A(r)&\stackrel{\scriptscriptstyle{r\to\infty}}
{\longrightarrow}&-1 \kma \\ 
f(r)&\stackrel{\scriptscriptstyle{r\to 0}}
{\longrightarrow}&const\cdot r 
&,&f(r)&\stackrel{\scriptscriptstyle{r\to\infty}}
{\longrightarrow}&1\kma 
\end{array}
\end{equation}
the magnetic flux  is $\Phi_M=2\pi/g$, the Dirac magnetic
flux quantum,  and the action is finite.

Since we have to consider fluctuations around these solutions
 a good numerical accuracy for the profile functions
$f(r)$ and $A(r)$ is required. As in previous publications
\cite{Baacke:1994bk,Baacke:2008zx} we have the method of Bais and Primack \cite{Bais:1975bq}.
The values for the classical string tension are
given in Table \ref{table:finalresults}.


\setcounter{equation}{0}
\section{Fluctuation operator and one-loop  string tension}
\label{flucsandtension}

The fluctuation operator is
defined in general form as
\begin{equation}
{\cal M} =  \frac{\delta^2S}{\delta \psi^*_i (x) \delta \psi_j (x')}
|_{\psi_k=\psi_k^{cl}}
\kma \end{equation}
where $\psi_i$ denotes the fluctuating fields and
$\psi_i^{cl}$ the ``classical'' background field
configuration; here these
will be the vortex and the vacuum configurations. If the fields
are expanded around the background configuration as
$\psi_i = \psi_i^{cl} + \phi_i$
and if the Lagrange density is expanded accordingly, then the
fluctuation operator is related to the second order Lagrange density
via
\begin{equation}
{\cal L}^{II} =  \frac{1}{2}  \phi^*_i {\cal M}_{ij}  \phi_j \; .
\end{equation}

In terms of the fluctuation operators
${\cal M}$ on the vortex and ${\cal M}^0$ on the
vacuum backgrounds, the effective action
is defined as
\begin{equation}
S_{eff} = \frac{i}{2} \ln \left\{ \frac{\det {\cal M}+i\epsilon}
{\det {\cal M}^0+i\epsilon} \right\} \; .
\end{equation}
As the background field is time-independent and also independent of
$z$ the fluctuation operators take the form
\be
\calm = -\partial_0^2+\partial_3^2 -\calm_\perp
\kma \ee
where $\calm_\perp$ is a positive-definite operator describing the transversal
fluctuations.
As is well known the  logarithm of the determinant can be written as the
trace  of the logarithm. One can do the trace over $p_0$, the
momentum associated with the time variable, by integrating over
$T\int dp_0/2\pi$, where $T$ is the lapse of time.
One then obtains
\begin{equation}
S_{\rm eff} =- i T \frac{1}{2} \sum \left[E_\alpha-E_\alpha^{(0)}\right]\kma
\end{equation}
where $E_\alpha$ are square roots of  the eigenvalues of the 
positive definite operator
\be
-\partial_3^2+\calm_\perp\kma
 \ee
and likewise $E_\alpha^{(0)}$ are those of the analogous operator
in the vacuum
\be
-\partial_3^2+\calm^0=-\partial_3^2-\vec \nabla_\perp^2 + {\bf m^2}
\pkt \ee
Here ${\bf m}^2=\diag(m_1^2,\dots,m_n^2)$ is the diagonal mass squared 
operator for the various fluctuations.

So the effective action becomes equal to the difference between 
the zero point energies of the
fluctuations around the vortex and in the vacuum, multiplied by $-T$.
We can also do the trace over the variable $p_3$
by integration over $L\int dp_3/2\pi$. We then obtain
\be
S_{\rm eff}=-i TL\sum_\alpha\int\frac{dk_3}{2\pi}
\frac{1}{2}
\left[\sqrt{k_3^2+\mu_\alpha^2}-\sqrt{k_3^2+\mu_\alpha^{(0)~ 2}}\right]
\kma \ee
where $\mu_\alpha^2$ are the eigenvalues of the operator $\calm^\perp$ and
$\mu_\alpha^{(0)~ 2}$ those of $-\vec \nabla_\perp^2+{\bf m}^2$. 
In the same way the classical action becomes
\be
S_{\rm cl}=-TL\sigma_{\rm cl}
\ee
where $\sigma_{\rm cl}$ is the classical string tension.
So our goal reduces to computing the
one-loop approximation to the string tension given by
\be
\sigma_{\rm 1-loop}=\sigma_{\rm cl}+\sigma_{\rm fl}
\kma \ee
where the fluctuation part of the string tension is 
given by
\be \label{fluctuationstringtension}
\sigma_{\rm fl}=\sum_\alpha\int\frac{dk_3}{2\pi}
\frac{1}{2}\left[\sqrt{k_3^2+\mu_\alpha^2}-\sqrt{k_3^2+\mu_\alpha^{(0)~ 2}}\right]
\pkt \ee
Of course all expressions are formal, the integrals do not exist before
a suitable regularization. Anyway we do not plan to compute any 
eigenvalues of the fluctuation operators but will reduce these
expressions to traces over  Euclidean Green' s functions, where
renormalization will be done properly. However the formal identities
will allow us to trace the way in which the counter terms in the
original Lagrangean enter the final expressions that are going to be
computed numerically.
 
The fluctuation operator has been derived previously
\cite{Baacke:1994bk}, in the
context of quantum corrections to the Abelian instanton; we 
here recall this derivation. The gauge and Higgs fields 
are expanded as
\begin{eqnarray}
A^\mu&=&A^\mu_{\rm cl}+a^\mu \kma  \\ 
\phi&=&\phi^{cl}+\varphi \; .
\end{eqnarray}
In the following we will drop the superscript 
${\rm cl}$, so the letters $A^\mu$ and
$\phi$ will denote the background field, and $a^\mu$ and $\varphi$ the 
quantum fluctuations.

In order to eliminate the gauge degrees of
freedom we introduce, as in Ref. \cite{Kripfganz:1989vm},
 the background gauge function
\begin{equation}
{\cal F}(a)=\partial_\mu a^\mu-\frac{ig}{2}\left(\phi
^\ast\varphi-\phi\varphi^\ast\right)
\pkt\end{equation}
We note that for the background field $\partial^\mu A_\mu=0$.
In the Feynman background gauge we get the gauge-fixing
Lagrange density
\begin{eqnarray}
{\cal L}_{GF}^{I\hspace{-.05cm}I}&=
&-\left(\frac{1}{2}{\cal F}^2(a)\right)
^{I\hspace{-.05cm}I} \nonumber \\ 
&=&-\frac{1}{2}(\partial_\mu a^\mu)^2
-\frac{ig}{2}a^\mu(\varphi\partial_\mu\phi
+\phi\partial_\mu\varphi
-\varphi^\ast\partial_\mu\phi-\phi
\partial_\mu\varphi^\ast) \\
&&+\frac{g^2}{8}\phi^2(\varphi-\varphi^\ast)^2  \; .
\nonumber \end{eqnarray}
The associated Faddeev-Popov Lagrangean becomes
\begin{equation}
{\cal L}_{FP} =\frac{1}{2} \eta^\ast(-\partial^2- g^2
\phi^2)\eta \; .
\end{equation}
In terms of the real components $\varphi= \varphi_1+i\varphi_2$
and $\eta = (\eta_1 + i \eta_2)$ the second order Lagrange
density now becomes
\begin{eqnarray} \label{lag2}
\left({\cal L}+{\cal L}_{GF}+{\cal L}_{FP}\right)^
{I\hspace{-0.05cm}I}&=&
-a_\mu\frac{1}{2}\left(-\Box+g^2\phi^2\right)
a^\mu\nonumber \\
&&+\varphi_1\frac{1}{2}\left[-\Box+g^2A_\mu A^\mu-
\lambda\left(3\phi^2-
v^2\right)\right]\varphi_1  \nonumber \\  
&&+\,\varphi_2\frac{1}{2}\left[-\Box+g^2A_\mu A^\mu-g^2\phi^2-
\lambda\left(\phi^2-v^2\right)\right]\varphi_2 \nonumber \\
&&+\,\varphi_2(gA^\mu\partial_\mu)\varphi_1
+\varphi_1(-gA^\mu\partial_\mu)\varphi_2  \\
&&+\,a^\mu(2g^2A_\mu\phi)\varphi_1
+\,a^\mu(2g\partial_\mu\phi)\varphi_2 \nonumber \\
&&+\eta_1\frac{1}{2}\left(-\Box- g^2\phi^2\right)\eta_1
+\eta_2\frac{1}{2}\left(-\Box-g^2\phi^2\right)\eta_2
\nonumber  
\kma\end{eqnarray}
where we have omitted the superscript from $\phi^{cl}$ and
$A_\mu^{cl}$).
We now specify the fluctuating fields
$ (\psi_1,\psi_2,\psi_3,\psi_4,\psi_5)$ as
$ (a^\perp_1,a^\perp_2,\varphi_1,\varphi_2,\eta_{12})$,
\be
\left(\begin{array}{l}\psi_1\\\psi_2\\\psi_3\\\psi_4\\\psi_5
\end{array}\right)=\left(\begin{array}{l}a^\perp_1\\a^\perp_2\\
\varphi_1\\\varphi_2\\\eta_{12}\end{array}\right)
\kma\ee
 where we have used
Euclidean notation for the transverse gauge field components.
We furthermore write (see also above)
\be
\calm =-\partial_0^2+\partial_3^2-\calm^\perp
\kma \ee
separating the trivial part from the one that is modified by the
background field.
With  these preliminaries we obtain the following
nonvanishing components of the overall fluctuation
operator $\calm^\perp_{ij}$:

\bigskip
\be\begin{array}{rcl@{\qquad}rcl}
{\cal M^\perp}_{11}&=&\displaystyle -\Delta^\perp + g^2 \phi^2&
{\cal M^\perp}_{22}&=&\displaystyle -\Delta^\perp + g^2 \phi^2 \\
{\cal M^\perp}_{13}&=
& 2 g^2 A^\perp_1 \phi&{\cal M^\perp}_{14}&=& 2 g \nabla_1 \phi \\
{\cal M^\perp}_{23}&=
& 2 g^2 A^\perp_2 \phi&{\cal M^\perp}_{24}&=& 2 g \nabla_2 \phi \\
{\cal M^\perp}_{33}&=&\displaystyle -\Delta^\perp+g^2{\bf A^\perp}^2 +
g^2 \phi^2
+\lambda (\phi^2-v^2)&
{\cal M^\perp}_{34}&=&\displaystyle -g {\bf A^\perp}\cdot {\bf \nabla}  \\
{\cal M^\perp}_{44}&=&\displaystyle -\Delta^\perp + g^2 {\bf A^\perp}^2 
+\lambda(3\phi^2-v^2)& {\cal M^\perp}_{43}&=& g {\bf A^\perp} \cdot 
{\bf \nabla} \\
{\cal M^\perp}_{55}&=&-\Delta^\perp+g^2 \phi^2 \; .&
&&
\end{array}\ee 
\bigskip

As discussed above it is understood that the contribution 
of the Faddeev-Popov
operator ${\cal M}_{55}$ enters with  a factor 
$-2$ into the definition of the one-loop string tension.
The fluctuation operators for the vortex and vacuum
background are now obtained by substituting the corresponding
classical fields.
The vacuum fluctuation operator becomes a diagonal
matrix of Klein-Gordon operators with masses
$(m_W,m_W,m_W,m_H,m_W)$. It is convenient
to introduce a potential $\cal V$ via
\begin{equation}
{\cal M} = {\cal M}^0 + {\cal V} \; .
\end{equation}
This potential will be specified below after partial
wave decomposition.


\setcounter{equation}{0}
\section{Partial wave decomposition}
\par
\label{partialwaves}
The fluctuation operator ${\cal M^\perp}$ can be decomposed 
into partial waves with respect to the polar angle $\varphi$,
and the string tension decomposes accordingly.
We introduce the following partial wave decomposition for
fields 
\begin{eqnarray}
\nonumber
\vec{a}&=&\sum_{n=-\infty}^{+\infty}
b_n(r)\left(\begin{array}{c}\cos\varphi\\
\sin\varphi
\end{array}\right)\frac{e^{in\varphi}}{\sqrt{2\pi}}+ic_n(r)\left(
\begin{array}{c}-\sin\varphi\\
\cos\varphi\end{array}\right)
\frac{e^{in\varphi}}{\sqrt{2\pi}} \kma \\
\nonumber\varphi_1&=&\sum_{n=-\infty}^{+\infty}
h_n(r)\frac{e^{in\varphi}}{\sqrt{2\pi}}\kma  \\
\varphi_2&=&\sum_{n=-\infty}^{+\infty}
\tilde{h}_n(r)\frac{e^{in\varphi}}{\sqrt{2\pi}}\kma  \\
\nonumber\eta_{12}&=
&\sum_{n=-\infty}^{+\infty} g_n(r)\frac{e^{in\varphi}}{\sqrt{2\pi}}
\; .
\end{eqnarray}

After inserting these expressions into the
Lagrange density and using the reality conditions for
the fields one finds that the following combinations are
real relative to each other and make the fluctuation operators
symmetric:
\begin{eqnarray}
\nonumber F^n_1(r)&=&\frac{1}{2}(b_n(r)+c_n(r))\kma  \\
\nonumber F^n_2(r)&=&\frac{1}{2}(b_n(r)-c_n(r)) \kma \\
F^n_3(r)&=&\tilde{h}_n(r) \kma \\
\nonumber F^n_4(r)&=&ih_n(r)  \kma\\
\nonumber F^n_5(r)&=&g_n(r)\pkt
\end{eqnarray}
Writing the partial fluctuation operators as
\begin{equation}
{\bf M^\perp} = {\bf M^\perp}_0 + {\bf V} \; ,
\end{equation}
the free operators ${\bf M^\perp}_0$ become diagonal matrices with
elements
\begin{equation}
M^\perp_{0,ii}= -\frac{d^2}{dr^2}-\frac{1}{r}\frac{d}{dr}
+\frac{n_i^2}{r^2} + m_i^2
\kma\end{equation}
where $(n_i) = (n-1,n+1,n,n,n)$ and $(m_i)=(m_W,m_W,m_W,m_H,m_W)$.
The potential ${\bf V}$ takes the elements

\bigskip
\be
\begin{array}{rcl@{\qquad}rcl}
{\bf V}_{11}^n&=&m_W^2\left(f^2-1\right) &{\bf V}_{12}^n&=&0\\
{\bf V}_{13}^n&=&\sqrt{2}m_W^{}f'&
{\bf V}_{14}^n&=&\displaystyle\sqrt{2}m_W^{}f
\frac{A+1}{r}\\
{\bf V}_{22}^n&=&{\bf V}_{11}^n&{\bf V}_{23}^n&=&{\bf V}_{13}^n\\
{\bf V}_{24}^n&=&-{\bf V}_{14}^n&
{\bf V}_{33}^n&=&\displaystyle
\frac{(A+1)^2}{r^2}+\left(\frac{m_H^2}{2}
+m_W^2\right)\left(f^2-1\right)\\
{\bf V}_{34}^n&=&
\displaystyle -2\frac{A+1}{r^2}n&{\bf V}_{44}^n&=&
\displaystyle\frac{(A+1)^2}
{r^2}+\frac{3}{2}m_H^2\left(f^2-1\right)  \\
{\bf V}_{55}^n&=&m_W^2\left(f^2-1\right)&{\bf V}_{i5}&=&0 \; .\\
\end{array}\ee
\bigskip 

Choosing the dimensionless variable $m_Wr$ one realizes that
the fluctuation operator only depends on the ratio
$m_H/m_W$ up to an overall factor $m_W^2$ which cancels in
the ratio with the free operator.

\setcounter{equation}{0}
\section{Computation of the effective string tension}
\label{effectivetension}

The method for computing the effective string tension used here
is based on  the Euclidean Green's function of the
fluctuation operator. This Green's function is defined by
\begin{equation}
({\nu^2+k_3^2+\cal M^\perp})~ {\cal G}(\vec x_\perp,\vec x_\perp',k_3,\nu) =
 {\bf 1}\delta(\vec x_\perp -\vec x_\perp').
\end{equation}
and similarly for the operator ${\cal M}^0$.
It contains the information on
the eigenvalues $\lambda_\alpha^2$ of the
fluctuation operator $\calm^\perp$ via
\begin{equation}
\int d^2x^\perp {\rm Tr}~ {\cal G}(\vec x^\perp,\vec x^\perp,k_3, \nu) =
\sum_\alpha \frac{1}{\lambda_\alpha^2 + k_3^2+\nu^2}.
\end{equation}
We define a function $F(k_3,\nu)$ as
\begin{equation}
\label{fk3nudef}
F(k_3,\nu) = \int d^2x^\perp~ 
{\rm Tr}~ ({\cal G}(\vec x^\perp,\vec x^\perp,\nu)-
{\cal G}^0(\vec x^\perp,\vec x^\perp, \nu))
\pkt\end{equation}
We then  find
\begin{equation}
-\int_{-\infty}^{\infty}\frac{d\nu \nu^2}{2\pi} F(k_3,\nu) =
\sum_\alpha\frac{1}{2}\left[\sqrt{k_3^2+\lambda_\alpha^2}
-\sqrt{k_3^2+(\lambda^{(0)}_\alpha)^2}\right]
\pkt \end{equation}
This expression is still to be integrated over $k_3$ and is by itself already
linearly divergent. The sum over eigenvalues becomes an integral 
$\int d^2 k^\perp$ and the difference of the energies behaves
asymptotically as $1/|k^\perp|$. So regularization is required. This will
be dicussed in the next section. Assuming that it has been achieved, we can do the
$\nu$ and $k_3$ integrations at once, using the fact that $F(k_3,\nu)$ only depends on $p=\sqrt{k_3^2+\nu^2}$, i.e., $F(k_3,\nu)=F(0,p)$. We then obtain
for the one-loop string tension
\be
\sigma_{\rm fl}=-\int_0^\infty\frac{dp  p^3}{4\pi}F(0,p)
\pkt\ee 

After these more formal considerations we will  present the
way in which we actually  compute $F(k_3,\nu)$. We first use the
partial wave decomposition to write
\begin{equation}          \label{fpdef}
F(k_3,\nu) \equiv F(p) = \sum_{n=-\infty}^{+\infty} F_n(p)
\kma \end{equation}
where
\begin{equation} \label{fnpdef}
F_n(p) = \int dr r {\rm Tr}({\bf G}_n(r,r,p)-
{\bf G}_n^0(r,r,p)) \; ,
\end{equation}
and where the partial wave Green functions are defined  by
\begin{equation} \label{greendgl}
({\bf M}_n +p^2) {\bf G}_n (r,r',p) ={\bf 1} \frac{1}{r}
\delta(r-r')  \; .
\end{equation}
For ${\bf M}_n^0$ the Green function is simply a
diagonal matrix with elements
\begin{equation}
{\bf G}^0_{n~ii}(r,r',p)
= I_{n_i}(\kappa_i r_<)K_{n_i}(\kappa_i r_>)
\kma \end{equation}
where $\kappa_i = \sqrt{m_i^2 + p^2}$. For the Green function
of the operator
${\bf M}_n$ the matrix elements similarly become
\begin{equation}
{\bf G}_{n~ij}(r,r',p) = f_{ni}^{\alpha-}(r_<)f_{nj}^{\alpha+}(r_>)
\end{equation}
where the functions $f_{ni}^{\alpha\pm}$ form a fundamental
system of linearly independent solutions of \eqn{greendgl}, regular as
$r \to 0$ for the minus sign and as $r \to \infty$
for the plus sign.
The correct normalization is obtained by imposing the
boundary conditions
\begin{eqnarray}
f_{ni}^{\alpha-}(r) &\simeq&  
\delta_i^{\alpha}I_{n_i} (\kappa_i r) \nonumber\kma \\
f_{ni}^{\alpha+}(r) &\simeq& \delta_i^\alpha K_{n_i} (\kappa_i r)
\kma \end{eqnarray}
as $r \to \infty$. Actually we have solved numerically
the differential equations for the
functions $h_i^{\alpha\pm}$ defined by
\begin{equation}
f_{ni}^{\alpha \pm} = B^\pm_{n_i}(\kappa_ir)
(\delta_i^{\alpha\pm} + h_{ni}^{\alpha\pm}(r))
\end{equation}
where $B^+_{n_i}=K_{n_i}$ and  $B^-_{n_i}=I_{n_i}$
   are the appropriate  Bessel functions, and with the boundary conditions
$h_{n~i}^{\alpha\pm}\to 0$ as $r\to \infty$. In this
way one keeps track of the free contribution 
$ \propto \delta_i^\alpha$ 
and
\begin{eqnarray}
&{\rm Tr}&\left[{\bf G}_n(r,r,\nu)-{\bf G}_n^0(r,r,\nu)\right]
\nonumber \\\label{Gcomp}
&=& \left[h_{ni}^{i-}(r)+h_{ni}^{i+}(r)
+h_{ni}^{\alpha-}(r)h_{ni}^{\alpha+}(r)\right]I_{n_i}(\kappa_i r)
K_{n_i}(\kappa_i r) \; ,
\end{eqnarray}
to be inserted into \eqn{fpdef}.


\setcounter{equation}{0}
\section{Renormalization}            \label{Renorm}
\par
Our numerical results are computed in such a way that we first compute
the contributions of the subsystems for fixed $n$, then sum over $n$
and finally integrate over $p$. The subtractions necessary to make
the summation over $n$ and the integration over $p$ convergent
are done in the partial waves. There are two ways of doing this:
either one plainly subtracts in the partial waves
the first order contributions 
 $\tr G_n^{(1)}=-\tr G_n^0 V_{ii}G_n^0$ and 
the second order contributions  
$\tr G_n^{(2)}=\tr G_n^0 V_{ij}G_n^0V_{ji}G_n^0$, or one does the subtractions
already in the functions $h_{n,i}^{\alpha\pm}$ as described in detail
in Refs. \cite{Baacke:1991sa}. For the Faddeev-Popov sector we have used
both methods and the results agree very well. In the gauge-Higgs sector
the singularity of the external gauge field at $r=0$ leads to specific
difficulties, to be discussed below. These are easier to handle
with the first method,  and so we decided to work with the
plain subtractions throughout. 

The essential problems occuring here are twofold: (i) the external
gauge field is not square integrable, due to its singularity at $r=0$
and (ii) there are cancellations between graphs of different order
in the external vertices, the well-known cancellation of the
quadratic divergence in the vacuum polarization involves seagull
terms with one vertex and graphs with two vertices. Such cancellations
occur in higher orders as well, so one has to be careful, some
contributions with two external vertices do not have to be subtracted,
because their divergences are cancelled in higher order. 

In the following we will treat the various contributions of first
and second order term by term, comparing the contributions which are
subtracted with the corresponding Feynman graphs. 
The subtracted contributions are then added back in covariant form.
In this way we preserve covariance, which would be violated if we
would introduce noncovariant cutoffs (e.g. in $p$ ).

\begin{figure}[htbp]
\parbox[c]{6cm}{
 \centering
\includegraphics[scale=0.22,angle=0]{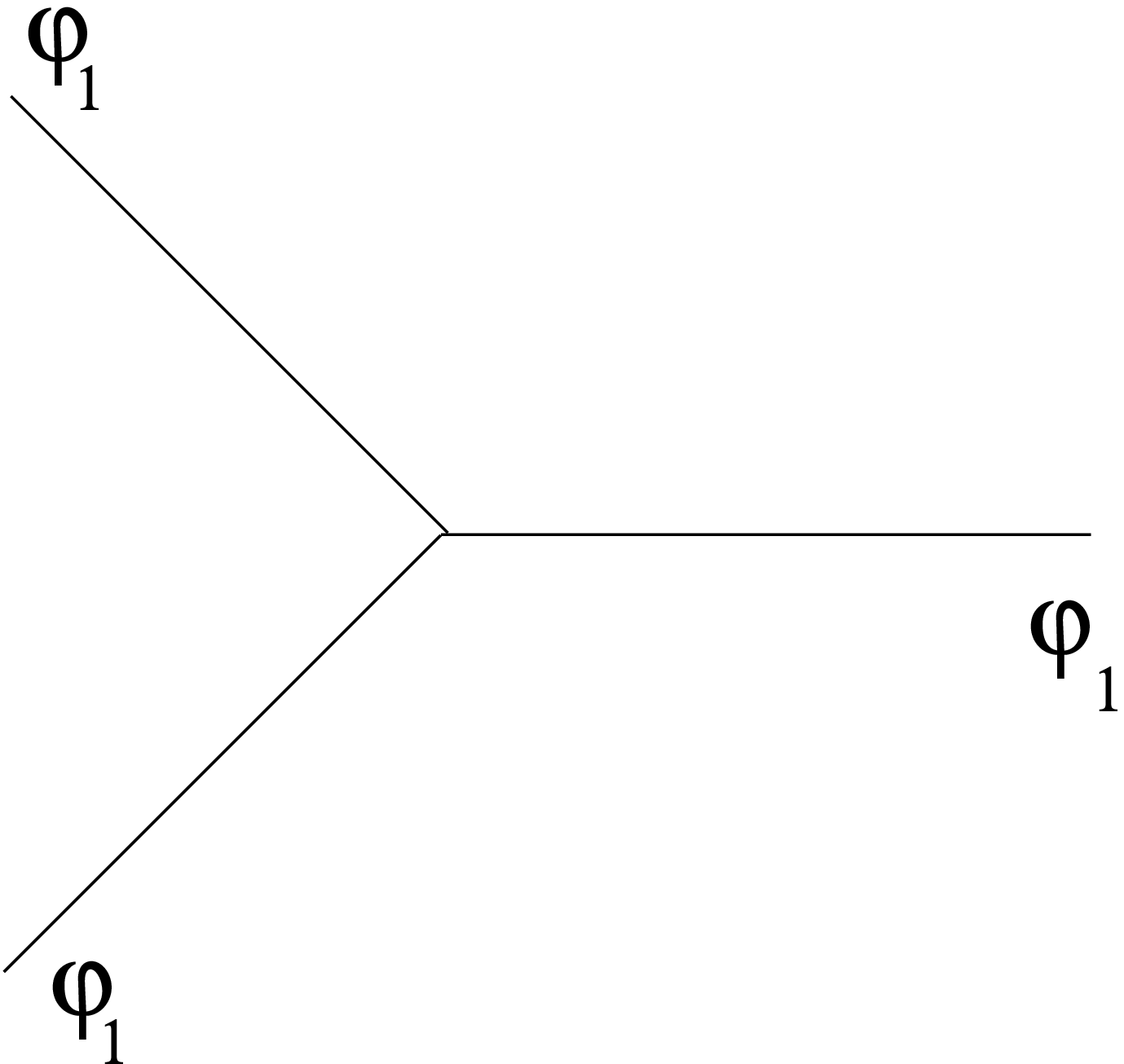}
\\
$-i\lambda v$}
\parbox[c]{6cm}{
 \centering
\includegraphics[scale=0.22,angle=0]{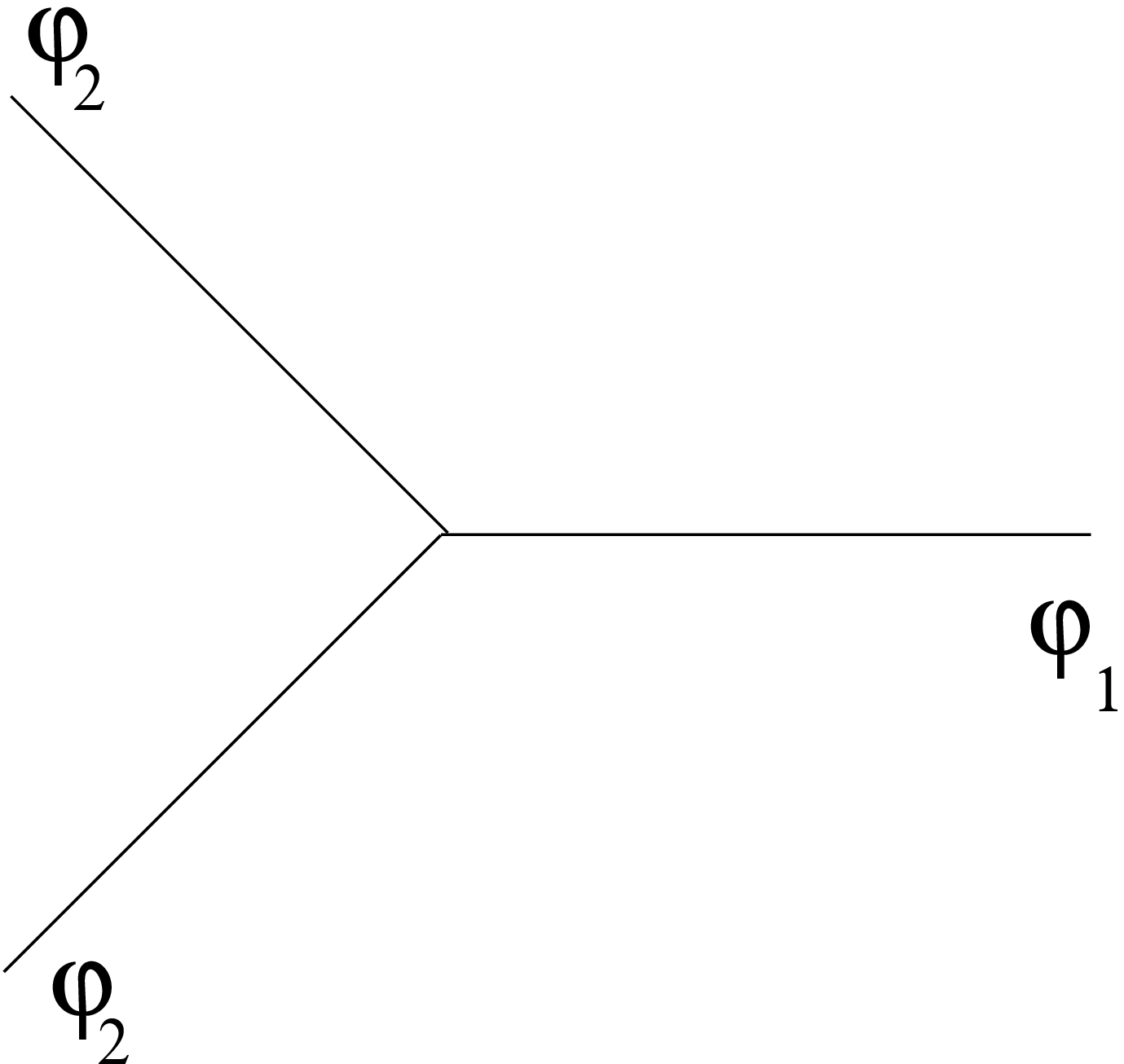}
\\
 $-i\lambda v$}
\\
\parbox[c]{6cm}{
 \centering
\includegraphics[scale=0.22,angle=0]{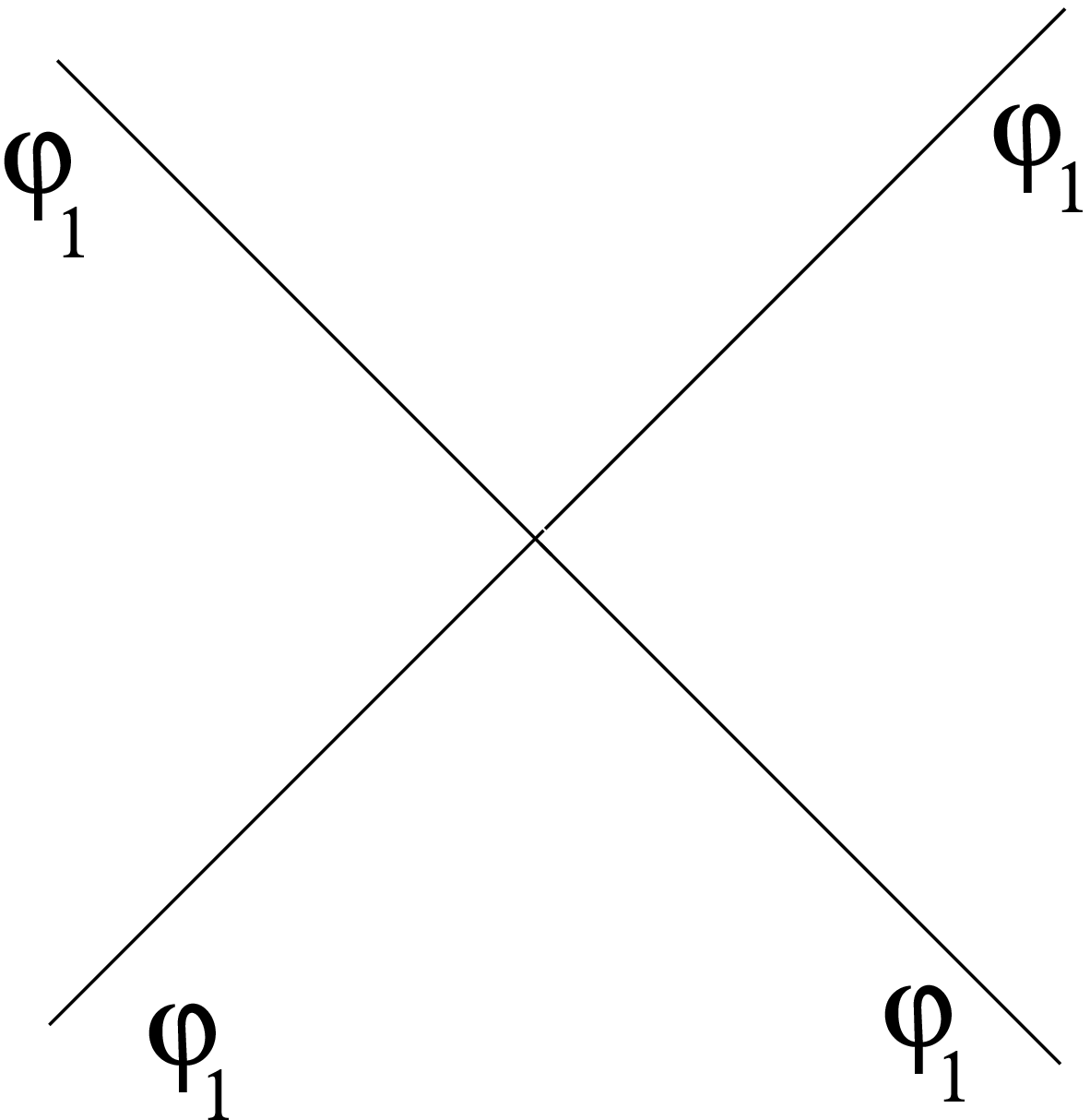}
\\
$-i\frac{\lambda}{4}$}
\parbox[c]{6cm}{
 \centering
\includegraphics[scale=0.22,angle=0]{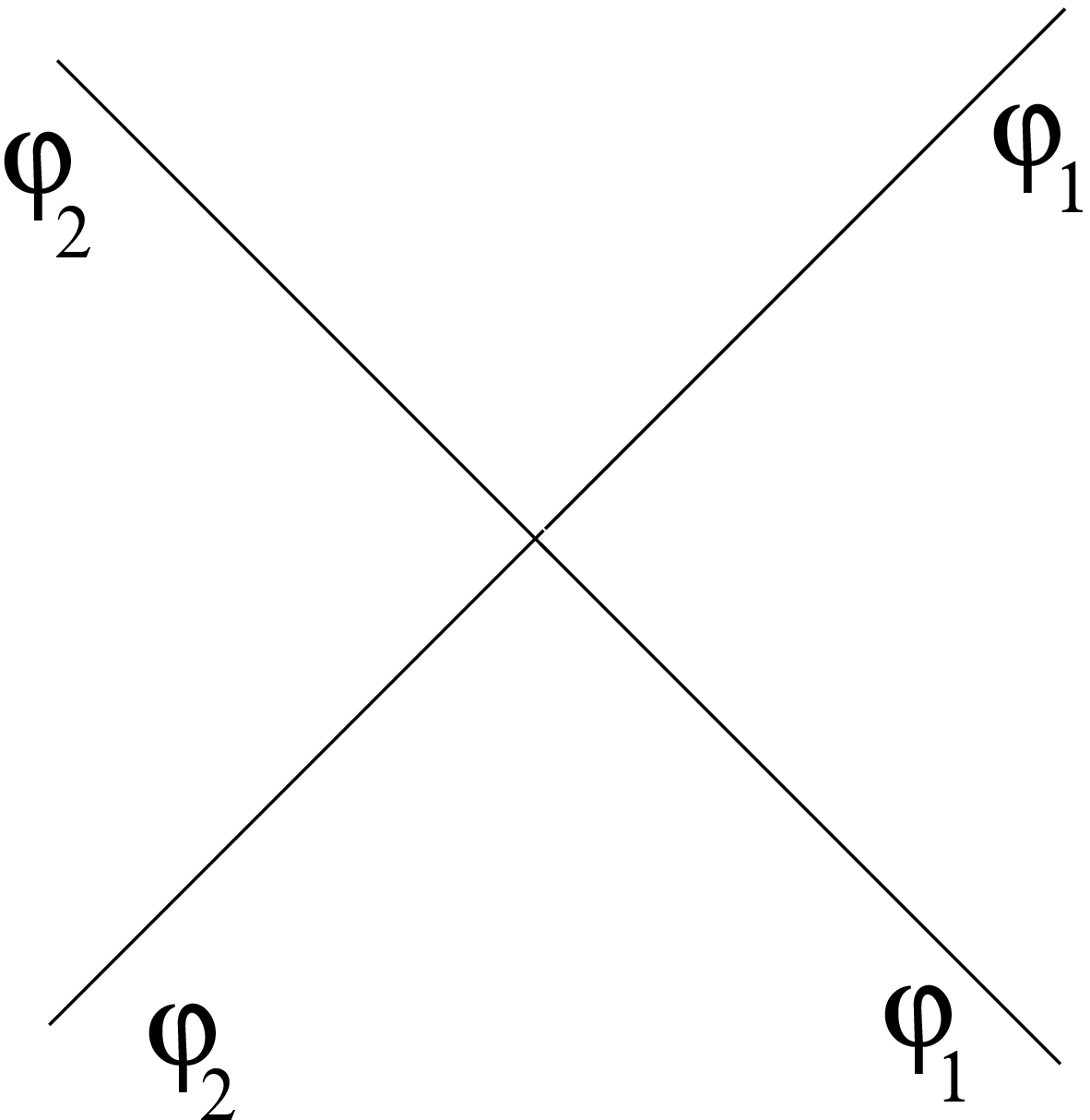}
\\
$-i\frac{\lambda}{2}$}
\vspace{4mm}\\
\parbox[c]{6cm}{
 \centering
\includegraphics[scale=0.25,angle=0]{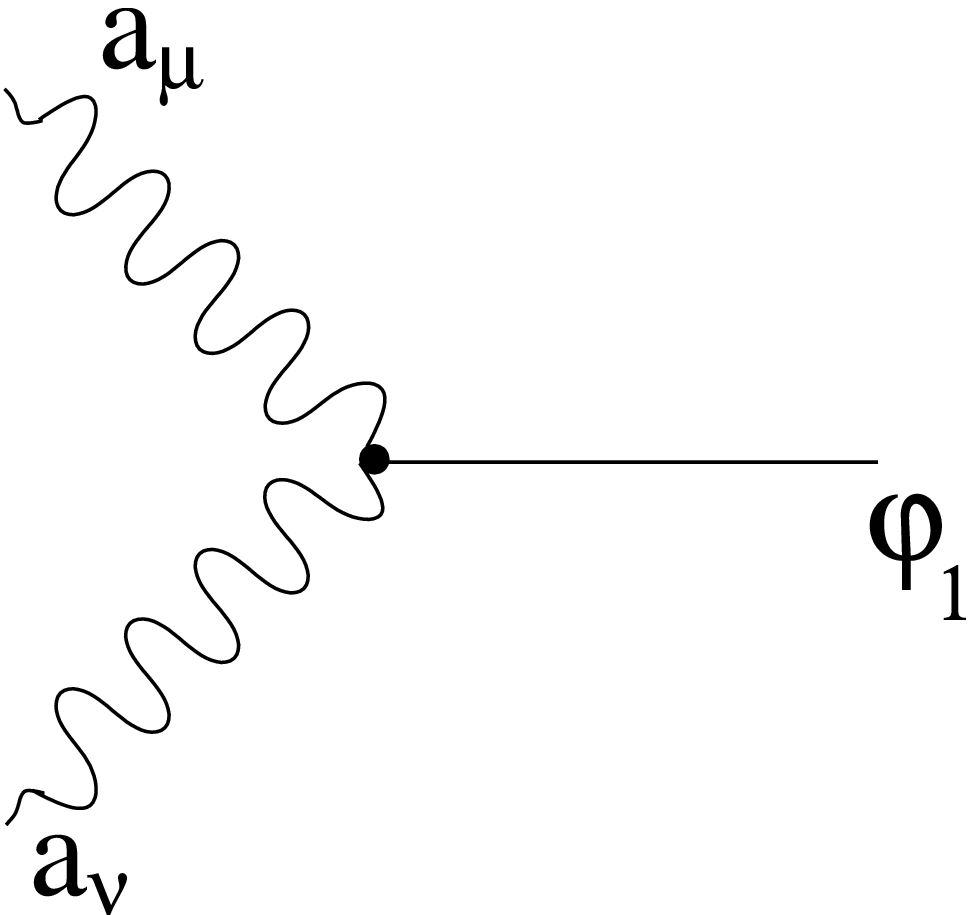}
\\
 $ig^2vg_{\mu\nu}$}
\parbox[c]{6cm}{
 \centering
\includegraphics[scale=0.25,angle=0]{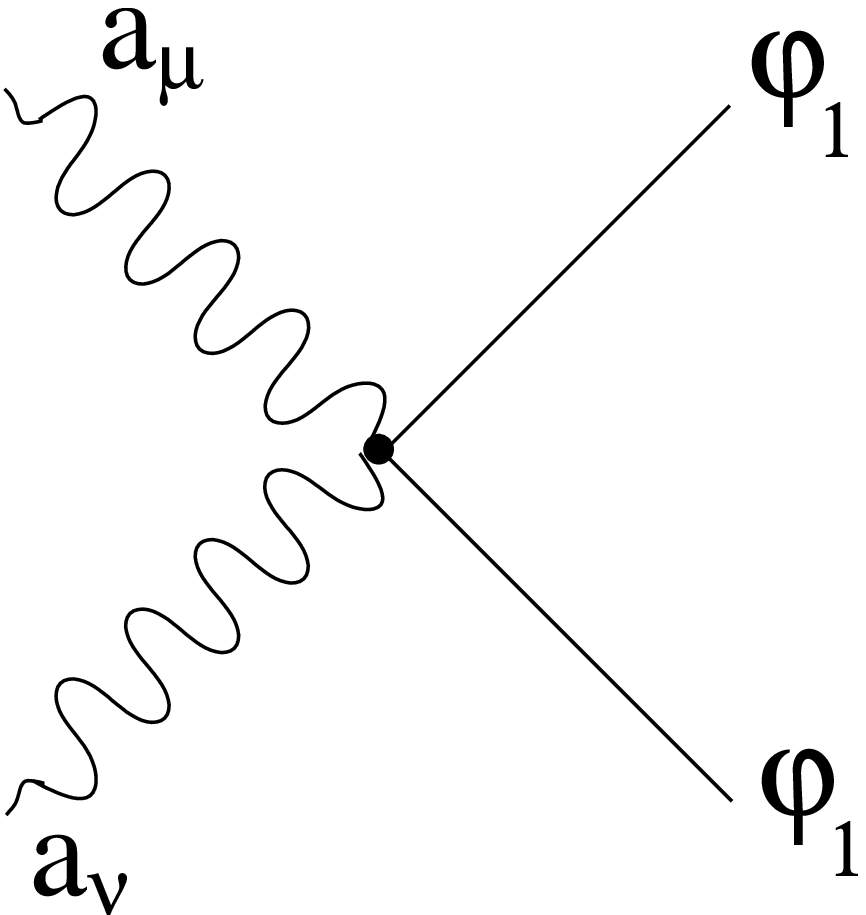}
\\
 $\frac{i}{2}g^2g_{\mu\nu}$}
\vspace{4mm}\\
\parbox[c]{6cm}{
 \centering
\includegraphics[scale=0.25,angle=0]{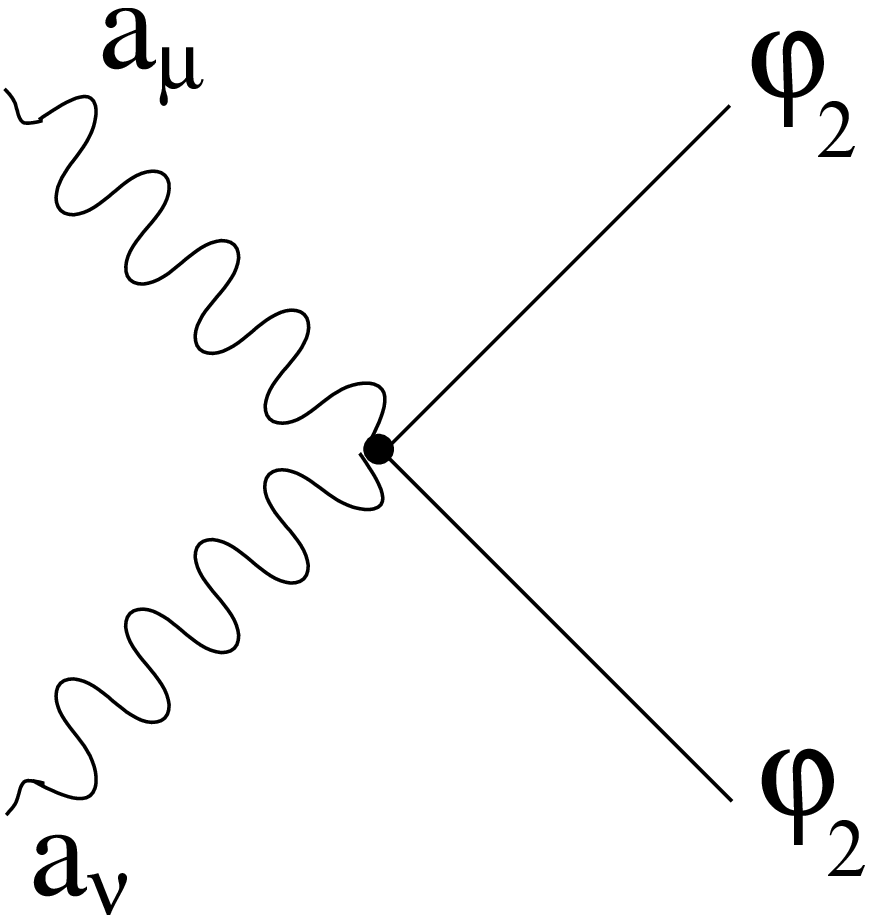}
\\
$\frac{i}{2}g^2g_{\mu\nu}$}
\parbox[c]{6cm}{
 \centering
\includegraphics[scale=0.25,angle=0]{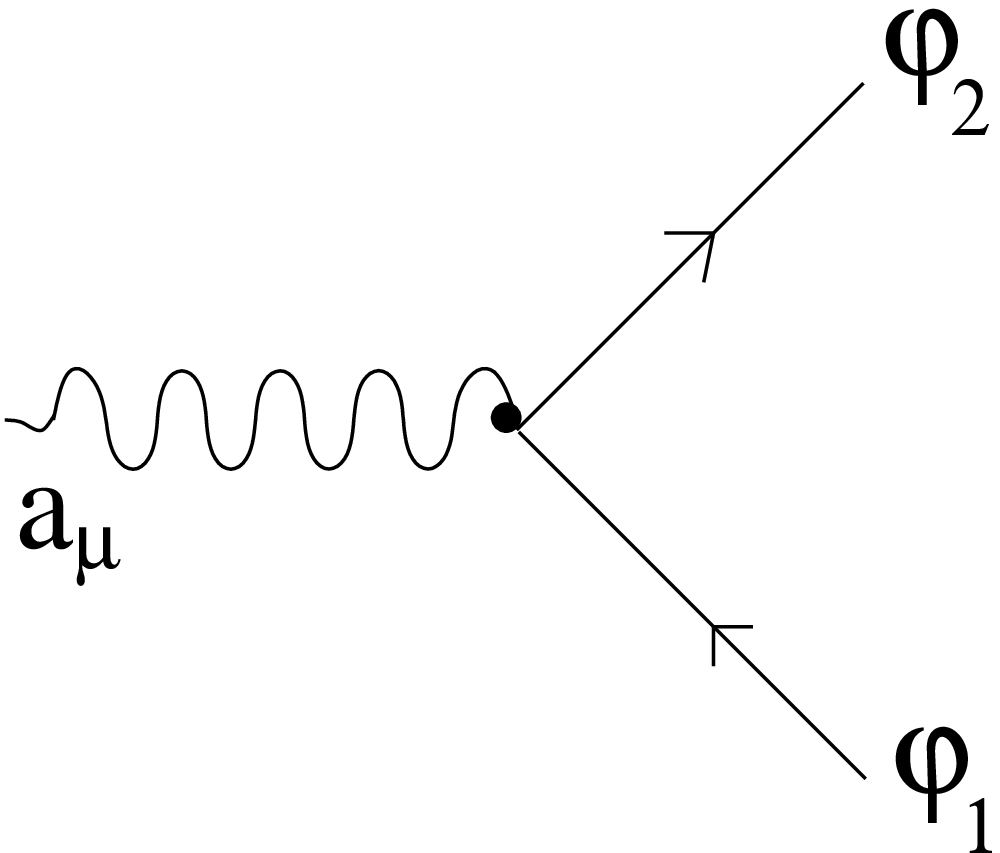}
\\
$2g p_1$}
\\\vspace{4mm}
\parbox[c]{6cm}{
 \centering
\includegraphics[scale=0.25,angle=0]{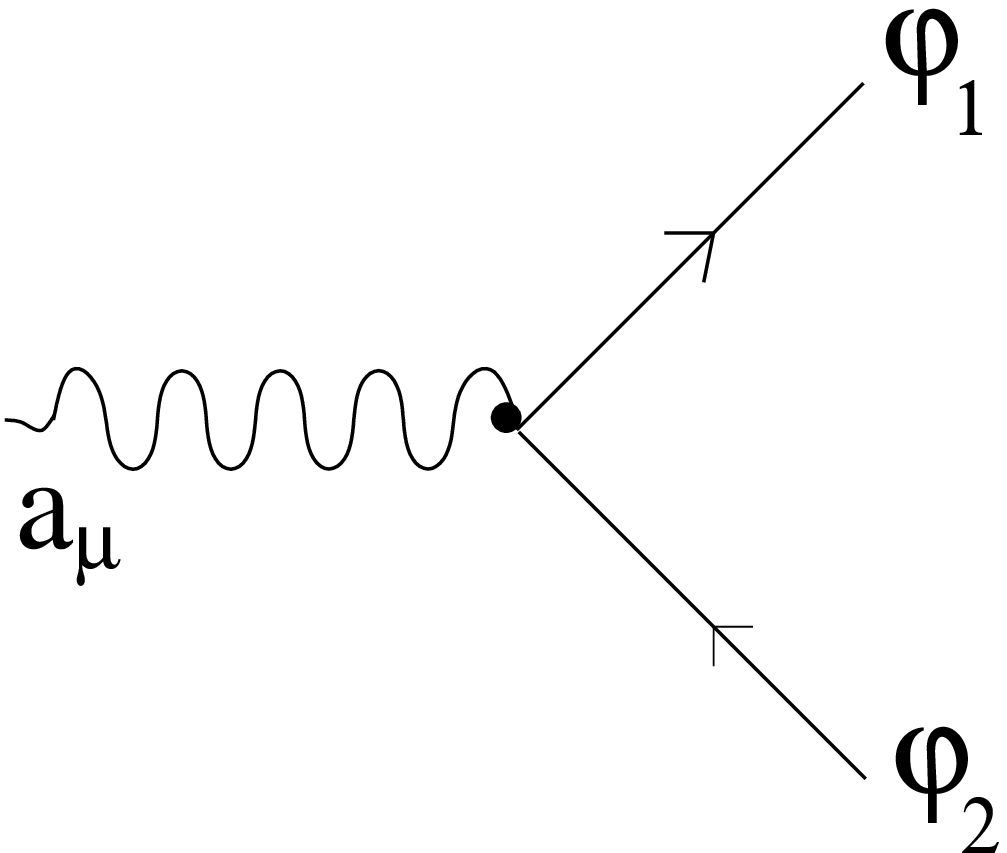}
\vspace{2mm}\\
$-2g p_1$}
\caption{ \label{fig:feynmanrules_gH}
Vertices for Feynman graphs derived from the Lagrangean
\eqn{L2vac}: gauge-Higgs sector. 
We have not included combinatorial factors for permutations
of the external lines.}
\end{figure}

\begin{figure}[htbp]
\parbox[c]{6cm}{
 \centering
\includegraphics[scale=0.22,angle=0]{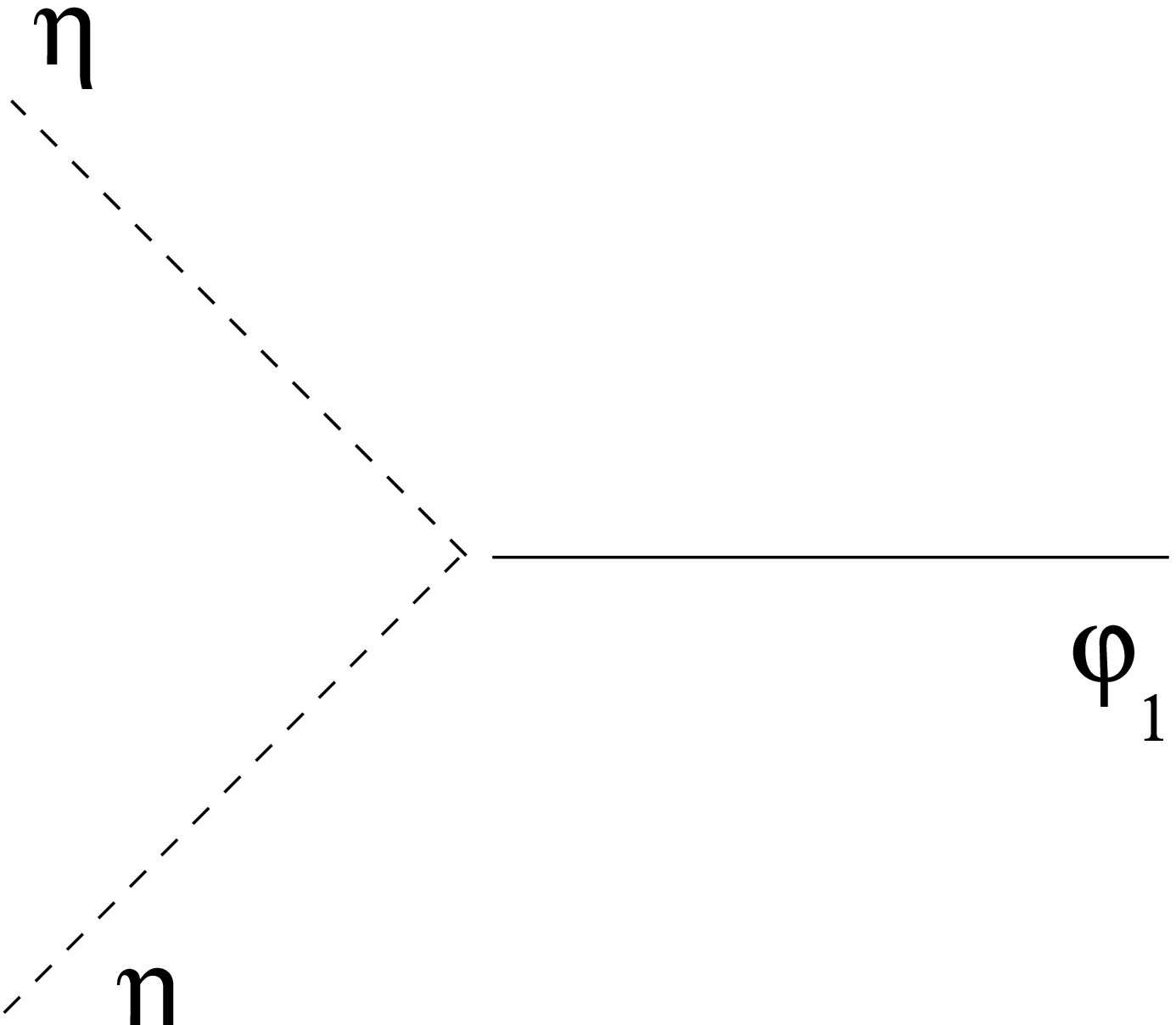}
\vspace{3mm}
\\ $i g^2v$}
\parbox[c]{6cm}{
 \centering
\includegraphics[scale=0.22,angle=0]{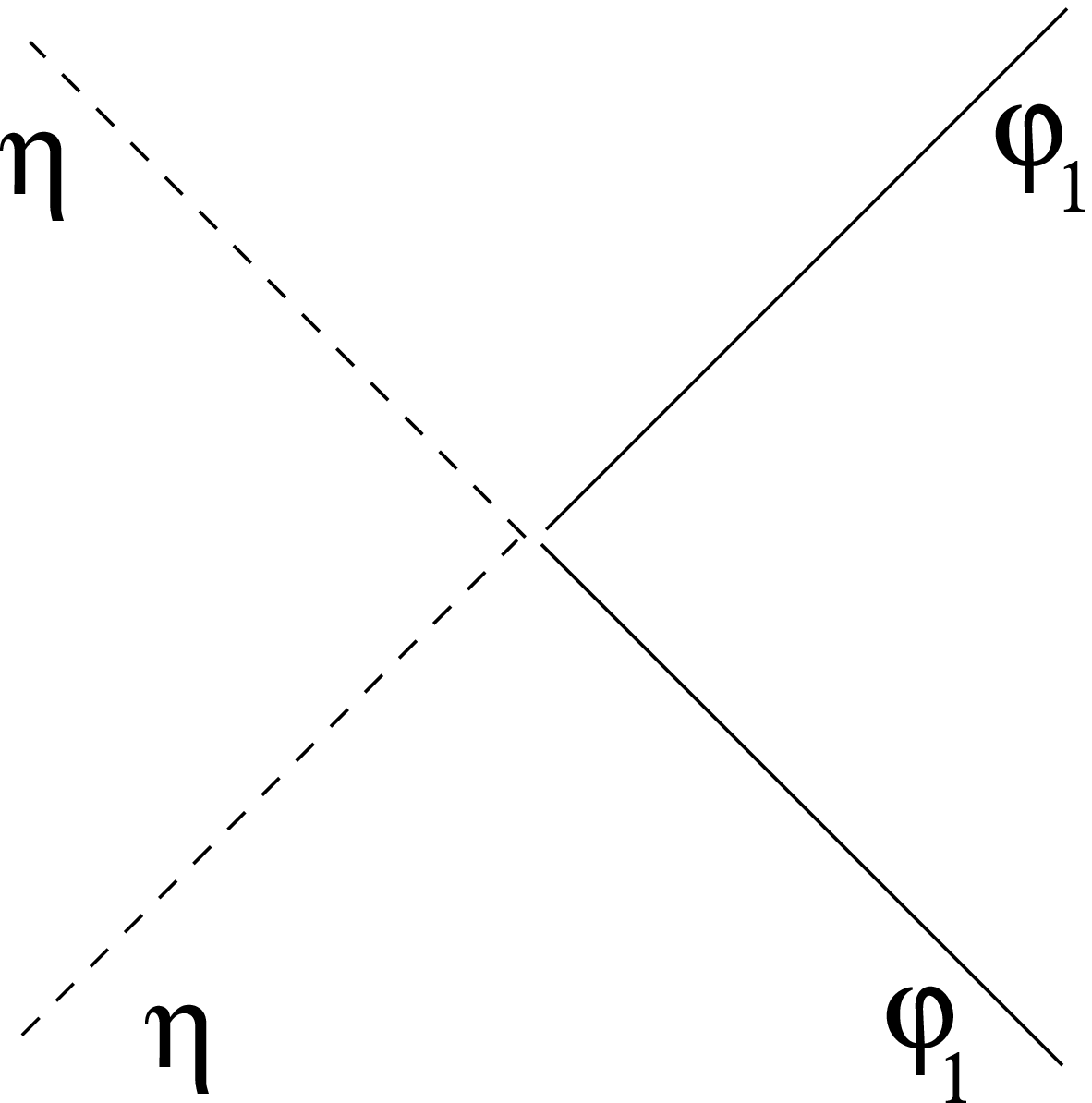}
\vspace{2mm}
\\ $-\frac{i}{2}g^2$}
\caption{\label{fig:feynmanrules_FP}
Vertices for Feynman graphs derived from the Lagrangean
\eqn{L2vac}: Faddeev-Popov sector, without combinatorial factors
for permutations of external lines}
\end{figure}

The Feynman rules are formulated in the vacuum sector, with a
gauge fixing analogous to the one for the vortex sector:
we define the background gauge function
\begin{equation}
{\cal F}_v(A)=\partial_\mu a^\mu-\frac{ig}{2}\left[(v+\varphi_1)
^\ast\varphi-(v+\varphi_1)\varphi^\ast\right]
\end{equation}
and the gauge fixing Lagrangean
\begin{eqnarray}
{\cal L}_{GF}^{I\hspace{-.05cm}I}&=
&-\left(\frac{1}{2}{\cal F}_v^2(A)\right)
^{I\hspace{-.05cm}I} \nonumber \\ 
&=&-\frac{1}{2}(\partial_\mu a^\mu)^2
+ g a^\mu(\varphi_2\partial_\mu\varphi_1
+(v+  \varphi_1)\partial_\mu\varphi_2) \\
&&-(\frac{g^2}{2}v^2+g^2v\varphi_1+\frac{g^2}{2}\varphi_1^2
)^2\varphi_2^2  \; .
\nonumber \end{eqnarray}
Then the Lagrangean, including the gauge fixing and
Faddeev-Popov terms, takes the form
\bea\nonumber
\call&=&-\frac{1}{2}\partial_\mu a_\nu\partial^\mu a^\nu
+\frac{1}{2} m_W^2 a_\mu a^\mu\\
&& \nonumber+\frac{1}{2}\partial_\mu \varphi_1\partial^\mu\varphi_1-\frac{1}{2}
m_H^2\varphi_1^2\\
&& \label{L2vac}+\frac{1}{2}\partial_\mu \varphi_2\partial^\mu\varphi_2-\frac{1}{2}
m_W^2\varphi_2^2\\
&&\nonumber + 2 g a^\mu\varphi_2\partial_\mu\varphi_1\\
\nonumber
&&+g^2a_\mu a^\mu v\varphi_1+\frac{1}{2}g^2a_\mu a^\mu
\left(\varphi_1^2+\varphi_2^2\right)
\\\nonumber
&&-\frac{\lambda}{4}\left(\varphi_1^4+\varphi_2^4+2\varphi_1^2\varphi_2^2\right)
-\lambda v \varphi_1^3-\lambda v \varphi_1\varphi_2^2\\
\nonumber 
&&-g^2 v \varphi_1\varphi_2^2-\frac{g^2}{2}\varphi_1^2\varphi_2^2
\\
\nonumber
&&+\sum_{i=1}^2\eta_i\frac{1}{2}
\left(-\Box-g^2(v+\varphi_1)^2\right)\eta_i
\pkt\eea
The vertices for the Feynman rules  are presented in Figs.
\ref{fig:feynmanrules_gH} and \ref{fig:feynmanrules_FP}.

We denote as Feynman graphs the vacuum graphs with the external
sources which are provided by the 
classical Higgs and gauge fields:
$\varphi_1(x)=v[f(r)-1]$, $\varphi_2=0$, $a_\mu=0$ for $\mu=0,1$ and
$a_i=\epsilon_{ik} x_k (A(r)+1)/gr^2$ for $i=3,4$.
Vacuum graphs with external fields $\varphi_2$ will
 not be displayed as they do not contribute. 

\begin{figure}
\begin{center}
\includegraphics[scale=0.4]{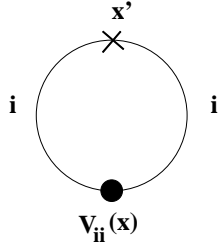}
\end{center}
\caption{\label{fig:Vii}
The one-vertex graph. The cross symbolizes the point where the
first order Green' s function is contracted.}
\end{figure}

\begin{figure}
\begin{center}
\includegraphics[scale=0.4]{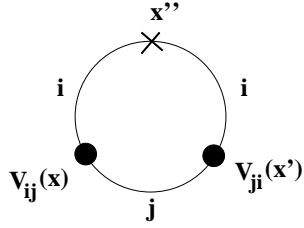}
\caption{\label{fig:VijVji}
The two-vertex graph. The cross symbolizes the point where the
second order Green' s function is contracted.}
\end{center}
\end{figure}

The vertex graphs originate from the expansion of the Green's function.
Starting with
\bea
(-\Delta_2+\calm_0)G(\bfx,\bfx')&=&-V(\bfx) G(\bfx,\bfx')\kma
\\
(-\Delta_2+\calm_0)G^0(\bfx,\bfx')&=&\delta^2(\bfx-\bfx')
\eea
we get the expansion
\be
G(\bfx,\bfx')=G^0(\bfx,\bfx')-\int d^2x''
G^0(\bfx,\bfx'')V(\bfx'')G^0(\bfx'',\bfx')+ ...
\ee
which we recall mainly in order to keep track of the signs.
The vertex graphs are obtained by contracting the Green's function
according to Eq. \eqn{fnpdef} in the partial waves,
or, after summation, according to Eq. \eqn{fk3nudef}. 
The first order and second order graphs
are depicted in Figs. \ref{fig:Vii} and \ref{fig:VijVji},
respectively. 

\subsection{Graphs with one vertex}
\label{secs:Vii}
\begin{figure}[htb]
\begin{center}
\includegraphics[scale=0.26]{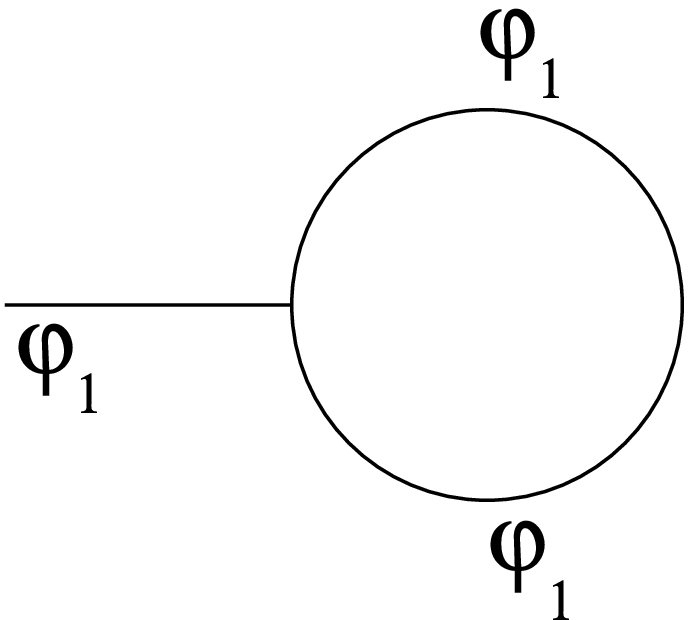}
\;\;\;
\includegraphics[scale=0.26]{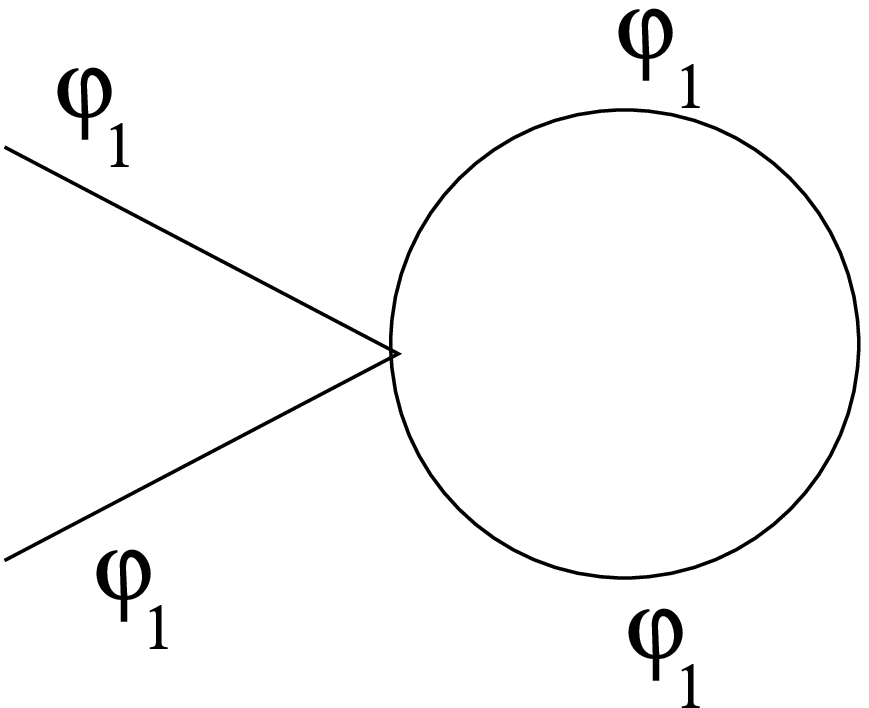}
\end{center}
\caption{\label{fig:V44feyn}
Feynman graphs corresponding to the one-vertex graph with
  $V_{44}^H$}
\end{figure}

The graphs with one vertex all come with in a combination 
analogous to the one displayed in Fig. \ref{fig:V44feyn}, except for
the graphs with the seagull vertex.
The external lines $v\phi_1$ and $\phi_1^2$ combine as
$v^2(f(r)-1)+v^2(f-1)^2/2=v^2(f^2-1)/2$. They then are in
correspondence with one of the vertex graphs with the external
potential $V_{ii}(r)$. In order to illustrate the
relation between the Feynman and vertex graphs we consider the graphs with a
Higgs field loop. The corresponding vertex graph is the one with
the vertex $V_{44}$. We use the subscript $44$ for identifying
this contribution.

The Feynman rules yield the contribution
\bea \nonumber
&&3i\lambda v^2\frac{m_H^2}{32\pi^2}(L_\epsilon+1-\ln\frac{m_H^2}{\mu^2})
LT\int d^2x 
\left[(f(r)-1)+\frac{1}{2}(f(r)-1)^2\right]
\\
&&=iLT\frac{3m_H^2}{64\pi^2}(L_\epsilon+1-\ln\frac{m_H^2}{\mu^2})
\int d^2x \left[f^2(r)-1\right]\equiv-iLT\Delta\sigma_{\rm {fl},44}  
\eea
where $L_\epsilon=2/\epsilon-\gamma+\ln 4\pi$ and where we have used
$\lambda v^2=m_H^2/2$.
The factor $LT$ comes from the trivial integrations over time and
along the string axis. The Feynman graph constitutes a correction to 
$iS$, where $S$ is the action, so as an  energy correction 
it receives a minus sign. 

The contribution we subtract is given, in the partial
waves, by
\be
F_{44}^n=-\int_0^\infty r\,dr \int_0^\infty r'\,dr'
V_{44}^H(r) G_n^0(r,r',\kappa_H)G_n^0(r',r,\kappa_H)
\pkt\ee 
The corresponding vertex graph is of the type shown in Fig.
\ref{fig:Vii}. As announced above, we do not include the
$(A(r)+1)^2/r^2$ term and denote the restricted potential
by $V_{44}^H$. We also leave out the index $n$ as this
potential does not depend on it.
The $r'$ integration can be carried out and one finds
\be
F_{n,44}=\frac{d}{d m_H^2}\int_0^\infty r\,dr
V_{44}(r)G_n^0(r,r,\kappa_H)
\pkt\ee
This can be summed up to yield
\bea 
\nonumber
F_{44}(p)&=&\frac{d}{dm_H^2}
2\pi\int_0^\infty r\,dr V_{44}^H(r)G^0(\bfx,\bfx,\kappa_H)
\\
&=&\frac{1}{2p}\frac{d}{dp}\int\frac{d^2k}{(2\pi)^2}\frac{1}{\bfk^2+p^2
+m_H^2}\int d^2 x V_{44}^H(r)
\eea
This is discussed in some more detail in Appendix \ref{G1}.
Integration over $dp p^3/4\pi$  leads, after integration by parts, to
\bea \nonumber
\Delta\sigma_{\rm{fl},44}&=&-\int_0^\infty
\frac{p^3dp}{4\pi}F_{44}(p)=
\int_0^\infty \frac{p\,dp}{4\pi}\frac{d^2k}{(2 \pi)^2}\frac{1}{
k^2+p^2+m_H^2}\int d^2 x V_{44}^H(r)
\\
&=&\frac{1}{2}\int \frac{d\nu dk_3d^2k}{(2 \pi)^4}\frac{1}{
\bfk^2+k_3^2+\nu^2+m_H^2}\int d^2 x V_{44}^H(r)
\pkt\eea
If we use dimensional regularization applied to the
Euclidean four momentum integration $d\nu dk_3 d^2k=d^4k$ we obtain
\bea \nonumber
\Delta \sigma_{\rm{fl},44}&=&
-\frac{m_H^2}{32\pi^2}\left[L_\epsilon
+1-\ln\frac{m_H^2}{\mu^2}\right]\int d^2 x V^H_{44}(r)
\\
&=&-3 \frac{m_H^4}{64\pi^2}\left[L_\epsilon
+1-\ln\frac{m_H^2}{\mu^2}\right]\int d^2 x (f^2(r)-1)
\eea
in agreement with the result from the Feynman graph.
In this case it was possible to check in detail the relation between the
subtracted part, the vertex graph and the Feynman graphs.
\begin{figure}[htb]
\begin{center}
\includegraphics[scale=0.26]{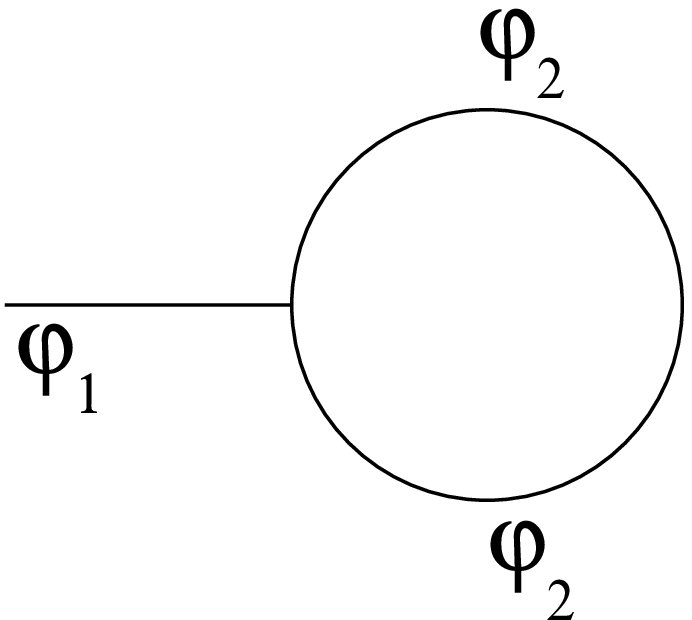}
\;\;\;
\includegraphics[scale=0.26]{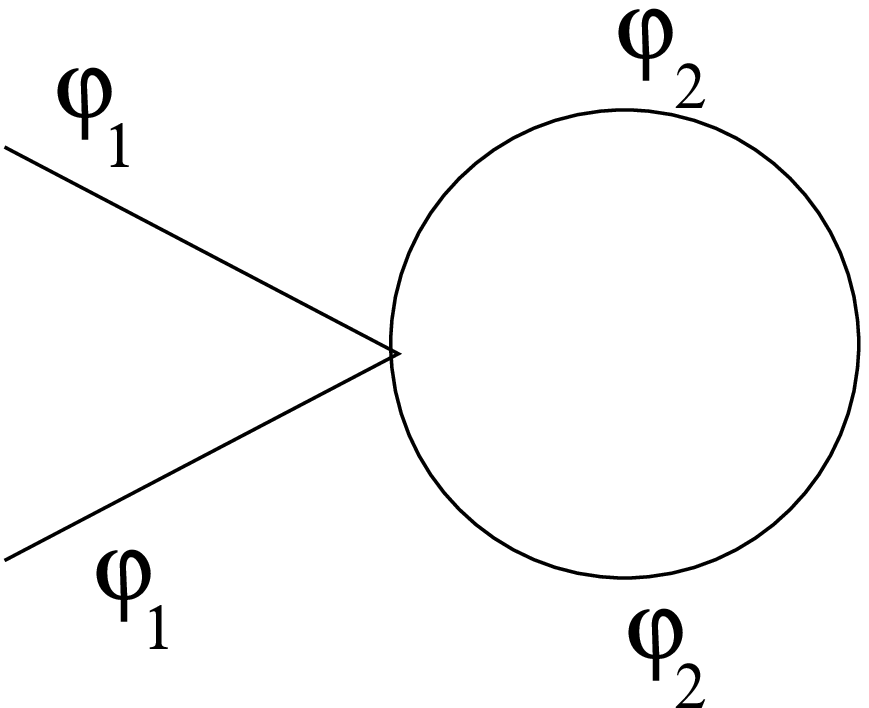}
\end{center}
\caption{\label{fig:V33feyn}
Feynman graphs corresponding to the one-vertex graph with
  $V_{33}^H$.}
\end{figure}

\begin{figure}[htb]
\begin{center}
\includegraphics[scale=0.26]{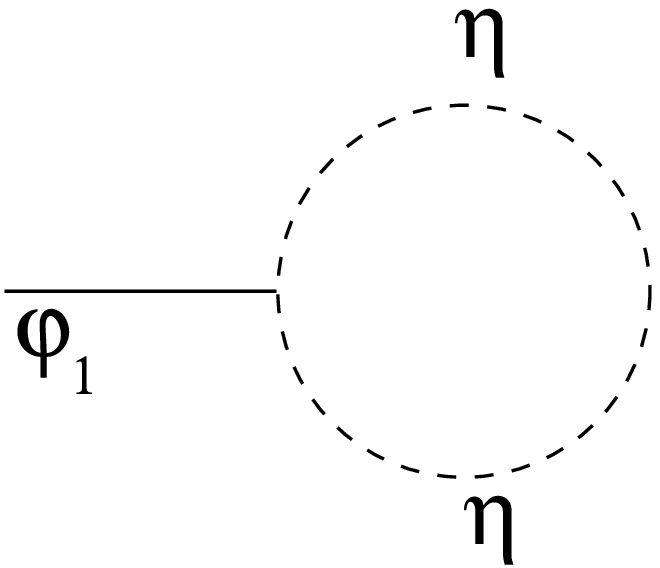}
\;\;\;
\includegraphics[scale=0.26]{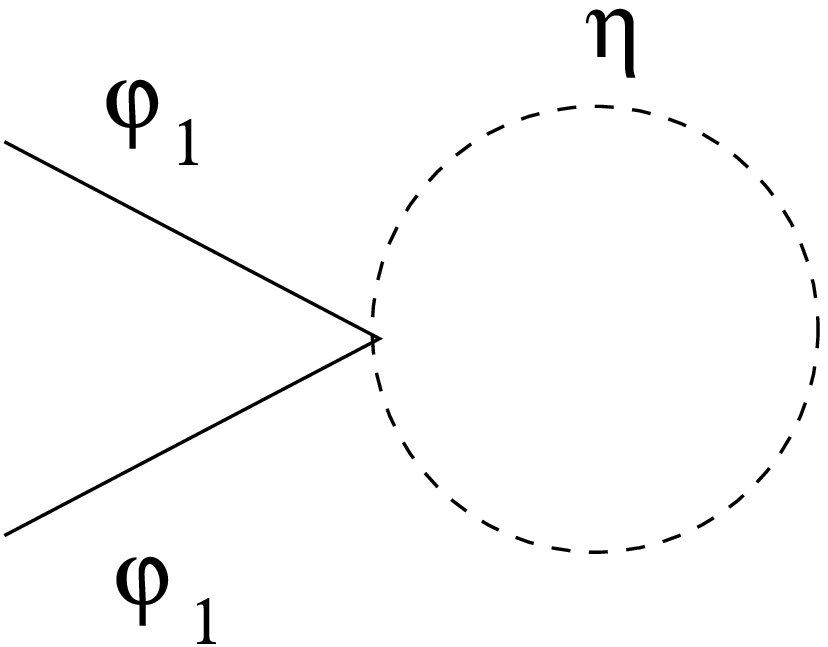}
\end{center}
\caption{\label{fig:V55feyn}
Feynman graphs corresponding to the one-vertex graph with
  $V_{55}$.}
\end{figure}

Obviously the procedure is analogous for the other
graphs with one vertex. We obtain, both by using the Feynman graphs
of Fig. \ref{fig:V33feyn} as also the vertex graph:
\bea 
\Delta \sigma_{\rm{fl},33}&=&
-\frac{m_H^2}{32\pi^2}\left[L_\epsilon
+1-\ln\frac{m_W^2}{\mu^2}\right]\int d^2 x V^H_{33}(r)
\\\nonumber
&=&-(m_H^2+2m_W^2) \frac{m_H^2}{64\pi^2}\left[L_\epsilon
+1-\ln\frac{m_W^2}{\mu^2}\right]\int d^2 x (f^2(r)-1)
\pkt \eea
For the contribution of $V_{11}$ and $V_{12}$ in the sector of
angular momentum $n$ one notes that they
\begin{figure}[htb]
\begin{center}
\includegraphics[scale=0.26]{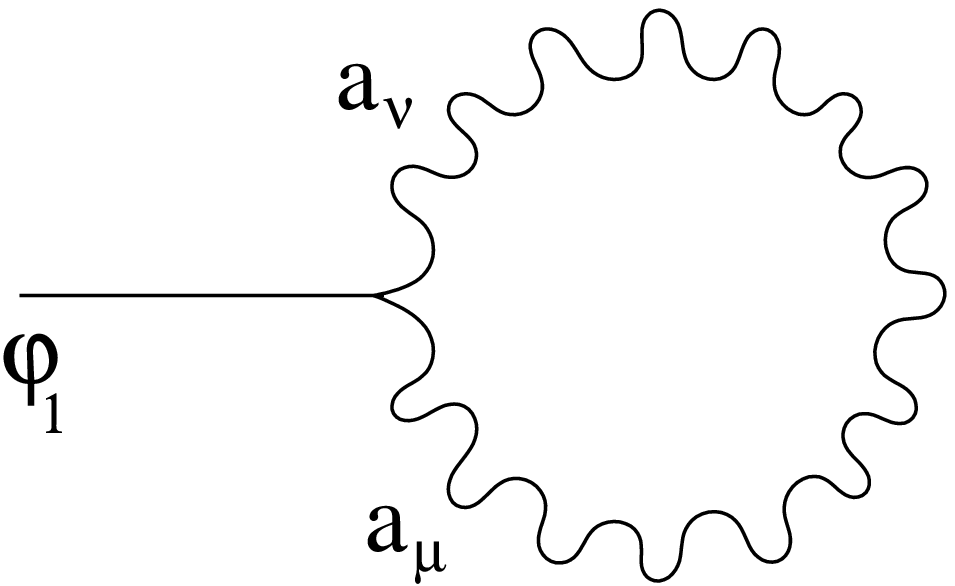}
\;\;\;
\includegraphics[scale=0.26]{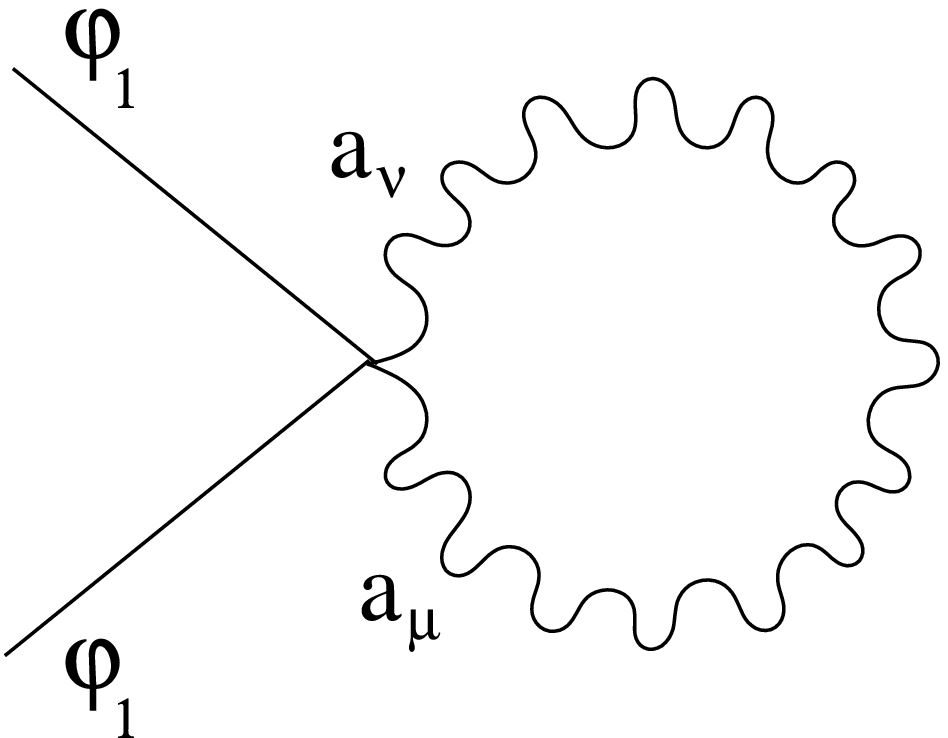}
\end{center}
\caption{\label{fig:V1122feyn}
Feynman graphs corresponding to the one-vertex graphs with
  $V_{11}$ and $V_{22}$.}
\end{figure}
involve the  Green's functions $G^0_{n\pm 1}$. In the sum the shifts
compensate each other and the result is, both from the Feynman
graphs in Fig. \ref{fig:V1122feyn} as from the vertex graphs,
\bea \nonumber 
\Delta \sigma_{\rm{fl},11+22}&=&
-\frac{m_W^2}{32\pi^2}\left[L_\epsilon
+1-\ln\frac{M_W^2}{\mu^2}\right]\int d^2 x [V_{11}(r)+V_{22}(r)]
\\
&=&-\frac{m_W^4}{16\pi^2}\left[L_\epsilon
+1-\ln\frac{m_W^2}{\mu^2}\right]\int d^2 x (f^2(r)-1)
\pkt\eea
These are cancelled exactly by the Faddeev-Popov graphs
of Fig. \ref{fig:V55feyn}.

\subsection{Graphs with two vertices: $V_{ii}\times V_{ii}$}
We have subtracted in the partial waves the second order
contributions to $\calf^{(2)}_n(p)$ of the type
\bea\nonumber
F_{n,ij,ji}(p)&=&\int dr'\,r'\int dr\,r \int dr''\,r''
V_{ij}(r)G^0_n(r,r',\kappa_j)V_{ji}(r')
\\ &&G^0_n(r',r'',\kappa_i) G^0_n(r'',r,\kappa_i)\kma
\eea
as represented graphically in Fig. \ref{fig:VijVji}.
The $r''$ integration can be done, see Appendix \ref{G2},
with the result
\be
F_{n,ij,ji}(p)=-\frac{d}{dm_i^2}\int dr'\,r' \int dr\,r\,  
V^n_{ij}(r)G^0_n(r,r',\kappa_j)V^n_{ji}(r')G^0_n(r',r,\kappa_i)
\pkt\ee
We recall that $\kappa_j^2=p^2+m_j^2$.
\begin{figure}[htb]
\begin{center}
\includegraphics[scale=0.26]{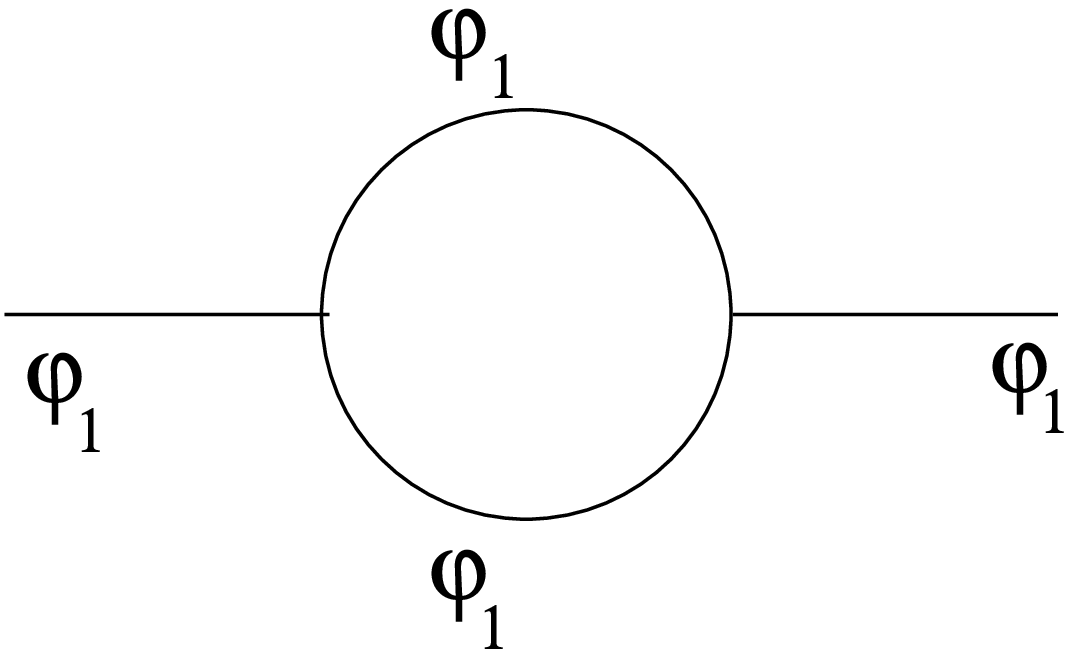}
\;\;\;
\includegraphics[scale=0.26]{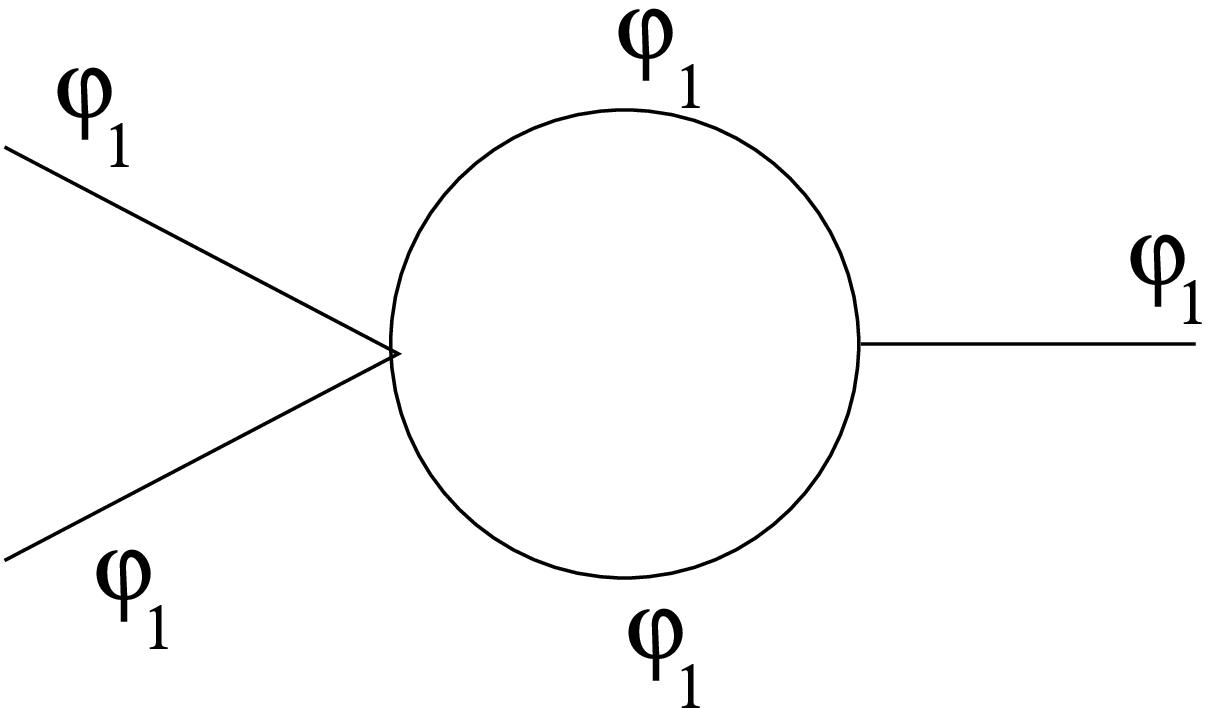}
\;\;\;
\includegraphics[scale=0.26]{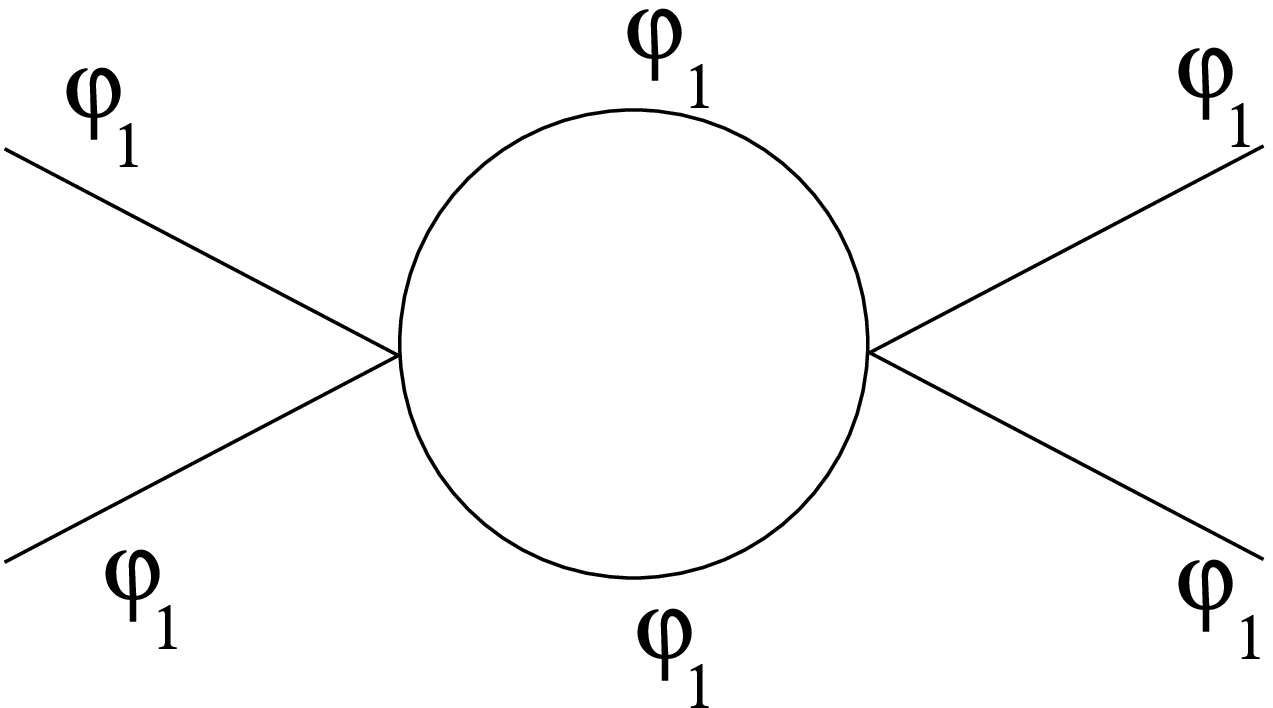}
\end{center}
\caption{\label{fig:V44V44feyn}
Feynman graphs corresponding to the vertex graph with
  $V_{44}^H\times V_{44}^H$}
\end{figure}
The one-to-one relation with Feynman graphs is most obvious for the
diagonal parts where $i=j$, and we will treat these
first. For $V_ {33}$ and $V_{44}$ we again discard the
gauge field parts $V_{33}^g=V_{44}^g=(A(r)+1)/r^2$, restricting them to the
Higgs field contributions $V_{33}^H$ and $V_{44}^H$ This will be justified later.
As now $\kappa_i=\kappa_j$ we can apply the 
derivative with respect to $m_i^2$
to both Green's functions and compensate this double action by
a factor $1/2$. Furthermore, we  can replace this derivative
by the derivative 
with respect to $d p^2=2p\,dp$, applied to the whole graph.
\begin{figure}[htb]
\begin{center}
\includegraphics[scale=0.26]{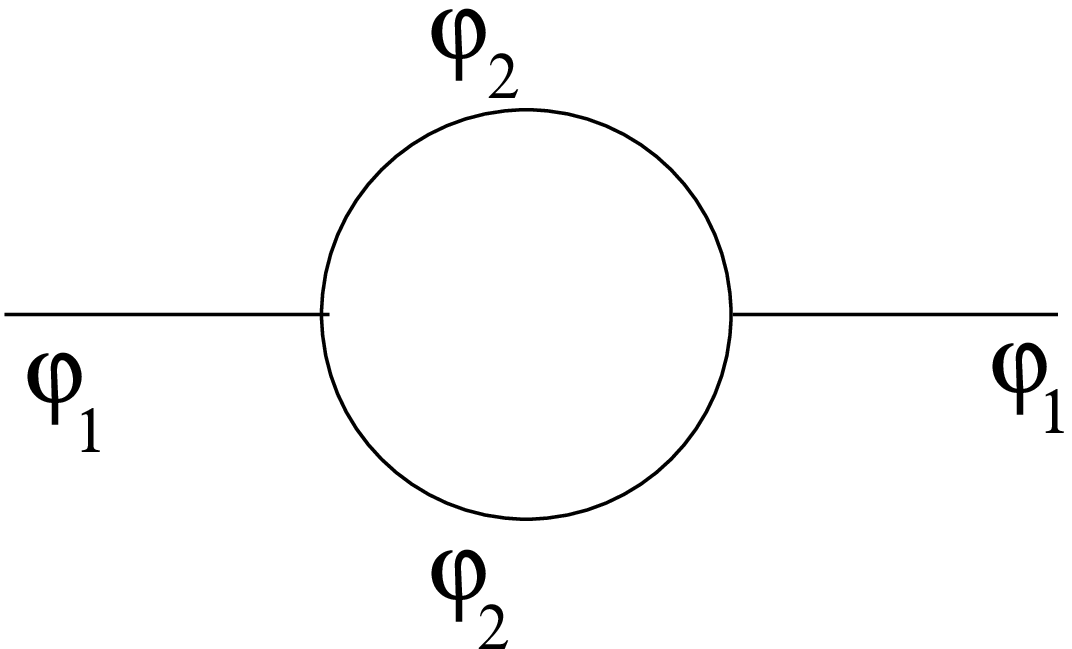}
\;\;\;
\includegraphics[scale=0.26]{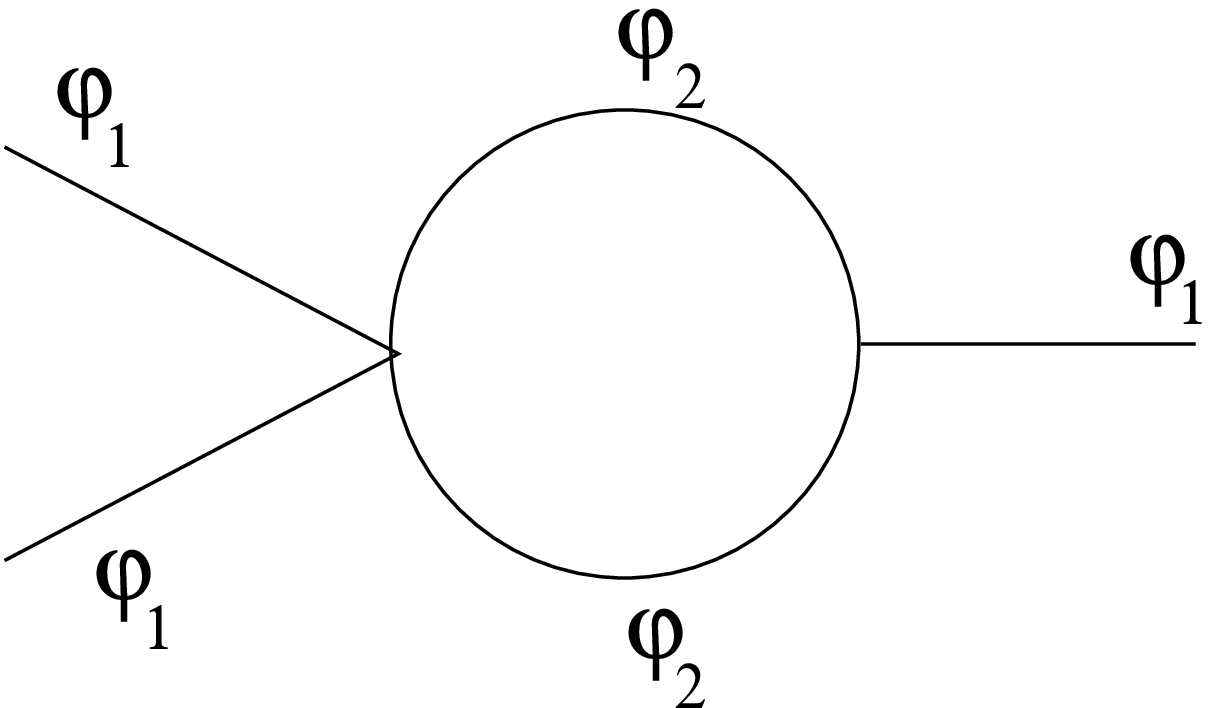}
\;\;\;
\includegraphics[scale=0.26]{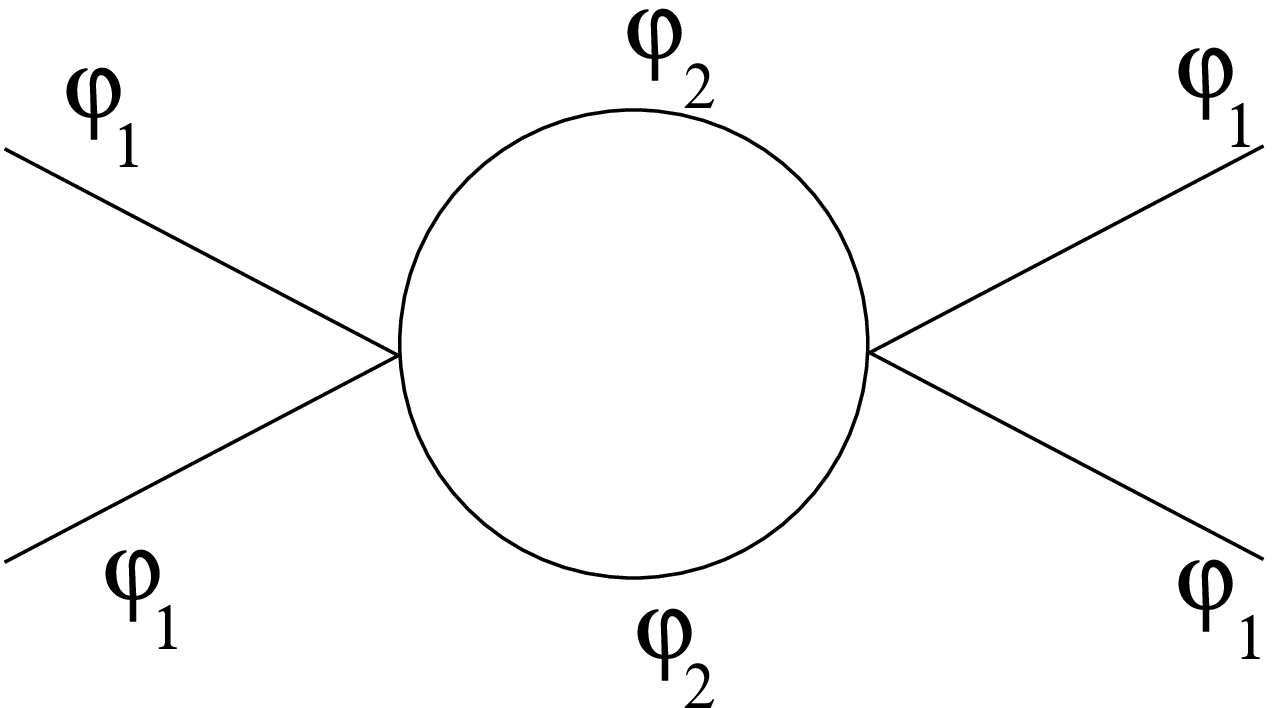}
\end{center}
\caption{\label{fig:V33V33feyn}
Feynman graphs corresponding to the vertex graph with
  $V_{33}^H\times V_{33}^H$}
\end{figure}
Then 
\bea \nonumber 
&&-\int \frac{dp}{4\pi}p^3 F_{n,ii,ii}(p)
\\\nonumber
&=& \frac{1}{2}
\int \frac{p^3dp}{4\pi}\frac{1}{2p}\frac{d}{dp}
\int dr'\,r'\int dr\, r \,
V^n_{ii}(r)G^0_n(r,r',\kappa_i)V^n_{ii}(r')G^0_n(r',r,\kappa_i)
\\
&=& -\int \frac{p\,dp}{8\pi}
\int dr'\,r'\int dr \, r \,
V^n_{ii}(r)G^0_n(r,r',\kappa_i)V^n_{ii}(r')G^0_n(r',r,\kappa_i)
\pkt\eea
As the potentials under consideration do not depend 
on $n$ we may sum over $n$ to obtain
\bea
-\int \frac{p\,d p}{8\pi}
\int d^2x\int d^2x' V_{ii}(r)G^0(\bfx,\bfx',\kappa_i)V_{ii}(r')
G^0(\bfx',\bfx,\kappa_i)
\kma\eea
and this is just the conventional Feynman graph with external sources.
As in the case with one vertex the integral over $pdp/2\pi$ can be
rewritten as an integral over $dk_3 d\nu/(2\pi)^2$, and the Green's function
$G^0$ in two dimensions involves the integral over $dk_\perp^2$, so 
altogether we have an integration over $d^4k/(2\pi)^4$, times a factor
$1/4$.
In momentum space we obtain
\be
-\frac{1}{4}\int \frac{d^2q}{(2\pi)^2}\left|\tilde V_{ii}(q)\right|^2
\int \frac{d^4k}{(2\pi)^4}\frac{1}{[k^2+m_i^2][(k+q)^2+m_j^2]}
\kma\ee
where $q=(0,0,\bfq)$ has only the transversal components.
\begin{figure}[htb]
\begin{center}
\includegraphics[scale=0.26]{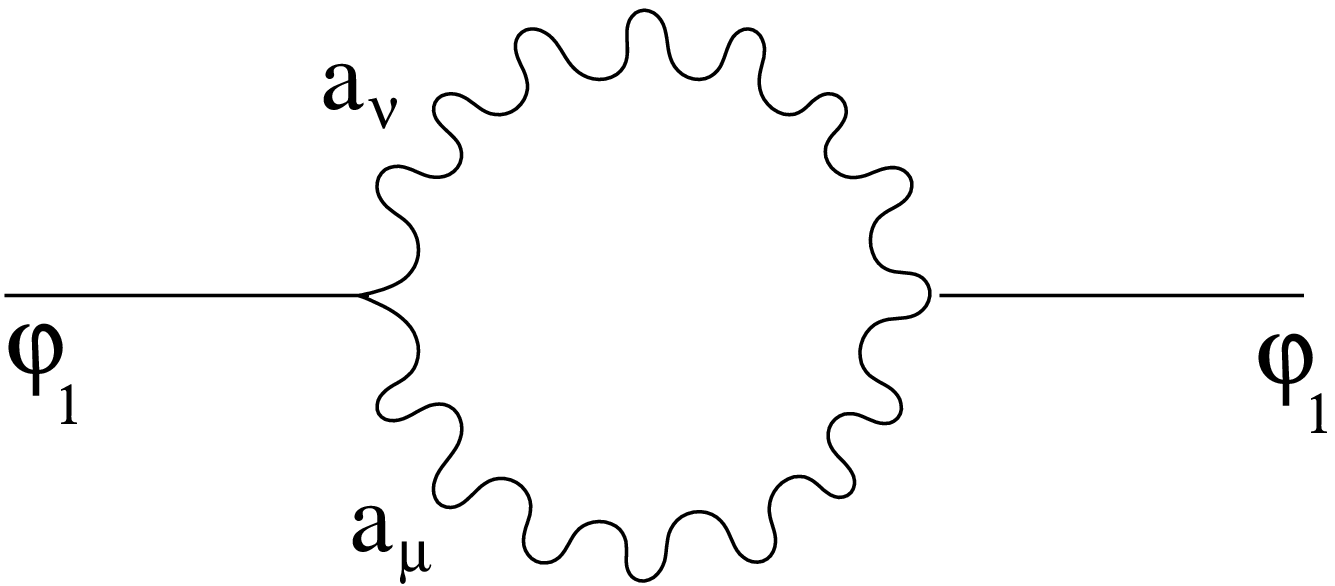}
\;\;\;
\includegraphics[scale=0.26]{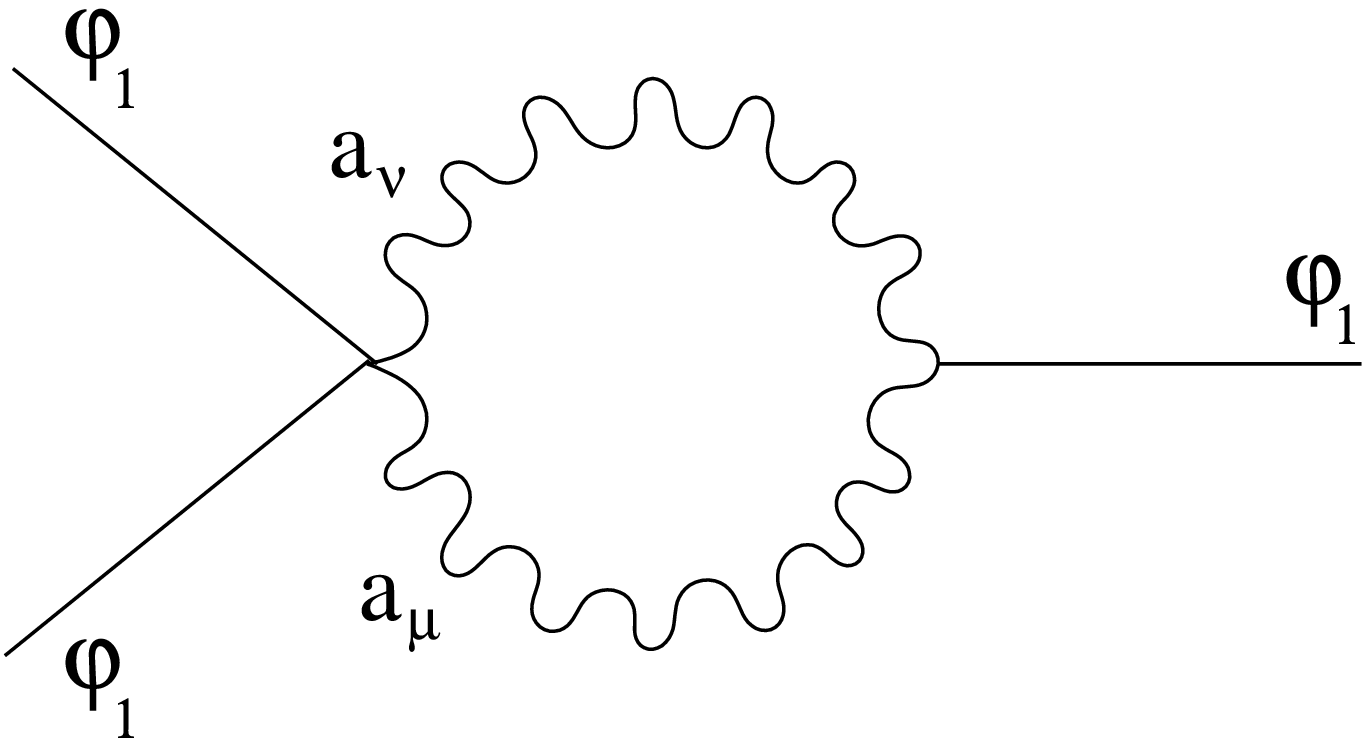}
\;\;\;
\includegraphics[scale=0.26]{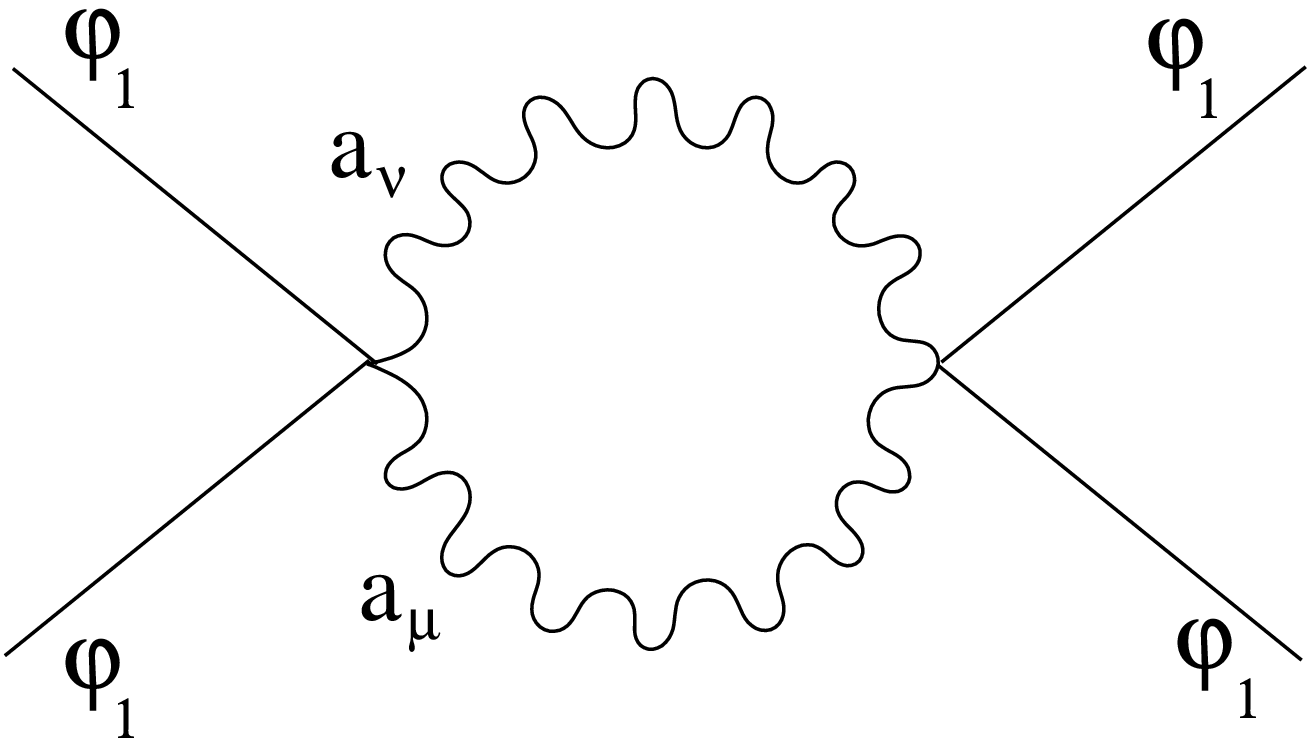}
\end{center}
\caption{\label{fig:V11V11feyn}
Feynman graphs corresponding to the vertex graphs with
  $V_{11}\times V_{11}$ and $V_{22}\times V_{22}$}
\end{figure}

The Euclidean Feynman integral is logarithmically
divergent. In dimensional regularization it is given by
\bea
&&\int \frac{d^{4-\epsilon} k}{(2\pi)^{4-\epsilon}}
\frac{1}{[k^2+m_i^2][(k+q)^2+m_j^2]}
\\ \nonumber
&&= \frac{1}{16\pi^2}\left\{L_\epsilon
 -\int_0^1 d\omega \ln \frac{\omega(1-\omega)\bfq^2+\omega m_i^2
+(1-\omega)m_j^2}{\mu^2}\right\}
\pkt 
\eea
\begin{figure}[htb]
\begin{center}
\includegraphics[scale=0.26]{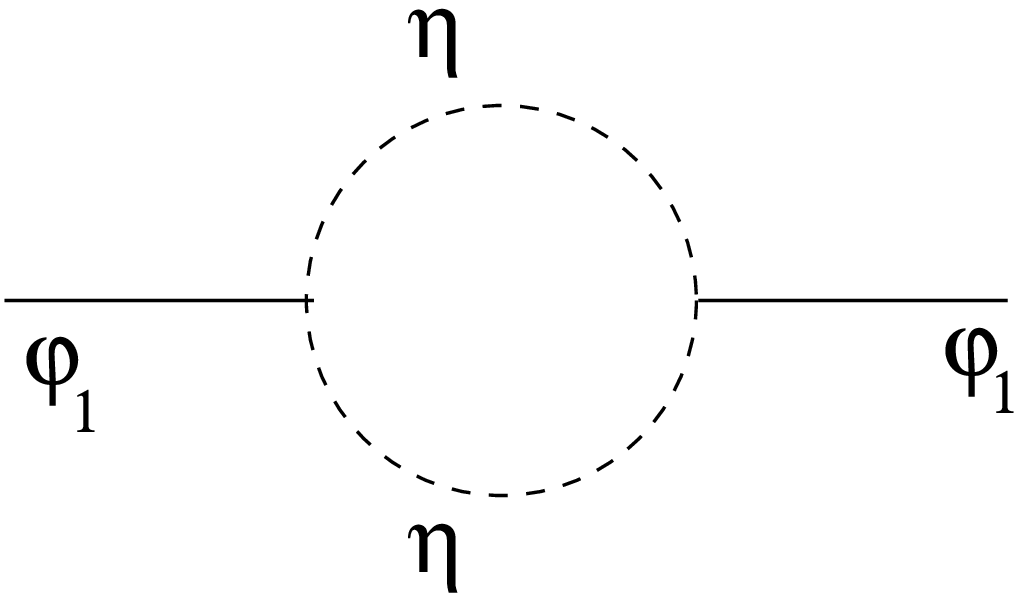}
\;\;\;
\includegraphics[scale=0.26]{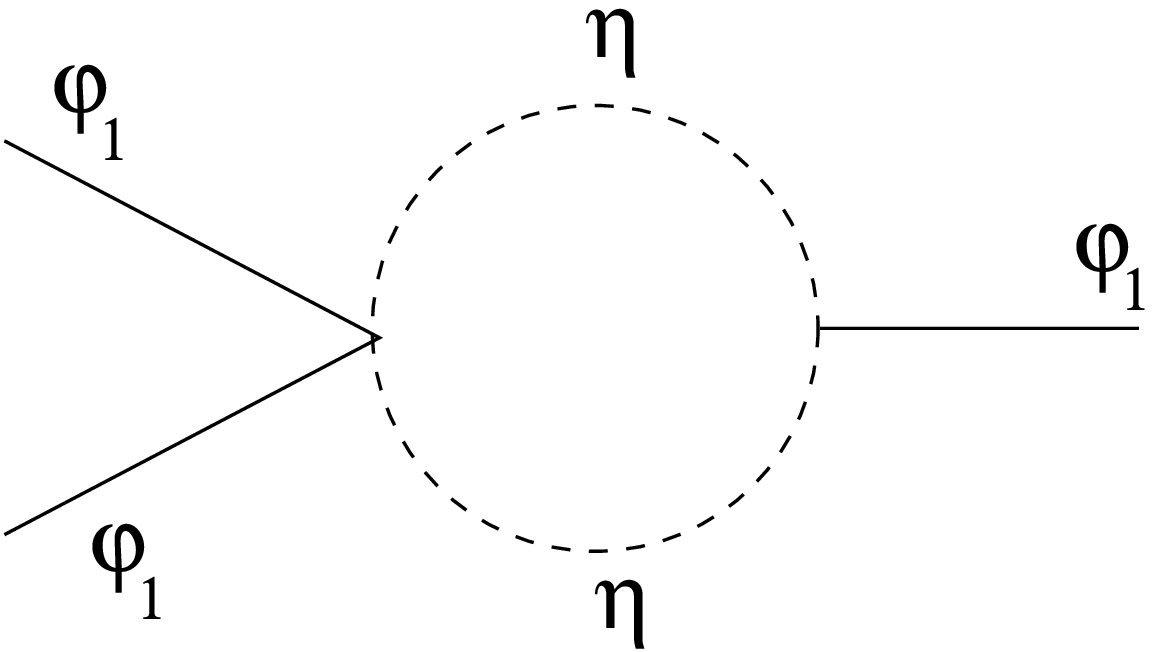}
\;\;\;
\includegraphics[scale=0.26]{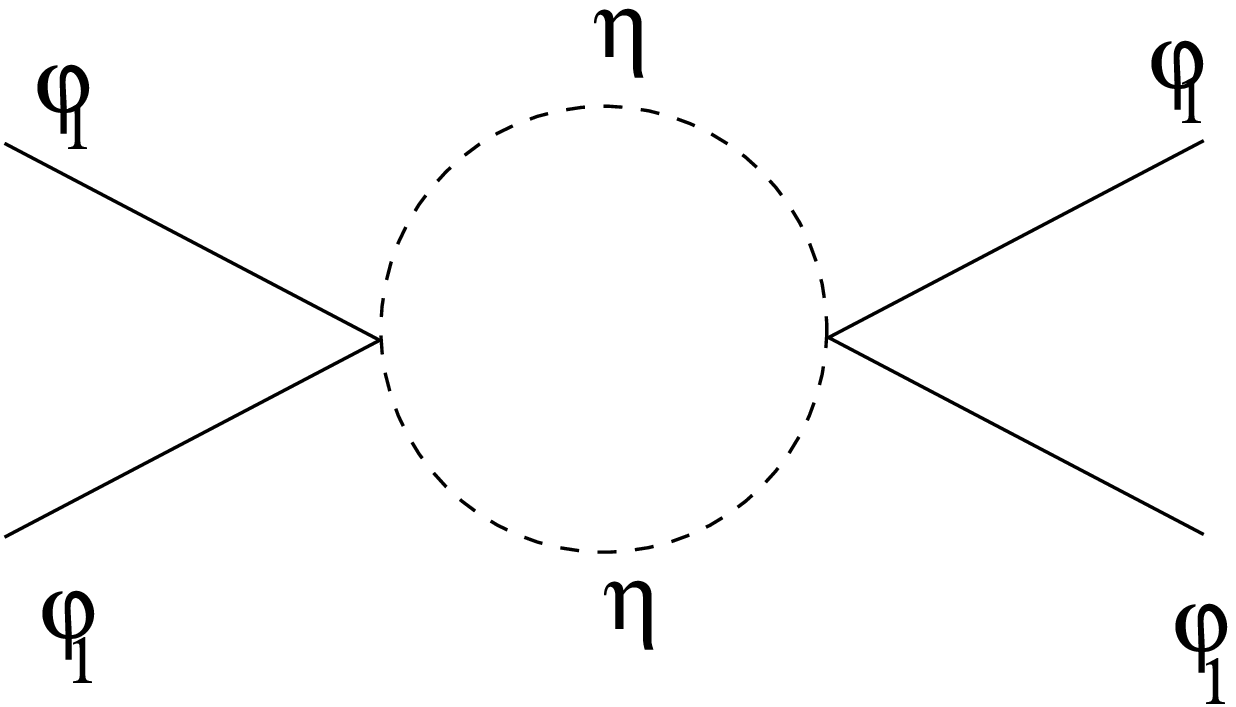}
\end{center}
\caption{\label{fig:V55V55feyn}
Feynman graphs corresponding to the Faddeev-Popov ghost
vertex graphs with   $V_{55}\times V_{55}$.}
\end{figure}

This is to be folded with the Fourier transform of 
$f^2-1$:
\be
\left[f^2-1\right]\tilde\,(\bfq)=2\pi\int_0^\infty dr\, r (f^2(r)-1)J_0(qr)
\equiv \chi_{f^2-1}(q)
\kma\ee
Altogether we find
\be
\Delta\sigma_{{\rm fl},44,44}=
-\frac{9m_H^4}{256\pi^2}\int \frac{d^2q}{(2\pi)^2}
\left\{L_\epsilon
 -\int_0^1 d\omega \ln \frac{\omega(1-\omega)\bfq^2
+m_H^2}{\mu^2}\right\}\left|\chi_{f^2-1}(q)\right|^2
\pkt\ee
The corresponding Feynman graphs are displayed in Fig.
\ref{fig:V44V44feyn}.
The external legs combine as
\bea\nonumber
&&18 \lambda^2 v^4\left[(f(r)-1)(f(r')-1)+(f(r)-1)^2(f(r')-1)/2
\right.\\
\nonumber &&
\left.+(f(r)-1)(f(r')-1)^2/2+(f(r)-1)^2(f(r')-1)^2/4\right]
\\
&=&(9/2)\lambda^2v^4(f^2(r)-1)(f^2(r')-1)
\\\nonumber
&=&(9/8)m_H^4(f^2(r)-1)(f^2(r')-1)
\pkt\eea
Using the $\overline{MS}$ scheme
the finite contribution is given by
\be
\Delta\sigma_{{\rm fl},44,44}=
\frac{9m_H^4}{256\pi^2}\int \frac{d^2q}{(2\pi)^2}
 \int_0^1 d\omega \ln \frac{\omega(1-\omega)\bfq^2
+m_H^2}{\mu^2}\left|\chi_{f^2-1}(q)\right|^2
\pkt\ee
The integral over the logarithm 
is discussed in Appendix \ref{secondorderkernels}. The Fourier transform
of the potential, which actually is a Fourier-Bessel transform,
is discussed in Appendix \ref{fourierbessel}.
The analogous diagonal contributions, displayed in 
Figs. \ref{fig:V33V33feyn}, \ref{fig:V11V11feyn} and
\ref{fig:V55V55feyn} are given by
\be
\Delta\sigma_{{\rm fl},33,33}=
\frac{\left(m_W^2+m_H^2/2\right)^2}{64\pi^2}\int \frac{d^2q}{(2\pi)^2}
 \int_0^1 d\omega \ln \frac{\omega(1-\omega)\bfq^2
+m_W^2}{\mu^2}\left|\chi_{f^2-1}(q)\right|^2
\kma
\ee
\be
\Delta\sigma_{{\rm fl},11,11}+
\Delta\sigma_{{\rm fl},22,22}=
\frac{m_W^4}{32\pi^2}\int \frac{d^2q}{(2\pi)^2}
 \int_0^1 d\omega \ln \frac{\omega(1-\omega)\bfq^2
+m_H^2}{\mu^2}\left|\chi_{f^2-1}(q)\right|^2
\kma\ee
and
\be
\Delta\sigma_{{\rm fl},55,55}
=-
\frac{m_W^4}{32\pi^2}\int \frac{d^2q}{(2\pi)^2}
 \int_0^1 d\omega \ln \frac{\omega(1-\omega)\bfq^2
+m_H^2}{\mu^2}\left|\chi_{f^2-1}(q)\right|^2
  \kma\ee
repectively. The latter two contributions,
those of the transverse gauge field and the Faddeev-Popov
fluctuations, cancel each other, as for the first order contributions.


\subsection{Graphs with two vertices:$V_{13}\times V_{31}$ and
$V_{23}\times V_{32}$}

\begin{figure}[htb]
\begin{center}
\includegraphics[scale=0.26]{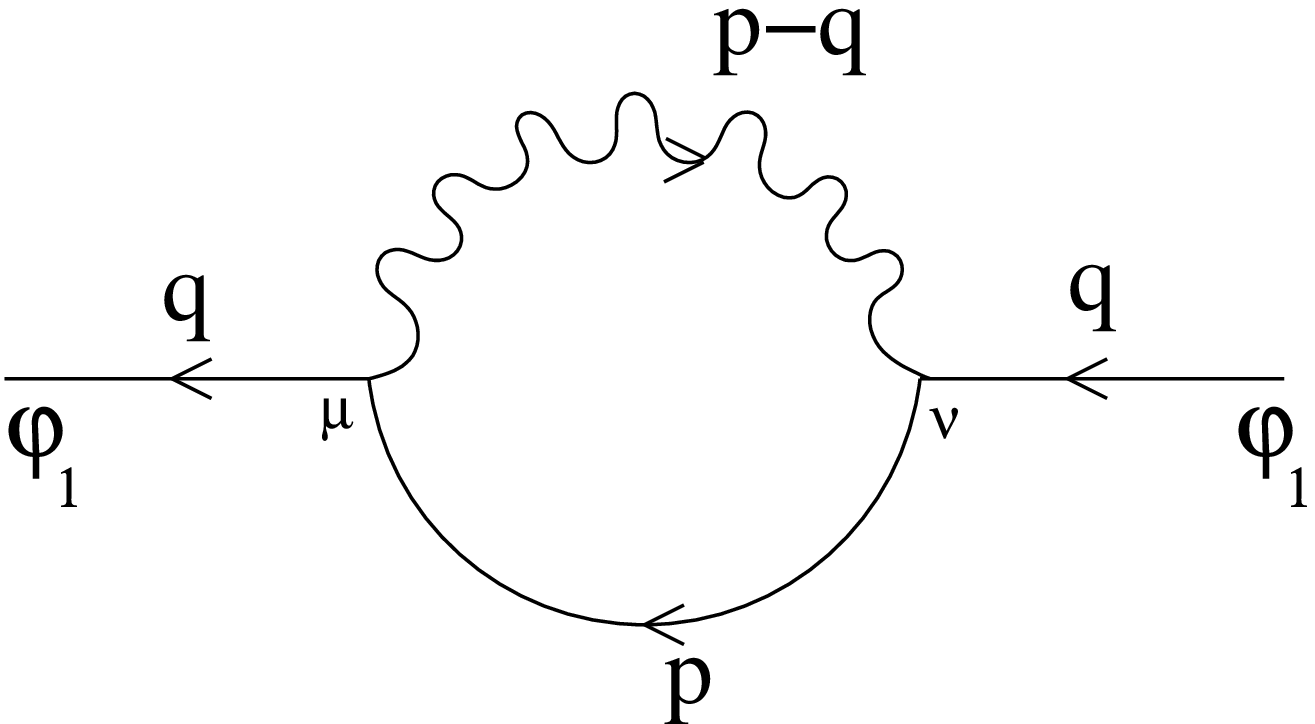}
\end{center}
\caption{\label{fig:V123V312feyn} Feynman graph coresponding to the
vertex graphs with $V_{13}\times V_{31}$ and $V_{23}\times V_{32}$.
}
\end{figure}

The vertex graphs with  $V_{13}\times V_{31}$ and
$V_{23}\times V_{32}$ correspond to the Feynman graph
of Fig. \ref{fig:V123V312feyn}. 
The Feynman integral is given by
\be
-\frac{ig^2}{8\pi^2}q^2\left\{L_\epsilon-\int_0^1 d\omega
\ln \frac{-\omega(1-\omega)q^2+m_W^2}{\mu^2}\right\}
\pkt\ee

The external legs are given by $v (f-1)$.
Denoting the Fourier transform by
\be
\left[v(f-1)\right]\tilde\,(\bfq)\equiv
v \chi_{f-1}(q)= 2\pi v \int_0^\infty dr \, r (f(r)-1)J_0(qr) 
\ee
we obtain for the Feynman graph including the  external legs 
\be
i LT \frac{m_W^2}{8\pi^2}\int \frac{d^2 q}{(2\pi)^2}
\bfq^2\left|\chi_{f-1}(q)\right|^2\left\{L_\epsilon-\int d\omega
\ln \frac{\omega(1-\omega)\bfq^2+m_W^2}{\mu^2}\right\}
\pkt\ee
The infinite part corresponds to the wave function
renormalization of the Higgs field.
In the $\overline{MS}$ scheme the finite correction 
to the string tension  is given by
\be
\Delta \sigma_{\rm{fl},(12)3(12)3}
=\frac{m_W^2}{8\pi^2}\int \frac{d^2 q}{(2\pi)^2}
\bfq^2\left|\chi_{f-1}(q)\right|^2\int d\omega
\ln \frac{\omega(1-\omega)\bfq^2+m_W^2}{\mu^2}
\pkt\ee


\subsection{Graphs with two vertices: $V_{14}\times V_{41}$
and $V_{24}\times V_{42}$}

\begin{figure}[htb]\begin{center}
\includegraphics[scale=0.26]{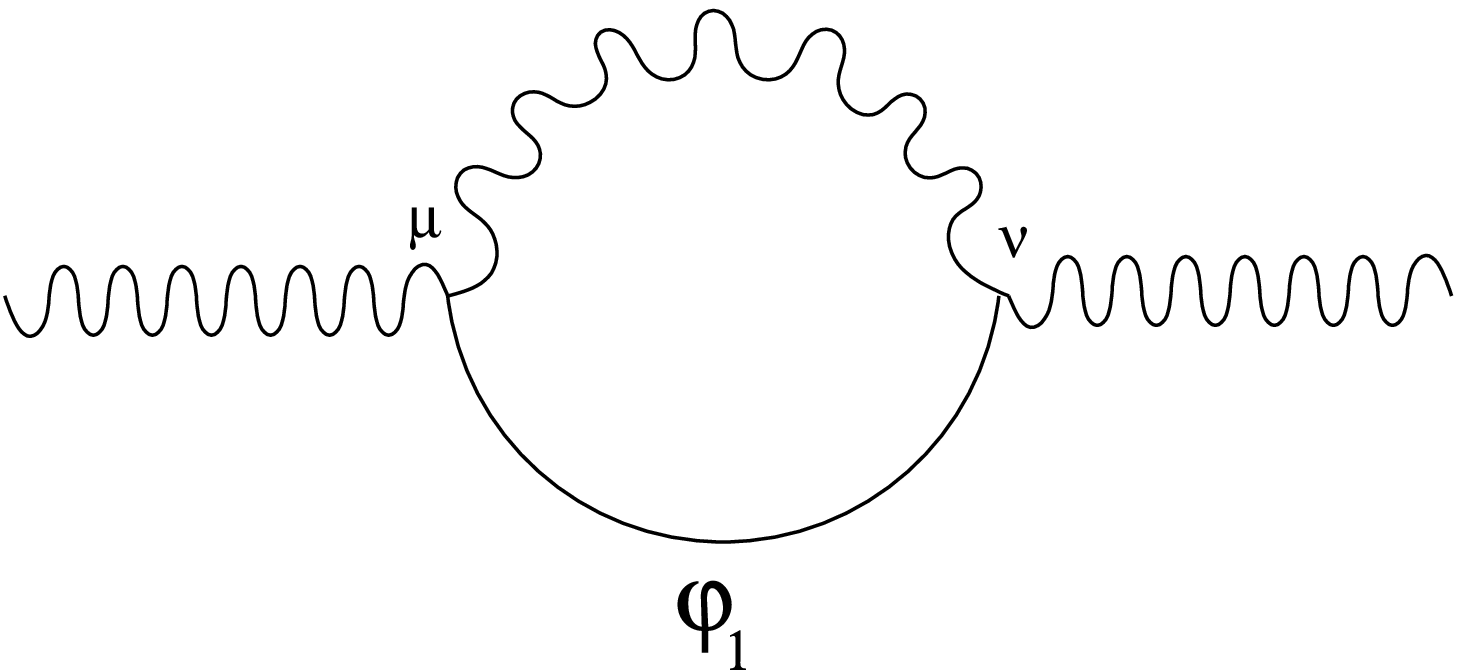}
\;\;\;\includegraphics[scale=0.26]{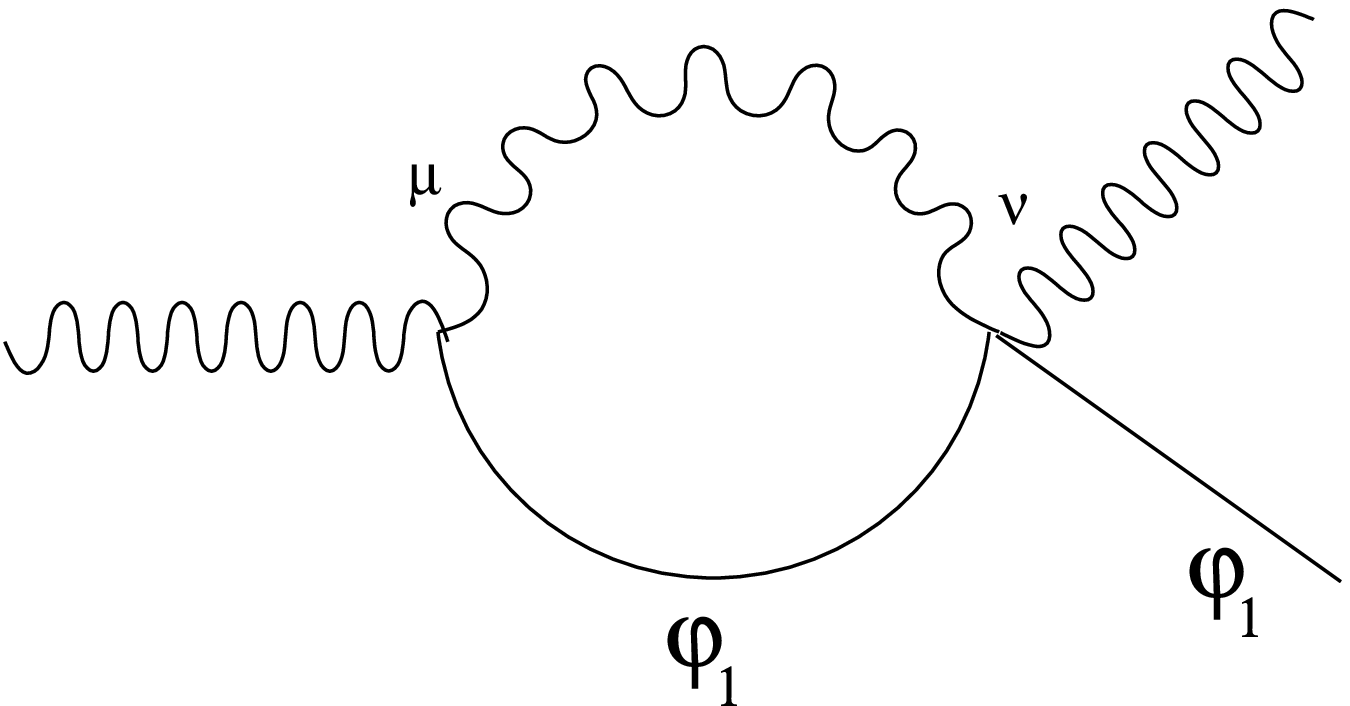}
\;\;\;\includegraphics[scale=0.26]{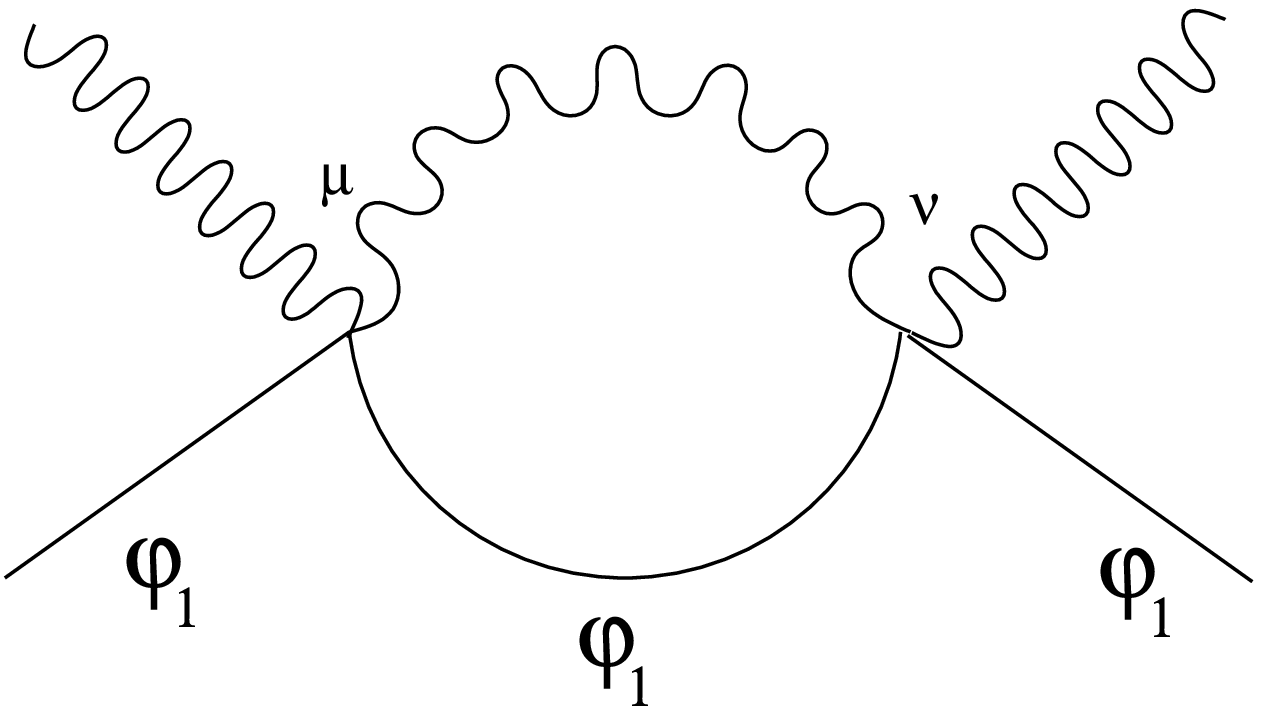}
\end{center}\caption{\label{fig:V124V412feyn}
The Feynman graphs corresponding to the vertex graphs with
$V_{14}\times V_{41}$ and $V_{24}\times V_{42}$.}
\end{figure}

These nondiagonal terms correspond to the Feynman graphs
of Fig. \ref{fig:V124V412feyn}. The external Higgs field legs
combine as
\be
1+(f(r)-1)+(f(r')-1)+(f(r)-1)(f(r')-1)=f(r)f(r')
\pkt
\ee
The Feynman integral is given by
\be
-ig_{\mu\nu}\frac{g^4}{4\pi^2}\left\{L_\epsilon-\int_0^1 d\omega
\ln \frac{\omega(1-\omega)\bfq^2+\omega m_W^2
+(1-\omega)m_H^2}{\mu^2}\right\}
\pkt\ee
The external legs combine to  $A_\mu \phi$, which restricts to a 
two-dimensional vector $A_i^\perp v\,f(r)$.
We write the Fourier transform as
\be
\left[A_i^\perp v f\right]_i \tilde\,=
2\pi \frac{v}{g}\epsilon_{ij}\hat q_j\int_0^\infty dr (A(r)+1)f(r)J_1(qr)
=\frac{v}{g}\epsilon_{ij}\hat q_j \chi_{Af}(q)
\pkt\ee
The Feynman graph with external legs then is given by
\be
iLT\frac{m_W^2}{8\pi^2}\int \frac{d^2q}{(2\pi)^2} \left|\chi_{Af}(q)\right|^2 
\left\{L_\epsilon -\int_0^1 d\omega\ln\frac{\omega(1-\omega)\bfq^2+
\omega m_W^2+(1-\omega)m_H^2}{\mu^2}\right\}
\pkt \ee
The infinite term corresponds to the coupling constant renormalization
in the term $g^2|\phi|^2A_\mu A^\mu/2$.
For the  finite correction to the string tension we obtain
\bea
&&\Delta\sigma_{{\rm fl},(12)4(12)4}=
\\\nonumber
&&\frac{m_W^2}{8\pi^2}\int \frac{d^2q}{(2\pi)^2} \left|\chi_{Af}(q)\right|^2 
\int_0^1 d\omega\ln\frac{\omega(1-\omega)\bfq^2+
\omega m_W^2+(1-\omega)m_H^2}{\mu^2}
\pkt\eea


\subsection{Graphs with two vertices: $V_{34}\times V_{43}$}
\label{secs:V34V43}
\begin{figure}\begin{center}
\includegraphics[scale=0.3]{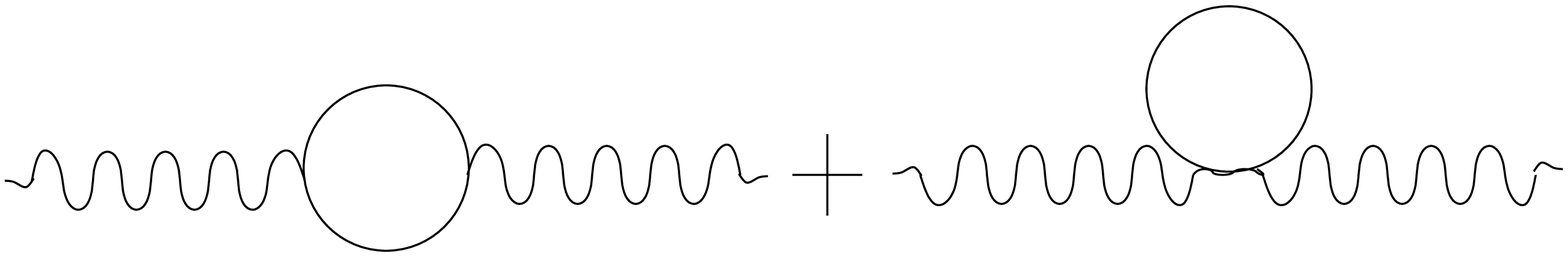}
\end{center}
\caption{\label{fig:V34V43feyn}Vacuum polarization Feynman graphs. They are
equivalent to the vertex graphs with $V_{34}\times V_{43}$ and
$V_{43}\times V_{34}$, and to the one-vertex graphs with 
$V_{33}^g$ and $V_{44}^g$, respectively.}
\end{figure}

The graphs containing $V_{34}\times V_{43}$ were the hardest
stumbling block for this computation. As is well known, the
quadratic divergence of the second order Feynman graph of 
Fig. \ref{fig:V34V43feyn} cancels with the one of the
 first order seagull diagrams displayed in the same figure. 
The problem we have here is
the fact that the quadratic divergence is proportional to
$A_\mu A^\mu$ and this has a nonintegrable singularity at
$r=0$. While the ultraviolet divergence appears only after
summation over $n$ and integration over $p$, the radial integrations
already appear in the partial waves. On the other hand,
all we have to ensure is the fact that the final ultraviolet 
divergence is a wave function renormalization and does not contain
terms proportional to $A_\mu A^\mu$. Furthermore our subtraction must
render the total result finite. The solution of this problem was found
essentially by trial and error and amounts to subtracting
in the partial waves a second order term
\bea\nonumber
&&\int r\, dr \int r'\,dr' V_{34}(r)
G^0_{n}(r,r',\kappa_W)\left[V_{43}(r')-\frac{r^2}{r'^2}V_{43}(r)
\right]\left[G_n^0(r',r,\kappa_H)
\right]^2
\\\nonumber&=&
\int r\, dr \int r'\,dr' 2n\frac{(A(r)+1)}{r^2}
G^0_{n}(r,r',\kappa_W) \\
&&\hspace*{15mm} \times 2n\frac{(A(r')-A(r))}{r'^2}\left[G_n^0(r',r,\kappa_H)
\right]^2
\pkt\eea 
and analogously for $V_{43}\times V_{34}$ or 
$\kappa_W\leftrightarrow \kappa_H$.
{\em This contains already the
 subtraction of 
first order divergences} that would arise from the gauge field
parts of $V_{33}$ and $V_{44}$. So the numerical results get finite
{\em without subtraction of 
those first order terms.} However, this subtraction does not 
correspond exactly to the subtraction of the Feynman graphs of Fig.
\ref{fig:V34V43feyn}, it only takes care in the correct
way of the divergent terms. The finite terms will be different. So
when we add back the subtracted term after covariant regularization, we have
to compute it {\em in the same way as we 
did the subtraction}, not by evaluating
the Feynman graphs with the external gauge field legs.
This we will describe now.

The Feynman integral corresponding to the second order diagram
is given, in our gauge, by
\be
g^2\int \frac{d^4 p}{(2\pi)^4}\frac{4p_\mu p_\nu}{(p^2-m_H^2
+i\epsilon)((p-q)^2-m_W^2+i\epsilon)}
\kma\ee
where $q_\mu$ is the external momentum.
The graph evaluates to
\be
i\frac{g^2}{4\pi^2}\int_0^1 d\omega\left[\frac{2 \calq^2}{4-\epsilon}
\left\{L_\epsilon +\frac{1}{2}-\ln\frac{\calq ^2}{\mu^2}\right\}+
q_\mu q_\nu \omega^2\left\{L_\epsilon-\ln\frac{ \calq^2}{\mu^2}
\right\}\right]
\kma\ee
with $\calq^2=\omega m_W^2+(1-\omega)m_H^2-\omega(1-\omega)q^2$.
Our gauge potential is transverse, so in the following
we do not have to keep track of the $q_\mu q_\nu$ terms. 
The external fields as we have used them in the subtraction
 can be written as
\bea \nonumber
&&\epsilon_{ik}\hat x_k\epsilon_{jl}\hat x_l\frac{A(r)+1}{gr}
\frac{A(r')-A(r)}{g r'}
\\ 
&&=\epsilon_{ik}\hat x_k\epsilon_{jl}\hat x_l
\left[\frac{A(r)+1}{gr}
\frac{A(r')-1}{g r'}-\frac{(A(r)+1)^2}{gr}\frac{1}{gr'}\right]
\pkt\eea
Taking the Fourier transform and setting $q=q'$ because of 
momentum conservation we have
\bea\nonumber
&&\frac{1}{g^2}
\epsilon_{ik}\hat q_k\epsilon_{jl}\hat q_l
(2\pi)^2\left[\int dr J_1(qr)(A(r)+1)\int dr' J_1(qr')(A(r')+1)
\right.\\
&&\left.- \int dr J_1(qr)(A(r)+1)^2\int dr' J_1(qr')\right]
\pkt\eea
Contracting with $g_{\mu\nu}$ we have
\bea
\nonumber &&
-\frac{1}{g^2}
(2\pi)^2\left[\int dr J_1(qr)(A(r)+1)\int dr' J_1(qr')(A(r')+1)
\right.\\
&&\left.- \int dr J_1(qr)(A(r)+1)^2\int dr' J_1(qr')\right]
\pkt\eea
At first we consider  the divergent part of the Feynman graph.
It is given by
\be
i\frac{g^2 }{16\pi^2}g_{\mu\nu}
L_\epsilon\left[\frac{1}{3} q^2-(m_H^2+m_W^2)\right]
\pkt\ee
This is to be multiplied with the external leg factor given above and to
be integrated with $d^2q/(2\pi)^2$. The part independent of $q^2$
vanishes upon this integration:
\bea \nonumber
&&\int q\,dq \left[\int dr J_1(qr)(A(r)+1)\int dr' J_1(qr')(A(r')+1)
\right.\\\nonumber
&&\left.- \int dr J_1(qr)(A(r)+1)^2\int dr' J_1(qr')\right]
\\\nonumber
&=& 
 \left[\int dr (A(r)+1)\int dr' (A(r')+1)
\right.\\
&&\left.- \int dr (A(r)+1)^2\int dr'\right]\frac{1}{r}
\delta(r-r')=0
\kma\eea
where we have used
\be
\int q\,dq J_n(qr)J_n(qr')=\frac{1}{r}\delta(r-r')
\pkt\ee
This cancellation implies, as it should, the absence of a 
divergent term proportional to
$A_\mu A^\mu$ in our subtraction. 
The remaining divergent term is proportional to $q^2$.
We note that now $q^2=-\bfq^2$. Furthermore,
$\int dr J_1(qr)=1/q$, $\int dq q^2 J_1(qr)=0$ and
$q J_1(qr)=-dJ_0(qr)/dr$.
Including all factors we find
\bea \nonumber
&&
i\frac{L_\epsilon}{16 \pi^2}\frac{1}{3}2\pi\int q^3 dq
\left[\int dr J_1(qr)(A(r)+1)\int dr' J_1(qr')(A(r')+1)
\right.
\\\nonumber
&&\left.- \frac{1}{q}\int dr J_1(qr)(A(r)+1)^2\right]
\\\nonumber
&&=i\frac{L_\epsilon}{16 \pi^2}\frac{1}{3}2\pi\int q\, dq 
\int dr\frac{dJ_0(qr)}{dr}(A(r)+1)\int dr' 
\frac{dJ_0(qr')}{dr'}(A(r')+1)
\\
&&=i\frac{L_\epsilon}{16\pi^2}\frac{1}{3}2\pi\int \frac{dr}{r}
\left[A'(r)\right]^2
\pkt\eea
This is just the kinetic term of the gauge field, multiplied with
the wave function renormalization factor
$(g^2/3)L_\epsilon/8\pi^2$.
In the $\overline{MS}$ scheme this term is dropped.
We are left with the finite part.
Denoting
\bea
2\pi\int dr J_1(qr)(A(r)+1)=\chi_A(q)
\kma\\
2\pi\int dr J_1(r)(A(r)+1)^2=\chi_{A^2}(q)
\kma\eea
the finite part is given by
\be
\Delta \sigma_{{\rm fl},3434} =
\frac{1}{8\pi^2}\int \frac{d^2q}{(2\pi)^2}
\int_0^1 d\omega
\calq^2\left[\ln\frac{\calq^2}{\mu^2}-1\right]
\left[\chi_A^2(q)-\frac{2\pi}{q}\chi_{A^2}(q)\right]
\kma\ee
with 
$\calq^2=\omega m_H^2+(1-\omega)m_W^2+\omega(1-\omega)\bfq^2$. 
The integration over $d\omega$ can be done analytically, we do not
display the somewhat lengthy result.


\subsection{Graphs with two vertices: gauge field contributions in $V_{33}$
and $V_{44}$.}
\label{gaugefieldinv33v44}
\begin{figure}[htb]
\begin{center}
\includegraphics[scale=0.35]{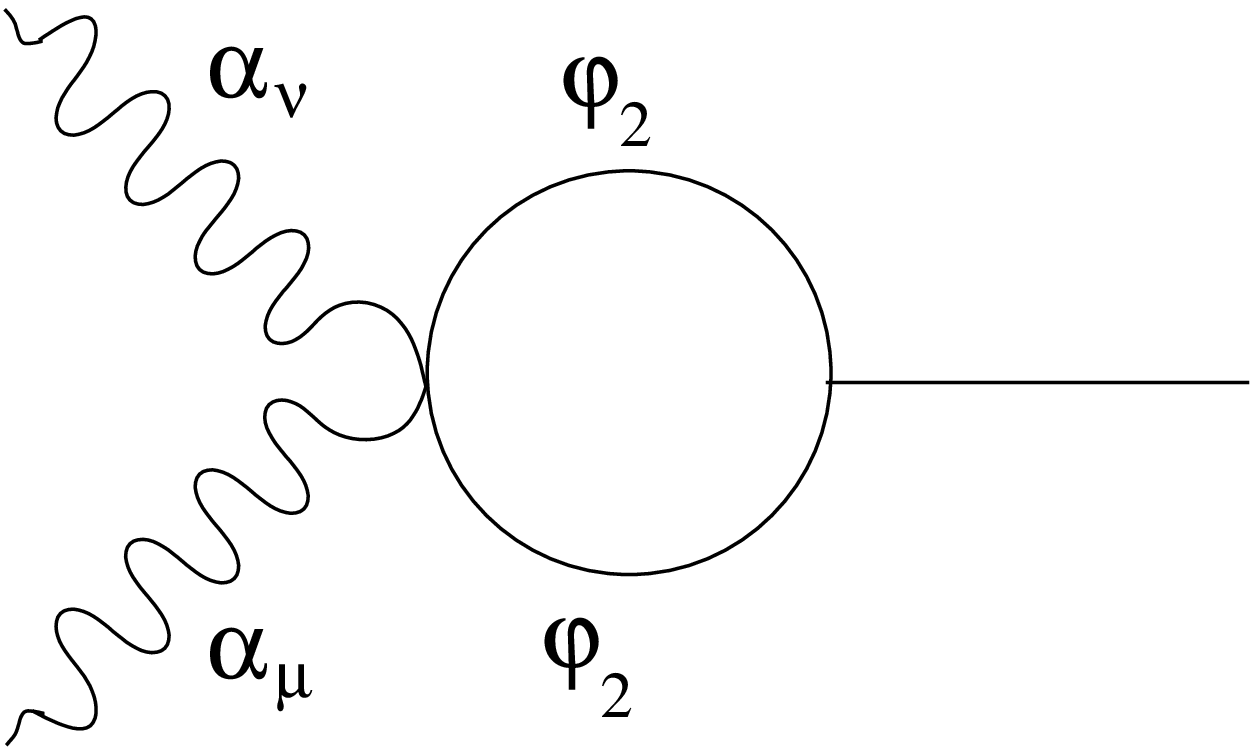}
\;\;\;
\includegraphics[scale=0.35]{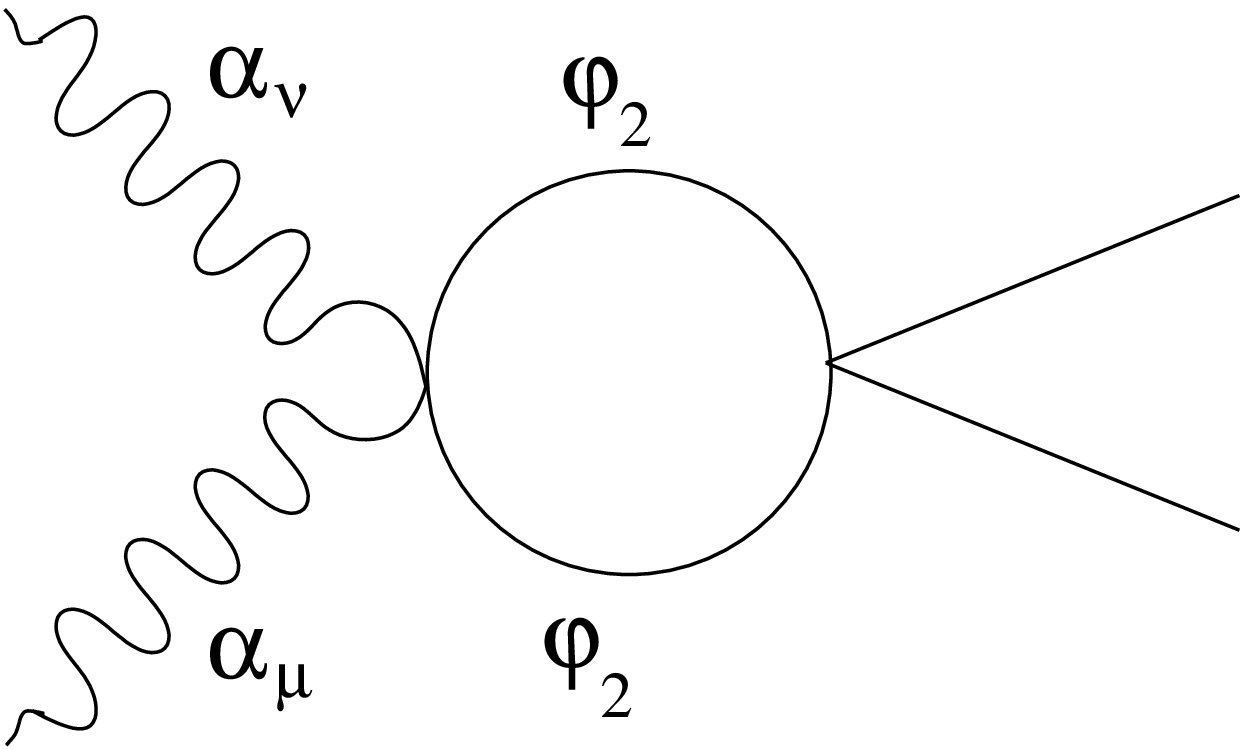}
\\
\includegraphics[scale=0.35]{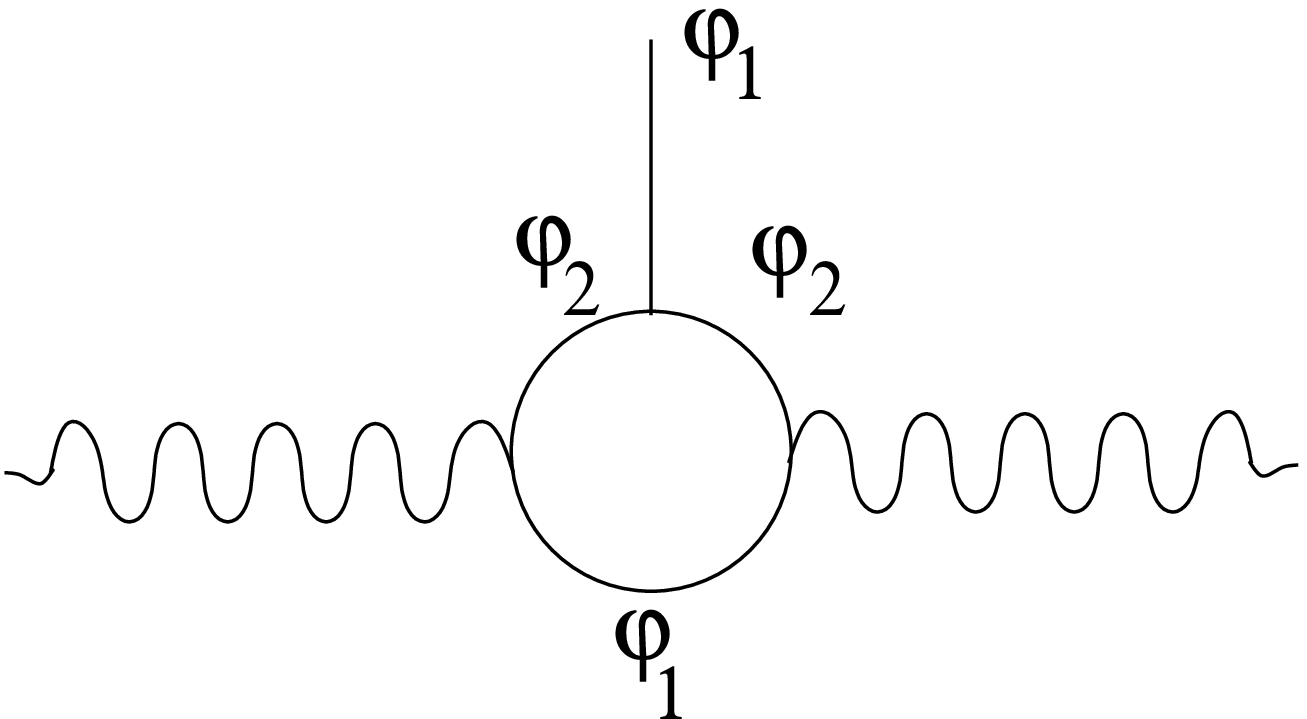}
\;\;\;
\includegraphics[scale=0.35]{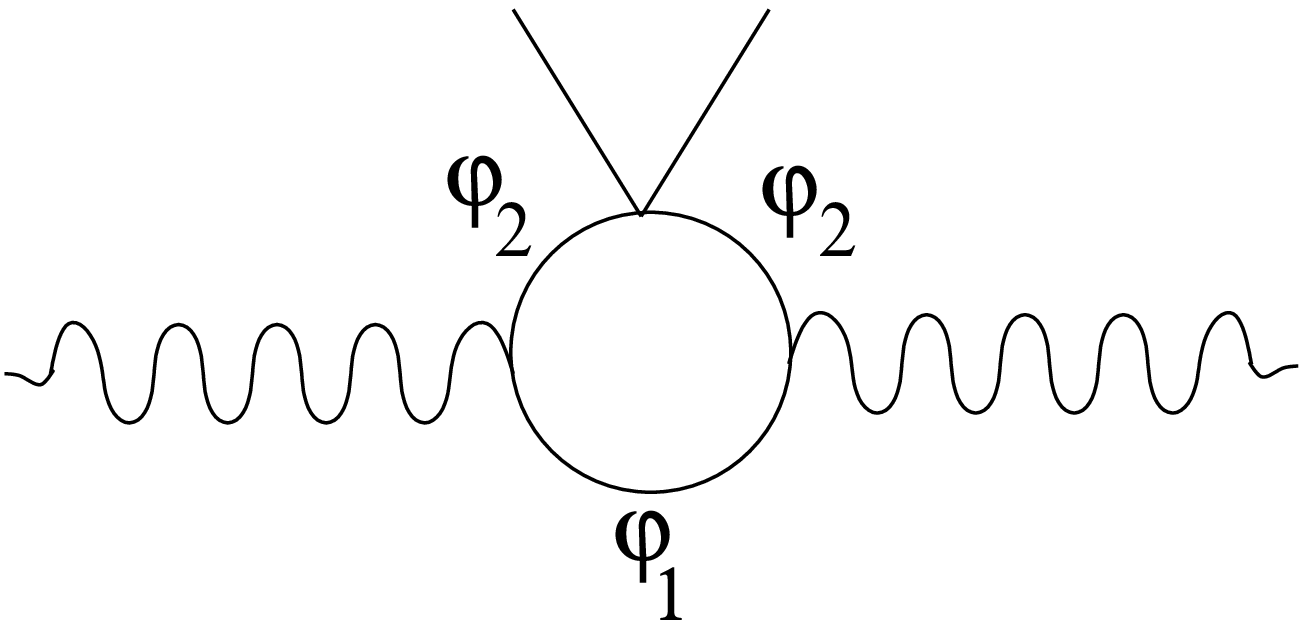}
\end{center}
\caption{\label{fig:V33gV33Hfeyn}
Feynman graphs corresponding to the vertex graph with
  $V_{33}^g\times V_{33}^H$ and the three-vertex graphs that cancel their
logarithmic divergences. The cancellation for the graphs
related to  $V_{44}^g\times V_{44}^H$ is analogous, with the replacement
 $\phi_1\rightarrow \varphi_2$ in the internal lines. }
\end{figure}

\begin{figure}[htb]
\begin{center}
\includegraphics[scale=0.30]{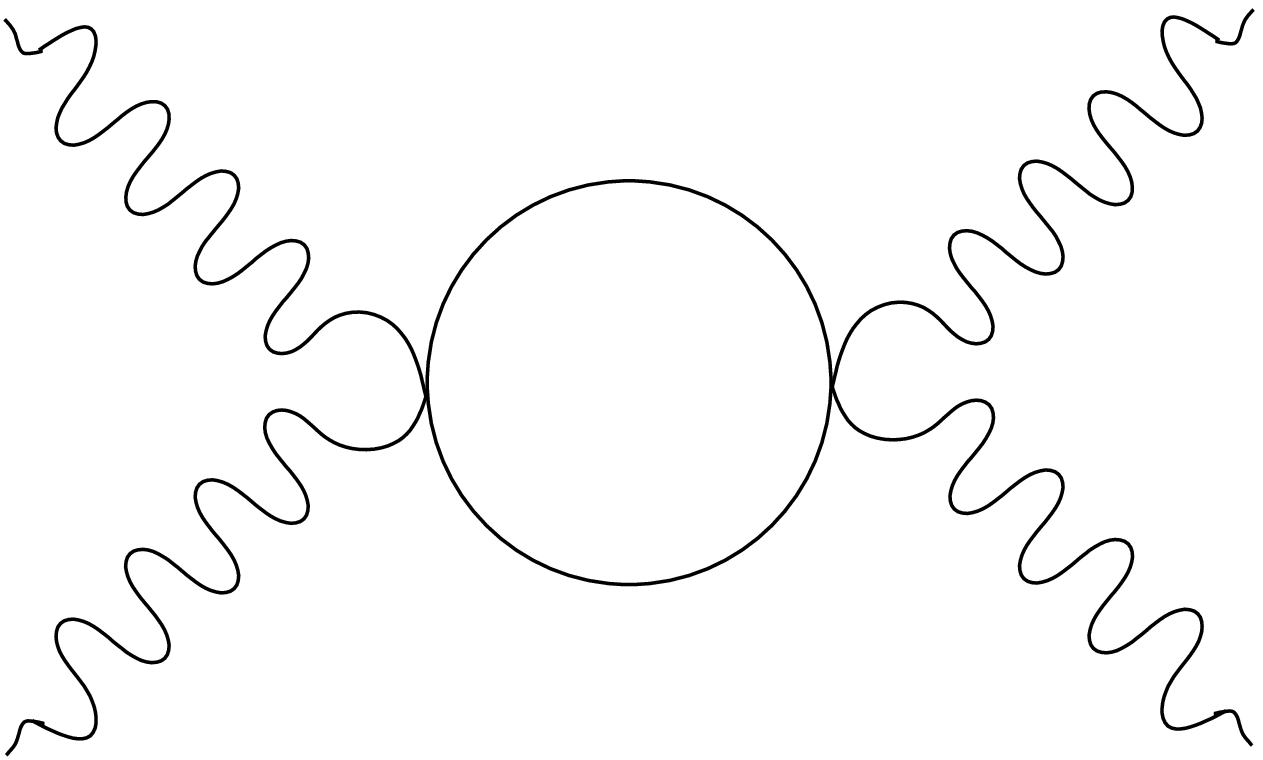}
\;\;\;
\includegraphics[scale=0.30]{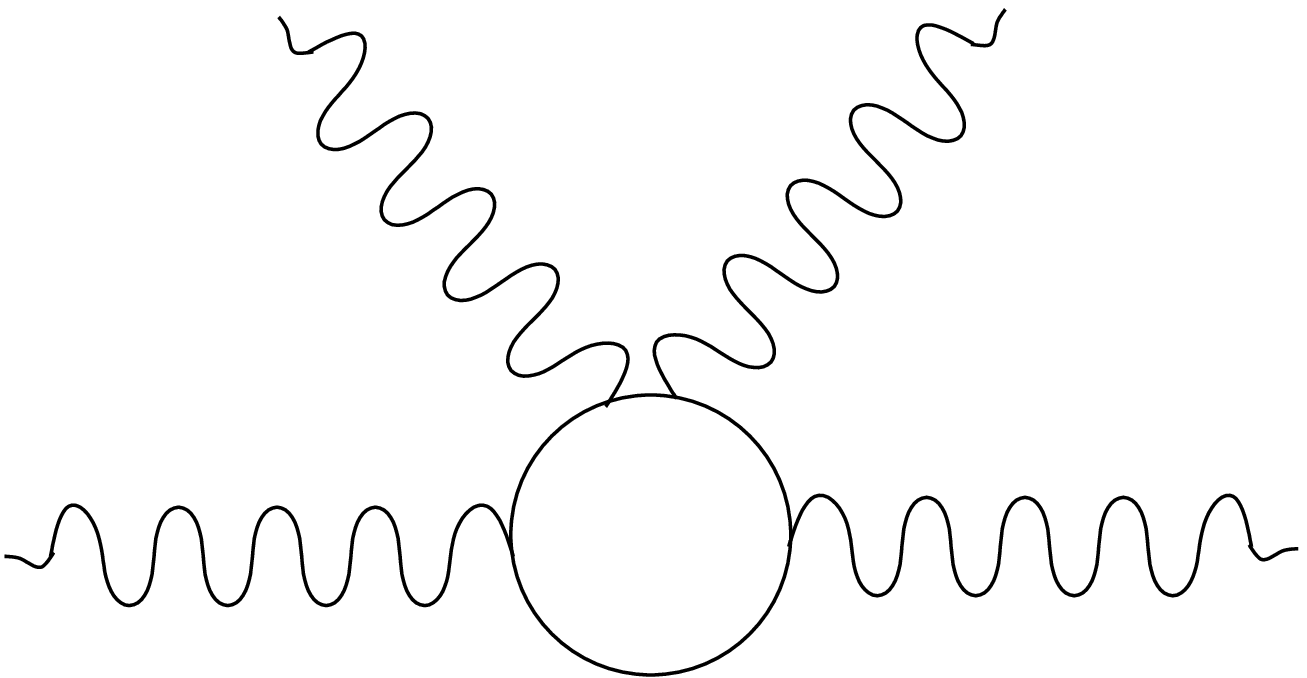}
\;\;\;
\includegraphics[scale=0.30]{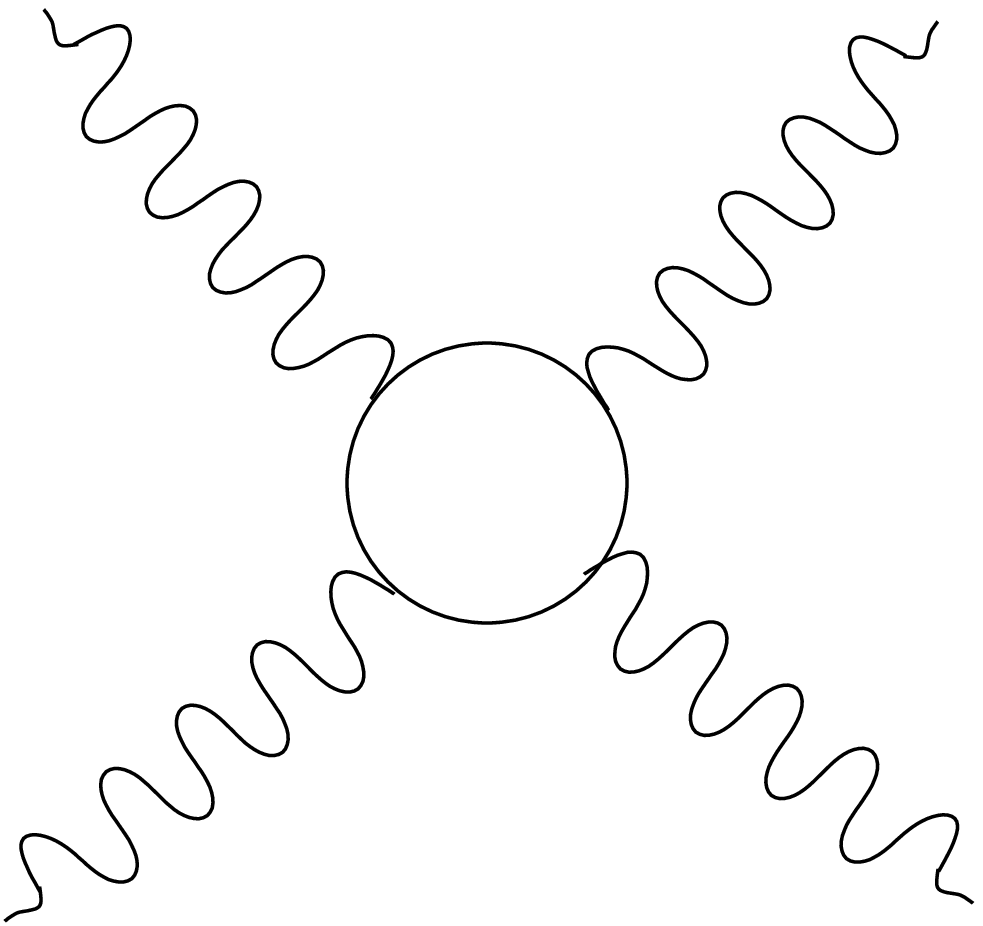}
\end{center}
\caption{\label{fig:V33gV33gfeyn}
Feynman graphs corresponding to the vertex graphs with
  $V_{33}^g\times V_{33}^g$ and $V_{44}^g\times V_{44}^g$ and the 
higher order graphs that cancel their
logarithmic divergences.  }
\end{figure}

There are some  contributions with two vertices which we have left
out. Denoting the gauge part of $V_{33}$ with $V_{33}^g$ 
with $V_{33}^g(r)=(A(r)+1)^2/r^2$ and analogously for
$V_{44}$ we have left out second order
 the diagonal parts $V_{33}^g\times V_{33}^H$
and $V_{33}^g\times V_{33}^g$,and likewise for the $44$ (Higgs) channel.
These diagrams with seagull type vertices combine with higher
order graphs to give finite results. So {\em 
they do not have to be subtracted}.
The corresponding Feynman graphs are displayed in Fig. 
\ref{fig:V33gV33Hfeyn}
and \ref{fig:V33gV33gfeyn}.
If one looks at the Feynman graphs there are graphs with more
than three vertices that are superficially divergent.
In fact these combine in such a way that their divergences are cancelled.
So with the subtractions described in the previous subsections
we have done the necessary steps towards computing the finite result.
Indeed with these subtractions the numerical results get finite, and we
have already presented the finite expressions by which these
graphs are to be replaced. 


\setcounter{equation}{0}
\section{Numerical results}
\label{numerics}

We have carried out the numerical programme as described in the
previous sections, for values of $\xi=m_H/m_W$ between
$0.5$ and $2$. 

 The programme starts with computing a fundamental
system of the mode functions
$h_{n,i}^{\alpha\pm}(r)$ for the $4\times 4$ gauge-Higgs and
the Faddeev-Popov sector. We have used $2000$ grid points in
$r$, up to $r_{\rm max}=30$, as we did already
for the classical profiles $f(r)$ and $A(r)$.
These are the basis for the partial wave Green's functions.
This computation is identical to the one performed already in
Refs. \cite{Baacke:1994bk,Baacke:2008zx}.

\begin{figure}[htb]
\vspace{12mm}
\begin{center}
\includegraphics[scale=0.50]{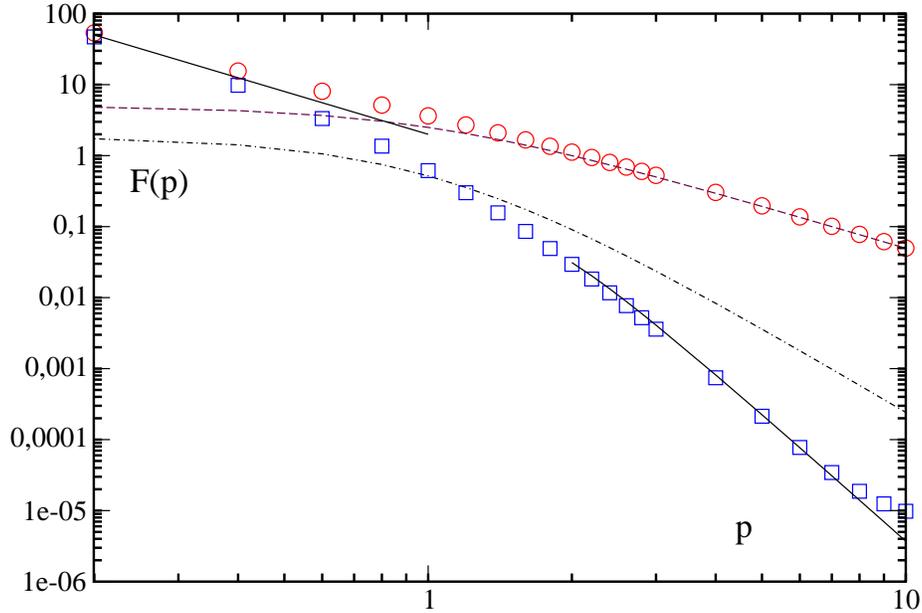}
\end{center}
\hspace{3mm}
\caption{\label{fig:fp1000} 
The integrand function $F(p)$ defined in Eq.
\eqn{fpdef}, for the gauge-Higgs sector: circles: the unsubtracted function;
dashed line: one vertex contribution; dash-dotted line:
two-vertex contribution; squares: subtracted function;
straight line at small $p$: translation mode pole $2/p^2$;
solid line at large $p$: asymptotic fit.}
\end{figure}

\begin{figure}[htb]
\hspace{10mm}
\begin{center}
\includegraphics[scale=0.50]{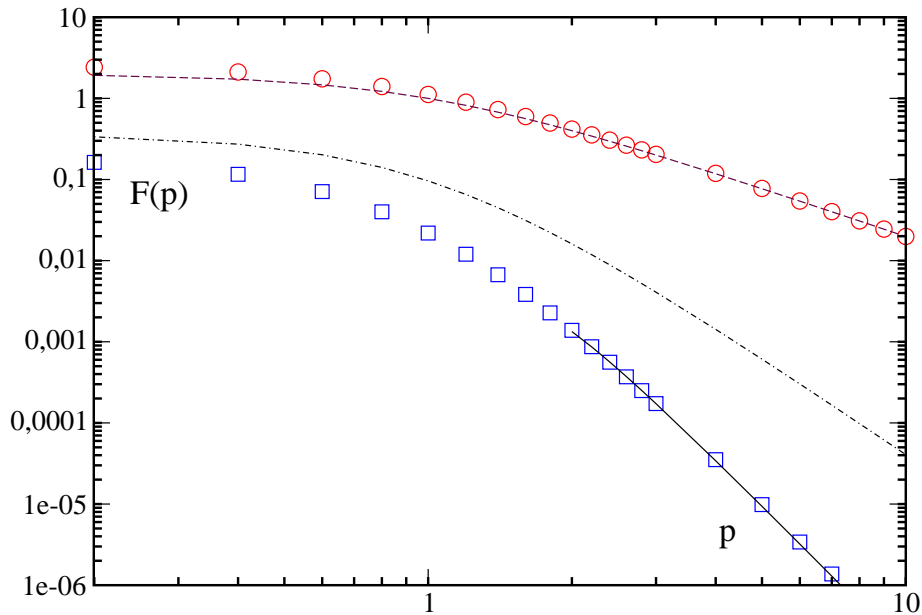}
\end{center}
\hspace{3mm}
\caption{\label{fig:fp1000FP} 
The integrand function $F(p)$ defined in Eq.
\eqn{fpdef}, for the Faddeev-Popov sector.
Symbols and lines as in Fig. \ref{fig:fp1000}, except that
there is no translation mode pole at small $p$.}
\end{figure}

We have computed the partial wave Green's function
and the related integrals up to $n=\bar n=35$. Contributions of higher $n$
were included by fitting the data between $\bar n-5$ and $\bar n$ using
power fits $A n^{-3}+B n^{-4}+C n^{-5}$, and by appending
the sum for $\bar n < n <\infty$ on the basis of these fits. 
The perturbative subtractions were done in the partial waves,
the sum over the unsubtracted and the subtracted partial wave contributions
constitute the unsubtracted and subtracted functions $F(p)$. 
These are displayed, for $\xi=1$ in Figs.
\ref{fig:fp1000} and \ref{fig:fp1000FP},
for the gauge-Higgs and the Faddeev-Popov sector, respectively.
All the general features of these figures are similar for other values
of $\xi$. In these figures we also present the perturbative contributions
of first and second order, summed separately. These can and have been
used for cross-checks against semi-analytical
results:

The numerical sum over the first-order subtractions 
can be checked against the result obtained
 by summing the partial wave contributions analytically.
 One finds
\be
\sum_{i=1,4} F_{ii}(p)=-\left[\frac{3m_W^2+m_H^2/2}{2\kappa_W^2}
+\frac{3m_H^2/2}{2\kappa_H^2}\right]\int r\, dr (f^2(r)-1)
\pkt\ee
The Faddeev-Popov contribution, which was
computed separately, behaves (including the factor -2)
as
\be
 F_{55}(p)=\frac{m_W^2}{\kappa_W^2}\int r\, dr (f^2(r)-1)
\pkt\ee
The integrals over $r$,  here and below, are
computed numerically using the classical profiles.

For the second-order contributions we cannot do the
sum over partial waves analytically, this involves the numerical
Fourier transforms of various functions of the classical
profiles; but we can easily
check  the asymptotic behaviour.
It is given by
\bea\nonumber
\sum_{ij}F_{ijij}(p)&=&\left[\left(m_W^4+(m_W^2+m_H^2/2)^2+9m_H^4/4\right)
\int r\, dr (f^2(r)-1)^2              \right.
\\\nonumber
&&+8 m_W^2\int r\, dr f'(r)^2
+8 m_W^2\int r\, dr f^2(r)\frac{(A(r)+1)^2}{r^2} 
\\
&& \left. +4\int r\, dr \frac{ \left [A'(r)\right]^2}{r^2}\right]\frac{1}{4p^4}
\pkt\eea
The first term collects the diagonal
contributions that can be derived analytically by summing up
the partial waves. The three other contributions are retrieved from
the Feynman graphs by doing the transverse loop momentum integrations
only. They collect the $1313$ and $2323$ contributions,
the $1414$ and $2424$ contributions, and the $3434$ contribution,
respectively.
Of course these can also be computed and verified separately.
The asymptotic behaviour of the 
Faddeev-Popov contribution is given by
\be
F_{5555}(p)=2\frac{m_W^4}{4p^4}\int r\, dr (f^2(r)-1)^2
\pkt\ee
These semianalytic results can be and have been checked against the numerical
subtractions, thus verifying prefactors and signs.

After the subtractions the function $F_{\rm sub}(p)$ behaves
as $p^{-6}$ and the integral over $p^4 dp$ can be done
in order to obtain the subtracted part of the string tension.
Finally we have to add back the 
subtracted terms in a covariantly regularized and
renormalized form given in the previous section.

As we have already mentioned, the integrands 
for the $p$ integration are displayed in Figs.
\ref{fig:fp1000} and \ref{fig:fp1000FP},
for the gauge-Higgs and the Faddeev-Popov sector, respectively.
The figures show the unsubtracted functions, the first order
and second order contributions and the subtracted functions.
Note that the functions differ, at large $p$, by several orders of 
magnitude, so these subtractions are quite delicate. 

The gauge-Higgs sector displays a
$2/p^2$ behaviour for small $p$, which is of course related to
the $2$ translation modes of the two-dimensional solution
(the ``instanton''). Here it merges into a continuum of transverse
string oscillations. The small-$p$ pole remains of course after subtractions
and causes the gauge Higgs sector to be much more important than
the Faddeev-Popov sector. Indeed, the  finite parts obtained 
after integration with
$p^3 dp/4\pi$ are much larger for the gauge-Higgs
than for the Faddeev-Popov sector. Obviously this important  contribution of 
the translation mode is related to the transversal 
quantum oscillation of the string. 

At large $p$ the subtracted integrand should behave as $p^{-6}$. For
the Faddeev-Popov sector this is realized in ideal form.
For the gauge-Higgs sector there are deviations for $p > 6$, which remain
even if higher partial waves are included. So they seem to be
caused by some very small numerical deficiencies in the low partial waves
Their origin is difficult to localize. As these contributions only
appear above $p\simeq 6$  we have done least-square fits 
of the form $A/p^6+B/p^8$ based on the data points between $p=2$ and
$p=6$. These fits then were used in order to append the integrals
from $p=6$ to $\infty$. The fits are displayed in Figs. 
\ref{fig:fp1000} and \ref{fig:fp1000FP} for the gauge-Higgs 
and Faddeev-Popov sectors, respectively. 

The integrals over 
the subtracted part of the function $F(p)$, including the asymptotic tail
based on the fits are given in Table \ref{table:finalresults}, 
they are denoted as $\Delta \sigma_{\rm sub}$.

Now that we have computed the subtracted integrals we have to add
back the regularized and renormalized  divergent contributions.
Those with one vertex are given, in unrenormalized form, by
\be
\Delta \sigma^{(1)}=
-\frac{m_W^2}{32\pi^2}\left\{ (L_\epsilon + 1-\ln\frac{m_W^2}{\mu^2})(1+2\xi^2)
 -\frac{3}{2}\xi^2\ln\xi^2\right\}\int d^2x (f^2(r)-1) 
\pkt\ee
One may choose the renormalization such as to omit these 
tadpole contributions entirely, as one would do in the 
$1+1$ dimensional theory, using normal ordering
the field operators. Here we use the strict
$\overline{MS}$ scheme and choose the renormalization scale as
$\mu^2=m_W^2$. Therefore we have to add back
\be
\Delta \sigma^{(1)}_{\rm fin}=
-\frac{m_W^2}{32\pi^2}\left\{ (1-\ln\frac{m_W^2}{\mu^2})(1+2\xi^2)
 -\frac{3}{2}\xi^2\ln\xi^2\right\}\int d^2x (f^2(r)-1) 
\ee  
As we separately present this contribution in Table \ref{table:finalresults}
the reader may easily change this contribution according to her/his
own preferences.

\begin{figure}[htb]
\vspace{15mm}
\begin{center}
\includegraphics[scale=0.5]{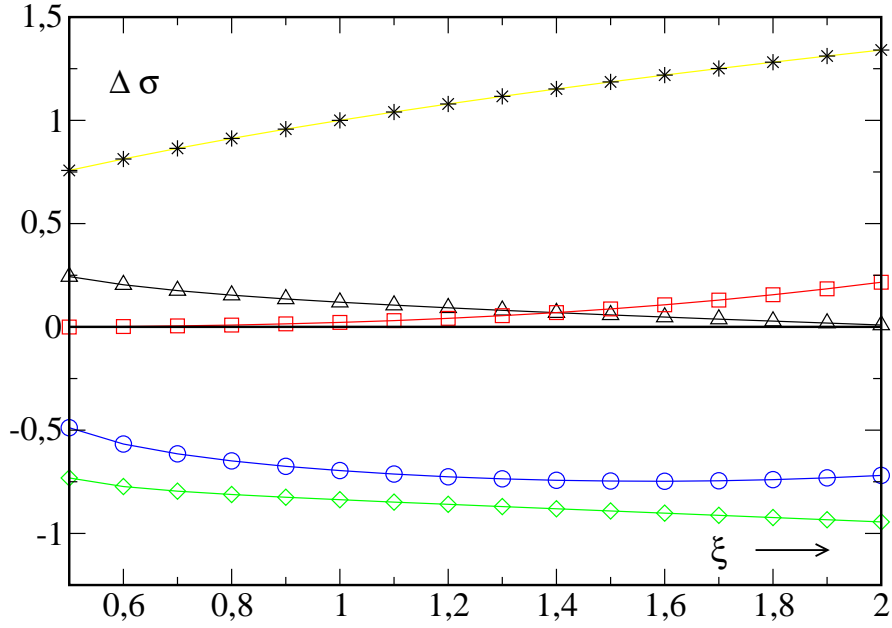}
\end{center}\vspace{5mm}
\caption{\label{fig:deltasigma}
The correction to the string tension as a function of
$\xi=m_H/m_W$. Squares:contributions with one vertex,
$\Delta \sigma^{(1)}$; triangles: contributions with two vertices
$\Delta\sigma^{(2)}$; diamonds: subtracted contribution
$\Delta\sigma_{rm sub}$, circles: total one-loop correction
$\Delta \sigma_{\rm tot}$, asterisks: the classical string tension
mutiplied by $g^2/\pi$. All string tensions are in units of $m_W^2$.}  
\end{figure}

The separate results for the graphs with one vertex,
$\Delta \sigma^{(1)}_{\rm fin}$, 
for the graphs with two vertices $\Delta \sigma
^{(2)}_{\rm fin}$, the subtracted contribution $\Delta \sigma_{\rm sub}$ 
and the total one-loop contribution $\Delta \sigma_{\rm tot}$ are  
listed  in Table \ref{table:finalresults} and
displayed in Fig. \ref{fig:deltasigma} . We should like to recall
that the subtracted contribution still includes some gauge field graphs
with two vertices, which by themselves are divergent, but
 whose divergences are cancelled by higher order
vertex graphs, as discussed in subsection \ref{gaugefieldinv33v44}.

\section{Summary}
\label{summary}
We have presented here an exact, albeit numerical, computation
of the one loop corrections to the string tension of the
the Nielsen-Olesen vortex in the $3+1$ dimensional Abelian Higgs model,
taking into account the fluctuations of gauge, the Higgs 
and the Faddeev-Popov fields.
One of the main complications arouse from the fact, that the divergences
of graphs with external gauge field loops require cancellations of Feynman
graphs with different numbers of external vertices. This
has required an extensive discussion of vertex graphs and
vacuum Feynman graphs. On the numerical side, we had
to adapt a previously developed
computation scheme to this new situation. We had to find
and have found  a way of 
implementing the necessary cancellations, which are relatively
straightforward when done analytically, into the numerical
procedure. 

The size of the corrections of course depends on the renormalization
scheme and renormalization conditions. We here have adapted
the $\overline{MS}$ scheme with $m_W^2$ as the renormalization scale.
The corrections are sizeable, but small with respect to the classical
string tension as long as the gauge coupling $g$ is smaller 
than unity, which may be considered as a reasonable assumption.
Unlike in the case of the fermionic corrections we do not have
at our disposal an extra parameter, the Yukawa coupling,
 that could render the corrections important for heavy fermions.
The parameters are fixed already at the classical level.

Of course, in the present situation of cosmic string phenomenology
a precise information on the string tensions and corrections to it
cannot be considered as very important. Indeed we consider
as our main result that we presented a  method for computing 
such corrections in a gauge theory with all its technical complications.
The situation may be different if these corrections are computed at
finite temperature, as it may be realistic in primordial cosmology. 
Such computations, using similar techniques, have been done
for bubble nucleation in Ref. \cite{Baacke:1993ne}.
In the high-temperature approximation to the electroweak theory
the corrections to the transition rate were found 
to be huge \cite{Kripfganz:1994ha,Baacke:1995bw}.

A further application of the methods presented here may be the
investigation of the r\^ole of quantum fluctuations for the
electroweak string, in particular in the context of its stabilization
\cite{Nagasawa:2002at}


\newpage

\begin{appendix}
\noindent
{\LARGE \bf Appendix}
\setcounter{equation}{0}

\section{Partial wave
 decomposition of the free \\Green's function}
\label{G0partial}
We shortly 
recall the partial wave decomposition of the free
Green's function, a decomposition that is used repeatedly in the
subtraction procedure. We here consider the Green's function in
two dimensions, with an effective mass $\kappa^2=m^2+p^2$;
the one in four dimensions is obtained
by replacing $p^2 = \nu^2+k_3^2$ and by further
integrations over $\nu$ and $k_3$. We further omit the subscript
$\perp$, replacing  $\bfx_\perp\to \bfx$ and
$\bfk_\perp\to \bfk$. The free Green's function $G_0(\bfx,\bfx',\kappa)$ 
is defined as
\be
G_0(\bfx,\bfx',\kappa)=
\int\frac{d^2k}{(2\pi)^2}\frac{e^{i \bfk\cdot (\bfx-\bfx')}}
{\bfk^2+\kappa^2}
\pkt\ee
It can be computed readily, using formulas
(9.1.18) and (11.4.44) of Ref. \cite{Abramowitz}:
\bea\nonumber
G_0(\bfx,\bfx',\nu)&=&\frac{1}{4\pi^2}\int_0^\infty k dk
\int_0^\pi d\phi \frac{2\cos(k R \cos\phi)}{k^2+\kappa^2}
\\ 
&=& \frac{1}{2\pi}\int_0^\infty \frac{dk~k J_0(kR)}{k^2+\kappa^2}
= \frac{1}{2\pi}K_0(\kappa R)
\kma\eea
with $R=|\bfx-\bfx'|$.
Furthermore this may be expanded, using Eq. (4) in section
7.6.1 of Ref. \cite{BatemanHTF} as
\be
G_0(\bfx,\bfx',\nu)=\frac{1}{2\pi}\sum_{k=-\infty}^\infty
K_n(\kappa r_>)I_n(\kappa r_<)e^{i n \phi}
\pkt\ee


\section{First order contributions to the partial waves}
\setcounter{equation}{0}\label{G1}
Using the formula for the free Green's function derived in
Appendix \ref{G0partial} the first order contribution to the
 Green's function in two dimensions is given by
\bea \nonumber
G^{(1)}(\bfx,\bfx',p)&=&
-\int d^2x'' G_0(\bfx,\bfx'',\kappa)V(r'')G_0(\bfx'',\bfx',\kappa)
\\\nonumber
&=& -\frac{1}{4\pi^2}\int dr'' r'' d\phi''\sum_{n=-\infty}^\infty
\sum_{n'=-\infty}^\infty K_n(\kappa r_>)I_n(\kappa r_<)
e^{in(\phi-\phi'')}
\\&&\hspace*{20mm}\times V(r'') K_{n'}(\kappa r'_>)I_{n'}(\kappa r'_<)
e^{in'(\phi''-\phi')}
\\\nonumber
&=& -\frac{1}{2\pi}\int dr'' r''\sum_{n=-\infty}^\infty
 K_n(\kappa r_>)I_n(\kappa r_<)
e^{in(\phi-\phi')}
\\\nonumber 
&&\hspace*{20mm}\times V(r'') K_n(\kappa r'_>)I_n(\kappa r'_<)
\eea
where $r_>=\max\{r,r''\}$, $r'_>=\max\{r',r''\}$ and similarly for
$r_<$ and $r'_<$. The ``potential'' $V(r)$ subsumes all vertex 
contributions appearing in first order.

We define
\be
G^{(1)}(\bfx,\bfx',p)=\frac{1}{2\pi}\sum_{n=-\infty}^\infty
e^{in(\phi-\phi')}G^{(1)}_n(r,r',p)
\ee
and therefore get
\be
G^{(1)}_n(r,r,p)=-\int dr'' r''
 K^2_n(\kappa r_>)I^2_n(\kappa r_<)
V(r'')
\pkt\ee
Using the integrals of Appendix \ref{integrals} this can be integrated
over $r~dr$
with the result
\bea\nonumber
\int_0^\infty dr ~r ~G^{(1)}_n(r,r,p)&=&-\int_0^\infty dr' r' V(r')
\\\nonumber
&&\hspace{-15mm}\times\frac{z}{2\kappa^2}\left\{
-2n\left[I_{n+1}(z)I_n(z)K_n^2(z)+K_{n+1}(z)K_n(z)I_n^2(z)\right]\right.
\\&&\hspace{-15mm}\left.+
z\left[K_{n+1}^2(z)I_n^2(z)-I_{n+1}^2(z)K_n^2(z)\right]\right\}
\kma\eea
with $z=\kappa r'$. Using the Wronskian relation
\be
I_n(z)K_{n+1}(z)+I_{n+1}(z)K_n(z)=\frac{1}{z}
\ee
this can be rewritten as
\bea \nonumber
\int_0^\infty dr ~r ~G^{(1)}_n(r,r,p)&=&-\int_0^\infty dr' r' V(r')
\frac{1}{2\kappa^2}\left\{-2n I_n(z)K_n(z)\right.
\\&&\left.+z\left[K_{n+1}(z)I_n(z)
-I_{n+1}(z)K_n(z)\right]\right\}\label{firstordercontribpartwav}
\\\nonumber
&=&-\int_0^\infty dr' r' V(r')\frac{1}{2\kappa}\frac{d}{d\kappa}
I_n(z)K_n(z)
\pkt\eea
This is the contribution of one partial wave to the sum over partial
waves, which is then to be integrated with respect to $\nu$ and
$k_3$ to yield the string tension in first order. 
However, the sum over partial
waves and the integrals are divergent;
the perturbative contribution has to be subtracted from
the full, nonperturbative contribution of each partial wave in order
to obtain covergent summations and integrations.
We may check  that formally the first order
perturbative contributions can be summed up and integrated
so as to reobtain the Feynman integral.
The sum over partial waves of this expression yields
\bea\nonumber
&&- \sum_{n=-\infty}^\infty \frac{1}{p}\frac{d}{dp}
\int_0^\infty dr' r' V(r')I_n(z)K_n(z)
\\\nonumber&=&
- 2\pi\frac{1}{2p}\frac{d}{dp}
\int_0^\infty dr' r' V(r')G_0(\bfx,\bfx,\kappa)
\\&=&
-\frac{1}{2p}\frac{d}{dp}
\int d^2 x V(|\bfx|)\int\frac{d^2k_\perp}{(2\pi)^2}\frac{1}
{\bfk_\perp^2+m^2+p^2}
\pkt\eea
Going back to section \ref{effectivetension} we see that we
have integrate this with $-\int p^3 dp/4\pi$.
We obtain
\bea\nonumber
\sigma^{(1)}&=&-\frac{1}{2}\int d^2 x V(|\bfx|)
\int\frac{p^3 dp d^2k_\perp}{(2\pi)^3}
\frac{1}{2p}\frac{d}{d\nu}\frac{1}{\bfk_\perp^2+p^2+m^2}
\\
&=&-\frac{1}{2}\int d^2 x V(|\bfx|)\int\frac{p\, dp d^2k_\perp}{(2\pi)^3}
\frac{1}{\bfk_\perp^2+p^2+m^2}
\\\nonumber
&=&-\frac{1}{2}\int d^2 x V(|\bfx|)\int \frac{d\nu d^3k}{(2\pi)^4}
\frac{1}{\bfk^2+\nu^2+m^2}
\kma\eea
where in the last step we have substituted $p^2=k_3^2+\nu^2$.
The result is indeed the first order vacuum graph.


\section{Second order contribution to the partial waves}
\setcounter{equation}{0}\label{G2}
Using similar steps as in Appendix \ref{G1} and using again the integrals
in Appendix \ref{integrals} and Wronskian relations, we find that
the second order contribution to the
partial waves is given by
\bea\nonumber
\int dr \, r G_n^{(2)}(r,r,p)&=&\sum_{ij}
\int dr' r' V^n_{ij}(r')\int dr'' r'' V^n_{ji}(r'')
I_{n_i}(\kappa_i r_<) K_{n_i}(\kappa_i r_>)
\\
&&\times\left(-\frac{1}{2\kappa_j}\frac{d}{d\kappa_j})\right)
I_{n_j}(\kappa_j r_<)K_{n_j}(\kappa_j r_>)
\kma\eea
where $r_<=\min(r',r'')$ and $r_>=\max(r',r'')$.
The computation is quite lengthy, as we start with an expression
where the values $r,r'$ and $r''$ appear in six different orderings. 
Here we  have written all relevant  indices, as the two propagators
have, in general, different masses and angular momenta.
We note that the differentiation with respect to 
$\kappa_j=\sqrt{p^2+m_j^2}$ can be written as a differentiation 
with respect to $p$ is the combinations $V_{ij}V_{ji}$ and
$V_{ji}V_{ij}$ are combined. If applied to the sum over
$i$ and $j$, this results in a double counting that has to be
compensated by a factor $1/2$.  So
\bea\nonumber
\int dr \, r G_n^{(2)}(r,r,p)&=&\frac{1}{2}
\left(-\frac{1}{2p}\frac{d}{dp}\right)\sum_{ij}
\int dr' r' V^n_{ij}(r')\int dr'' r'' V^n_{ji}(r'')
\\&&\times I_{n_i}(\kappa_i r_<) K_{n_i}(\kappa_i r_>)
I_{n_j}(\kappa_j r_<)K_{n_j}(\kappa_j r_>)
\pkt\eea


\section{Some integrals}
\label{integrals}
\setcounter{equation}{0}
In our calculations we repeatedly need some integrals over Bessel functions
which we write down here, without derivation. They can be obtained
from Eq. (11.3.29) of Ref. \cite{Abramowitz} by taking the limit
$k \to l$ (see there). The formulas are: 
\bea
&&\hspace{-10mm}\int_0^r dr r I_n^2(\kappa r)=\frac{z}{2\kappa^2}
\left\{-2 n I_{n+1}(z)I_n(z)+
z\left[I_n^2(z)-I_{n+1}^2(z)\right]\right\}
\kma\\
&&\hspace{-10mm}\int_r^\infty dr r K_n^2(\kappa r)=\frac{z}{2\kappa^2}
\left\{-2 n K_{n+1}(z)K_n(z)-
z\left[K_n^2(z)-K_{n+1}^2(z)\right]\right\}
,\\\nonumber
&&\hspace{-10mm}\int^r dr r I_n(\kappa r)K_n(\kappa r)=
\frac{z}{2\kappa^2}
\left\{n \left[K_{n+1}(z)I_n(z)-K_n(z)I_{n+1}(z)\right]\right.
\\
&&\left.\hspace{9mm}+z \left[I_n(z)K_n(z)+I_{n+1}(z)K_{n+1}(z)\right]\right\}
\kma\eea
with $z=\kappa r$. Note the limits of integration,
the third integral is an indefinite one.


\section{Some useful sums}
\label{somesums}
\setcounter{equation}{0}

We define the two-dimensional Green's function
$G_2^{(0)}(\bfx,\bfx',\nu^2)$ as the solution of the equation
\be
\left[-\Delta_2+m^2+\nu^2\right]G^2_{(0)}=\delta^2(\bfx-\bfx')
 \pkt\ee
We readily obtain
\be
G_2^{(0)}(\bfx,\bfx',\nu^2)=
\int\frac{d^2k}{(2\pi)^2}\frac{e^{i\bfk(\bfx-\bfx')}}
{\bfk^2+m^2+\nu^2}
\pkt\ee
The integral may be done explicitely with the result
\be
G_2^{(0)}(\bfx,\bfx',\nu^2)=\frac{1}{2\pi}K_0(\kappa R)
\kma\ee
with
\be
R=|\bfx-\bfx'|=\sqrt{r^2+r'^2-2rr'\cos(\varphi-\varphi')}
\pkt\ee
Furthermore the Gegenbauer expansion of the modified Bessel
function yields
\be
G_2^{(0)}(\bfx,\bfx',\nu^2)=\frac{1}{2\pi}\sum_{n=-\infty}^\infty
e^{in(\varphi-\varphi')}I_n(\kappa r_<)K_n(\kappa r_>)
\pkt\ee
The limit $\bfx\to\bfx'$ of the Green' s function does not exist.
However, what we need is 
\be
-\frac{1}{2\kappa}\frac{d}{d\kappa}
\sum_{n=-\infty}^\infty I_n(\kappa r)K_n(\kappa r)
\pkt
\ee
We note that 
\be
-\frac{1}{2\kappa}\frac{d}{d\kappa}K_0(\kappa R)=
\frac{R}{2\kappa} K_1(\kappa R)
\pkt\ee
The limit $R\to 0$ exists and one obtains
\be
-\frac{1}{2\kappa}\frac{d}{d\kappa}
\sum_{n=-\infty}^\infty I_n(\kappa r)K_n(\kappa r)
=\frac{1}{2\kappa^2}\pkt
\ee


\section{Fourier-Bessel transforms}
\label{fourierbessel}
\setcounter{equation}{0}

For the evaluation of the perturbative contributions we need the
Fouriertransform of the external sources. Let us consider at first
a scalar field $\Phi(r_\perp)$, independent of Euclidean time
$\tau$ and $z$. In our application $\phi$ will be
$f(r_\perp)-1$, $f^2(r_\perp)-1$ etc..
We have
\bea\nonumber
\tilde \Phi (q_0,q_z,q_\perp)&=&\int d^4x e^{-ip\cdot x}\Phi(r_\perp)
\\&=& (2\pi)^2\delta(p_0)\delta(p_z)\int_0^\infty dr r
\int_{-\pi}^\pi d\varphi \Phi(r)e^{-iq_\perp r \cos \varphi}
\\&=& (2\pi)^3\delta(p_0)\delta(p_z)\int_0^\infty r dr \Phi(r)J_0(q_\perp r)
\pkt\eea 
In the following we will omit the trivial factor
$(2\pi)^2\delta(p_0)\delta(p_z)$ originating from the $\tau$ and $z$ 
integrations and write
\be
\tilde \Phi (q)=\chi_\Phi(q)=2\pi \int_0^\infty r dr \Phi(r)J_0(q r)
\ee
implying $q \sim q_\perp$.
The relation between the Fourier-Bessel transformation and the
inverse transformation implies the relation
\be
\int q dq J_l(q r)J_l(qr')=\frac{1}{r}\delta(r-r')
\ee
for arbitrary $l$. We deduce the Parseval equation
\be
\int \frac{d^2q}{(2\pi)^2}|\chi_\Phi(q)|^2=
\int \frac{dq q}{2\pi}|\chi_\Phi(q)|^2=
2\pi \int_0^\infty r dr |\Phi(r)|^2=\int d^2x  |\Phi(r)|^2
\pkt\ee
which can be used as a numerical cross check.
If one disregards $V_{34}=V_{43}$ and the gauge field parts of $V_{33}$ and
$V_{44}$ these Fourier-Bessel transforms are unproblematic.
We shortly discuss the various components.

In the diagonal we have the Higgs field parts which are proportional
to $f^2(r)-1$. This function goes to zero exponentially as 
$r\to \infty$, so the Fourier transform exists. It is found to decrease
exponentially at large $q$.

Considering the contribution of $V_{13}=V_{23}=\sqrt{2}m_W f'$ the function
$f'(r)$ decays exponentially as $r\to \infty$ and goes to a constant
as $r\to 0$. So $\tilde V_{13}(q)$ exists. It is found to decrease as $1/q^2$. 

The function $V_{14}(r)=\sqrt{2}m_W f (A+1)/r$ goes to a constant as
$r\to 0$ and decreases exponentially as $r\to \infty$. So again the
Fourier transform exists, and again it decreases as $1/q^2$.

All the Fourier transforms discussed up to now have to be folded with
the Feynman integral kernels (see Appendix \ref{secondorderkernels}) which behave
as $\ln q^2$ at large $q$. One easily convinces oneself, that the integrals
over the kernels times the squared Fourier transforms are well convergent.

The Fourier transform of the gauge field
\be
A^\perp_i=\epsilon_{ij} \hat x_j \frac{A(r)+1}{r}
\ee
is problematic. The function is not square integrable, due to the singularity
at $r=0$ its norm diverges logarithmically. So the 
Fourier transform is not square integrable either.
Nevertheless it can be computed.
The Fourier transform must be of the form
\be
\tilde A^\perp_i(q)=\epsilon_{ij}\hat q_j \chi_A(q)
\pkt\ee
We have
\bea\nonumber
\chi_A(q)&=&\epsilon_{ij}\hat q_j \tilde A^\perp_i(q) 
\\
&=& \epsilon_{ij}\epsilon_{ik}
\int d^2x \frac{A(r)+1}{r} 
e^{-i \vec q \vec r}\hat q_j\hat x_k
\\\nonumber
&=& \int_0^\infty r~ dr  \frac{A(r)+1}{r}\int^\pi_{-\pi} 
e^{-i  q  r \cos \varphi}\cos\varphi
\\\nonumber
&=& 2\pi \int_0^\infty r~ dr  \frac{A(r)+1}{r}
J_1(qr)
\pkt\eea
The function $\chi_A(q)$ behaves as $\beta q$ as $q\to 0$
with
\be
\beta=\pi \int_0^\infty r~dr  (A(r)+1) 
\kma\ee
and like $2\pi / q$ as $r\to \infty$.
This asymptotic behaviour makes the square norm of the Fourier
transform logarithmically divergent, as a reflection of the
logarithmic divergence of the norm of the field $A^\perp_i(r)$. 
The handling of the graphs with two external gauge field legs is discussed
in some detail is subsection \ref{secs:V34V43}.


\section{The second order kernels}
\setcounter{equation}{0}
\label{secondorderkernels}
The finite parts of the second order graphs all contain an integral
\be
\cali(m_a^2,m_b^2,q^2)=\int_0^1d\omega \log[\omega(1-\omega)q^2
+\omega m_a^2+(1-\omega)m_b^2]
\pkt\ee
We generally assume $m_a^2>0,m_b^2>0$. If furthermore $q^2>0$ then
we define
\be
\omega_\pm=\frac{1+m_a^2-m_b^2}{2q^2 }\pm
\sqrt{\frac{(1+m_a^2-m_b^2)^2+4m_b^2q^2}{4q^4}}
\pkt\ee
One easily realizes that $\omega_+>1$ and $\omega_- <0$.
With this definition  $\cali$ is given by
\bea \nonumber
\cali(ma^2,m_b^2,q^2)&=&\omega_+\ln\omega_+ + \omega_- \ln(-\omega_-)+
(1-\omega_+)\ln(\omega_+-1)
\\
&&+(1-\omega_-)\ln(1-\omega_-)+\ln q^2 - 2
\pkt\eea 
This expression is symmetric in $m_a^2$ and $m_b^2$.

If $q^2=0$ the integral is given, for $m_a^2\neq m_b^2$ by
\be
\cali(m_a^2,m_b^2,0)=\frac{m_a^2\ln m_a^2-m_b^2\ln m_b^2}{m_a^2-m_b^2}-1
\pkt\ee
Finally, if $m_a^2=m_b^2$ we have
\be
\cali(m_a^2,m_a^2,0)=\ln m_a^2
\pkt\ee

For small $q^2$ the integral behaves as $c_1+c_2 q^2$ where the constant
$c_1$ is of course identical to $\cali^2(m_a^2,m_b^2,0)$.
For large $q^2$ the integral behaves as $\ln q^2 +{\rm const.} + O(1/q^2)$.

The correction to the gauge boson propagator, 
the term $\Delta \sigma_{{\rm fl},
3434}$ contains an integral
\be
\calj=\int_0^1 d\omega
\calq^2\left[\ln\frac{\calq^2}{\mu^2}-1\right]
\kma\ee
with 
$\calq^2=\omega m_H^2+(1-\omega)m_W^2+\omega(1-\omega)\bfq^2$. 
Using the same conventions for $\omega_-$ and $\omega_+$ with
$m_a=m_w,m_b=m_H$. It is given by
\bea \nonumber
\calj&=&q^2\left[\left(9\omega_-+9\omega_+-18\omega_-\omega_+-6\right)
\ln(-\omega_-\omega_++\omega_-+\omega_+-1)\right.
\\\nonumber&&
+\left(9\omega_-+9\omega_+-18\omega_-\omega_+-6\right)\ln q^2
\\&&      +3\omega_-^2\left(3\omega_+-\omega_-\right)\ln\frac{1-\omega_-}{-\omega_-}+
\\\nonumber
&&      +3\omega_+^2\left(3\omega_--\omega_+\right)
\ln\frac{\omega_+-1}{\omega_+}
\\\nonumber&&  \left.    -3\omega_-^2+18\omega_-\omega_+-6\omega_--3\omega_+^2
      -6\omega_++4\right]/18
\pkt\eea

\end{appendix}

\newpage
\begin{table}
\begin{center}
\begin{tabular}{|r|r|r|r|r|r|}
\hline
$\xi$& $\Delta \sigma^{(1)}_{\rm fin}$&$\Delta \sigma^{(2)}_{\rm fin}$&
$\Delta\sigma_{\rm sub}$&$\Delta\sigma_{\rm tot}$&$g^2\sigma_{cl}/\pi$
\\
\hline
$0.5$&.244&-.001&-.731&-.488&.75742\\
$0.6$&.204&.002&-.773&-.567&.81306\\
$0.7$&.176&.005&-.795&-.615&.86441\\
$0.8$&.154&.009&-.811&-.649&.91232\\
$0.9$&.135&.014&-.825&-.675&.95737\\
$1.0$&.119&.022&-.837&-.700&1.0000\\
$1.1$&.105&.030&-.848&-.713&1.0405\\
$1.2$&.092&.041&-.859&-.726&1.0792\\
$1.3$&.080&.054&-.870&-.736&1.1163\\
$1.4$&.069&.069&-.881&-.743&1.1518\\
$1.5$&.058&.086&-.891&-.747&1.1860\\
$1.6$&.048&.107&-.902&-.747&1.2190\\
$1.7$&.038&.130&-.913&-.745&1.2509\\
$1.8$&.028&.155&-.923&-.740&1.2818\\
$1.9$&.018&.184&-.934&-.731&1.3116\\
$2.0$&.009&.216&-.944&-.719&1.3406\\
\hline
\end{tabular}
\end{center}
\vspace*{3mm}
\caption{\label{table:finalresults} The correction to the string tension as a function of
$\xi=m_H/m_W$. We present the finite part of the
 contributions with one vertex,
$\Delta \sigma^{(1)}_{\rm fin}$, the finite contributions from
graphs with two vertices $\Delta\sigma^{(2)}_{\rm fin}$, 
the sum of higher order terms
$\Delta\sigma_{\rm sub}$, and the total one-loop correction
$\Delta \sigma_{\rm tot}$. We also include the classical string tensions.
All entries are in units of $m_W^2$.} 
\end{table}


\bibliography{novzpe}
\bibliographystyle{h-physrev4}

\end{document}